\documentclass[aps,twocolumn,pra,superscriptaddress]{revtex4}
\usepackage{epsfig,graphicx,times}
\usepackage{amstext}
\usepackage{amsmath}
\usepackage{lipsum}
\usepackage{comment}
\usepackage{amssymb}
\usepackage{float}
\usepackage{graphicx}
\usepackage{latexsym}
\usepackage{bm}
\usepackage{epstopdf}
\usepackage{appendix}
\usepackage[colorlinks,citecolor=blue, linkcolor=blue,hyperindex,CJKbookmarks]{hyperref}
\newcommand{\Tr}{\text{Tr}}

\begin{document}
\title{A General Quantum Speed Limit for Non-Hermitian Systems}

\author{Zhanxi Wang}
\affiliation{Center for Quantum Sciences and School of Physics, Northeast Normal University, Changchun 130024, China}
\author{Xiaozhe Hao}
\affiliation{Center for Quantum Sciences and School of Physics, Northeast Normal University, Changchun 130024, China}
\author{X. X.  Yi\footnote{yixx@nenu.edu.cn}}
\affiliation{Center for Quantum Sciences and School of Physics, Northeast Normal University, Changchun 130024, China}
\affiliation{Center for Advanced Optoelectronic Functional Materials Research, and Key Laboratory for UV Light-Emitting Materials and Technology of Ministry of Education, Northeast Normal University, Changchun 130024, China}

\date{\today}

\begin{abstract}
The quantum speed limit (QSL) refers to the maximum speed of a quantum system to evolve  from  an initial state to its orthogonal states.  The bound on the QSL for Hermitian systems, for example the  Mandelstam-Tamm (MT) and Margolus-Levitin (ML) as well as Sun-Zheng(SZ) bound, was studied  respectively from the perspectives of average value and  variance of the system Hamiltonian as well as  the geometry  of the system. While the compactness of the MT-type, ML-type and SZ-type bounds has been examined well for Hermitian systems,
a compact QSL for non-Hermitian systems has not been well studied.
In this work, based on the biorthogonal basis
theory we derive two distinct and tighter bounds on the QSL for
non-Hermitian systems, which correspond to the MT and ML bounds for Hermitian systems.  We show that the shortest evolution time corresponding to the two bounds of the non-Hermitian system  can  be attained by certain initial states, showing the
compactness  and tightness of our  bounds. These initial states dubbed fastest initial states(FIS) are different from that in Hermitian systems. A bound close to QSL  for non-FIS is presented and
comparison of our bound with others in literature is performed.
To illustrate our results, we present  a minimal  non-Hermitian system to show  QSL, and the condition for the shortest evolution time is derived
analytically using the present theory.
\end{abstract}
\maketitle

\section{Introduction} 
For a quantum system, the evolution speed of quantum state
is limited, known as the quantum speed limit(QSL), which sets a bound on the time
required for the quantum system to evolve into a state that is orthogonal to
its initial state.
Due to the nature of quantum mechanics, only orthogonal states can
be distinguished, so QSL actually
determines the maximum rate of operation of the system to access
orthogonal states in unit time \cite{Agnew2014, Martinez2023, Cai2024}.
This bound is relevant to various fields, including
quantum computing \cite{Ashhab2012, Basilewitsch2024, Bremermann1962,
Bremermann1967, Caneva2009, Giovannetti2012, Lloyd2002, Lloyd2000,
Deffner2017b}, quantum information \cite{Shanahan2018, delCampo2013}, {machine learning}
\cite{Zhang2018} and others \cite{Wu2014, Bukov2019,
Marvian2015, Sun2019, Okuyama2018}. Recent studies have demonstrated
that the QSL is universal across diverse systems \cite{Cimmarusti2015,
Fogarty2020, Mukhopadhyay2018, Deffner2020}, including ultracold gases
\cite{DelCampo2021}, quantum oscillators at high temperatures
\cite{Deffner2017} and quantum spin systems \cite{Hamma2009}. It  manifests itself in many respects such as quantum many-body system entanglement
\cite{Giovannetti2003, Liu2015, Xu2014}, the  decoherence times \cite{Pires2016, Beau2017},
the quantum representation of the Wigner function \cite{Shanahan2018,
Uzdin2016, Hu2020, Deffner2017, GarciaPintos2019}  and  quantum thermodynamics
\cite{Deffner2010}.
The concept of QSL is also valid and helpful in the study of the evolution of operators \cite{GarciaPintos2022,
Carabba2022, Mohan2022}, superoperators \cite{Hornedal2022},
various classical \cite{Nicholson2020} and non-Hermitian quantum
systems \cite{Hornedal2022}, which has been extensively verified
by  experiments \cite{Ness2021} and applied into
quantum science and technology \cite{Lloyd2000,
Murphy2010}.

The study of QSL was initiated by Mandelstam and Tamm,
who derived the QSL bound in relation to the energy variance of
the system \cite{Mandelstam1945, Fleming1973, Vaidman1992, Anandan1990}.
This bound is known as the MT bound as
$
\Delta E \cdot \tau \geqslant \frac{\pi}{2 },
$
where
$\Delta E^2 = \langle \hat{H}^2 \rangle - \langle \hat{H} \rangle^2$,
and $\tau$ is the time required for a quantum state to evolve into
a state orthogonal to it. This result reveals the fundamental connection
between energy uncertainty and evolution, indicating that the speed of
quantum evolution is fundamentally constrained by energy dispersion.
In 1998, Margolus and Levitin proposed an alternative QSL bound based
on the mean value of Hamiltonian \cite{Margolus1998}, referred to as the ML bound as
$\langle E \rangle\cdot
\tau \geqslant \frac{\pi}{2},
$
where $\langle E \rangle = \langle \hat{H} \rangle$.
In addition to the MT and ML bounds based on the variance and  mean value of the system Hamiltonian,  Sun and Zheng\cite{Sun2019} presented  a distinct bound of the quantum speed limit for  quantum system   by
employing the gauge invariant and geometric natures of quantum mechanics.
These findings emphasize that the  average
energy and energy variance as well as the geometry of quantum system establishes a fundamental constraint on the speed of quantum
evolution \cite{Minganti2019, Ashida2020, Bender2007, Roccati2022,
Mostafazadeh2002b, Bender2002, Carmichael1993}.

Most recently,  significant attention
has also been directed toward QSL for non-Hermitian systems.
Non-Hermitian quantum mechanics incorporates the interaction between
the system and its environment \cite{delCampo2013, Ashida2020, Deffner2013,
Campaioli2018}, making it helpful for describing open
quantum systems \cite{Cao2023}.
To generalize QSL from unitary to  non-unitary evolution governed by non-Hermitian
Hamiltonian, a widely adopted approach involves deriving the QSL
by defining a metric between two quantum states in differential manifold
spaces \cite{Pires2016, Deffner2013, Sun2019}.  In Ref.~\cite{Pires2016},
the authors demonstrate that the QSL family can be derived by
constructing a compact Riemannian metric, revealing that different
metrics lead to distinct QSL bounds. Similarly, in Refs.~\cite{Sun2021b,
Sun2019}, the authors derive a unified result for the ML- and MT-type
bounds through  quantum phase accumulation, and its
specific form for non-Hermitian systems is illustrated in Ref.~\cite{Cui2012}. These studies
show that the bound  remains tight with unitary evolution, but it is not the case for non-unitary evolution.  Furthermore,  researchers have explored
alternative approaches to explore  QSL for non-Hermitian systems,
including the non-Hermitian uncertainty relations \cite{Srivastav2024},
superoperator method \cite{Hornedal2022}, auxiliary system frameworks
\cite{Hasegawa2023}, and perturbation techniques \cite{Yadin2024}. These
investigations reveal that the effects of environment-system couplings, such as non-Markovian
processes \cite{Zhang2015, Liu2016, Ashida2020, Weidemann2021, Impens2021},
can significantly affect the  evolution of quantum system \cite{Deffner2013}.

Although the theory of the QSL for non-Hermitian systems has been explored from
various perspectives, several unresolved issues remain. While the physical
significance of the QSL has been discussed from different
angles \cite{Pires2016, Deffner2013, Sun2019, Sun2021b, Sun2019}, determining
the tightest bound that is attainable  in non-Hermitian systems keeps being a challenge. Specifically, deriving a tight and  practically attainable
QSL bound with appropriate metrics and methods is still difficult
\cite{Pires2016, Campaioli2019}. Moreover  most
studies have focused on extending the MT-type bound, while the ML-type bound
has made less progress for non-Hermitian systems. These results suggest that extending the ML-type bound is interesting and
highly desired.

In this work, we propose a general bound of QSL for non-Hermitian systems, which extended  both the
ML- and MT-type bounds for Hermitian systems.
Specifically, through our generalized ML- and MT-type bounds,
we find that there exists a fastest initial state (FIS)
whose evolution time to its orthogonal state is the
shortest among all possible initial states.
Except the FIS, the initial states that approach closely our ML-type and MT-type bounds are also examined. We claim that the bound can not only be achieved for FIS, but it
remains tight as well, though in this case it is only asymptotically attainable with initial states
in a high-dimensional Hilbert space. To verify and illustrate our theory, we derived  the QSL   and the
shortest evolution times with different parameters for a
non-Hermitian three-level system, and  compare our results with
those in Refs.~\cite{Sun2021b, Sun2019}.

                        
\section{Results}

\subsection{Setup and Assumptions}
We analyze and calculate the QSL of non-Hermitian systems by utilizing
the inner product between the initial and final states in terms of the
non-Hermitian biorthogonal basis.
In non-Hermitian quantum mechanics, the biorthogonal
basis $\{|\psi_n\rangle, |\widetilde{\psi}_n\rangle\}$, defined by the Hamiltonian and its Hermitian
conjugate, can be introduced as the orthonormal basis of the non-Hermitian system
\cite{Wong1967, Jing2024, Sun2022, Mostafazadeh2002}. Here $|\psi_n\rangle$ is the
eigenstate of the non-Hermitian Hamiltonian $\hat{H}$, and $|\widetilde{\psi}_n\rangle$ is the
eigenstate of the Hermitian conjugate $\hat{H}^\dagger$ of $\hat{H}$.
The biorthogonal relation holds:
\begin{align}
    \langle \widetilde{\psi}_n | \psi_m\rangle = \delta_{nm} , \quad \langle \psi_m | \widetilde{\psi}_n\rangle = \delta_{nm},
    \nonumber
\end{align}
while neither $\langle \psi_n | \psi_m\rangle = \delta_{nm}$ nor $\langle \widetilde{\psi}_n | \widetilde{\psi}_m\rangle = \delta_{nm}$ is satisfied.
Consider a non-degenerate non-Hermitian system with Hamiltonian $\hat{H}$,
we can define a new Hamiltonian $\hat{H}' = \hat{H} -
(\omega_{\text{min}} - i\gamma_{\text{min}}) \mathbb{I}$,
 {where $\omega_{\text{min}}$
and $\gamma_{\text{min}}$ are the minimum
real and imaginary parts of the eigenvalues of $\hat{H}$,
respectively.
The shift of Hamiltonian  does not change the
dynamics of the system, which ensures  nonnegativity of  both
the real  and imaginary parts of the eigenvalues
$\omega_n'$ and $\gamma_n'$ of $\hat{H}'$.  We denote the minimum  of  $\omega_n'$ as $\omega_0'$,
obviously $\omega_0'=0$.}
In Refs.~\cite{Wong1967, Brody2013},
the author defines the dual spaces
$\mathcal{V} = \text{Span}(\{|\psi_n\rangle\}_{n=0}^{N})$
and $\widetilde{\mathcal{V}} = \text{Span}(\{|\widetilde{\psi}_n\rangle\}_{n=0}^{N})$
over the complex field $\mathbb{C}$ using the eigenstates $|\psi_n\rangle$
and $|\widetilde{\psi}_n\rangle$ of $\hat{H}'$ and $\hat{H}'^\dagger$ to discuss
the non-Hermitian dynamics of the system. The set $\{|\psi_n\rangle, |\widetilde{\psi}_n\rangle :
n = 0, 1, 2, \ldots, N\}$ forms a complete set of biorthogonal
normalized basis. With this set of basis, we choose the
initial state of the system as
\begin{align}\label{int}
    |\psi(0)\rangle = \sum_{n=0}^{N} c_{n} |\psi_n\rangle,
\end{align}
where $c_{n}$ is a complex coefficient that satisfies
the normalization condition, $\mathcal{K}(0) = \langle
    \tilde{\psi}(0)|\psi(0)\rangle = \sum_{n=0}^{N}
    |c_{n}|^2 = 1$. Here, $|\psi(0)\rangle \in \mathcal{V}$
and $|\tilde{\psi}(0)\rangle \in \widetilde{\mathcal{V}}$.
The biorthogonal basis and model used in this work are
introduced in detail in Appendix \ref{appendix 0}.

\subsection{ML-type Quantum Speed Limit Bound}

Firstly, we consider the ML-type bound,
which is the bound associated with the mean of the non-Hermitian Hamiltonian.
To derive the ML-type bound, we begin by the inner product
$S(t)$ between the initial and final states, as introduced in
\cite{Levitin2009},
\begin{align}
    S(t) = \langle \tilde{\psi}(0) | \psi(t) \rangle
    =\frac{1}{\mathcal{K} (t)} \sum_{n=0}^{N} e^{-i(\omega_n' - i \gamma_n') t} |c_n|^2,\nonumber
\end{align}
where $\mathcal{K} (t)=\langle \tilde{\psi}(t)|\psi(t)\rangle$, whose real and imaginary parts are:
\begin{align}    \label{eq2}
    \text{Re}[S(t)] &= \frac{1}{\mathcal{K} (t)}\sum_{n=0}^{N} e^{-\gamma_n' t} \cos(\omega_n' t) |c_n|^2, \\ \label{eq3}
    \text{Im}[S(t)] &= \frac{1}{\mathcal{K} (t)}\sum_{n=0}^{N} e^{-\gamma_n' t} \sin(\omega_n' t) |c_n|^2.
\end{align}
Throughout this work  we will denote $\tau$ as the time when the system evolves  into a state  orthogonal to its initial one, i.e.,  $S(\tau) = 0$  We denote $\tau$ as the evolution time to the state orthogonal to the initial one, i.e., $S(\tau) = 0$, which requires:
\begin{align}
    \text{Re}[S(\tau)] = 0 \quad \text{and} \quad \text{Im}[S(\tau)] = 0.
    \nonumber
\end{align}

We apply the inequality \cite{Margolus1998}:
\begin{align}    \label{eq4}
    \cos x \geqslant 1 - \frac{2}{\pi}(x + \sin x), \quad x \geq 0,
\end{align}
where equality holds if and only if $x = 0$ or $x = \pi$.
Since $e^{-\gamma_n'\tau} \geqslant 0$, $|c_n|^2 \geqslant 0$ and $\omega_n' \geqslant 0$, we obtain:
\begin{align*}
    \mathcal{K}(\tau)\text{Re}[S(\tau)]
    &= \sum_{n=0}^{N} e^{-\gamma_n'\tau} \cos(\omega_n'\tau) |c_n|^2 \nonumber \\
    &\geqslant \sum_{n=0}^{N} e^{-\gamma_n'\tau} |c_n|^2 \nonumber\\
    &\quad - \frac{2}{\pi}\sum_{n=0}^{N} e^{-\gamma_n'\tau} \omega_n'\tau |c_n|^2 \nonumber\\
    &\quad - \frac{2}{\pi}\sum_{n=0}^{N} e^{-\gamma_n'\tau} \sin(\omega_n'\tau) |c_n|^2.
\end{align*}
Substituting the abbreviations Eq.(\ref{eq4}) and the definition of $\text{Im}[S(\tau)]$ into Eq.(\ref{eq2}), and using $S(\tau)=0$, we derive the ML-type bound: (throughout this work, we set $\hbar=1$)
\begin{align}\label{ml2}
            {
                \mathcal{F}_{\text{ML}} (\tau) = \frac{\Tr [\hat{\Omega } e^{-\hat{\Gamma } \tau} \rho(0)] }{\Tr [e^{-\hat{\Gamma } \tau} \rho(0)]} \cdot \tau\geqslant \frac{\pi}{2},
            }
\end{align}
where $\hat{\Omega } = \sum_{n=0}^{N} \omega_n'| \psi_n \rangle \langle \widetilde{\psi}_n |$ and $\hat{\Gamma } = \sum_{n=0}^{N} \gamma_n' | \psi_n \rangle \langle \widetilde{\psi}_n |$ is the Hermitian and non-Hermitian part of the Hamiltonian $\hat{H}^\prime \equiv \hat{\Omega } - i\hat{\Gamma },$ respectively.
Details of derivation can be found in Appendix \ref{appendix A}.

The ML-type bound of the QSL for non-Hermitian systems given Eq.(\ref{ml2}) is one of the main results in this work.
For Hermitian system, $\gamma_{n}'=0$,
this reduces  to $\langle H^\prime\rangle \cdot \tau\geq \frac {\pi}{2},$ i.e.,
the result in Ref.~\cite{Margolus1998}.
Comparing the bound in Eq.(\ref{ml2}) and the ML bound(i.e., $\langle H^\prime\rangle \cdot \tau\geq \frac {\pi}{2}$), we find that $\mathcal{F}_{\text{ML}} (\tau)$ here play the role of $\langle H^\prime\rangle \cdot \tau$ in Hermitian systems. In fact, $e^{-\hat{\Gamma } \tau} \rho(0)$ behaves like a state at time $\tau$ and $\mathcal{F}_{\text{ML}} (\tau)/\tau$ in this sense is the average of the Hermitian part $\hat{\Omega }$ of the Hamiltonian.

\subsection{Saturation Conditions and Fastest Initial States for ML-type Bound}

For the bound Eq.(\ref{ml2}) to be attained, the following conditions must be satisfied:
\begin{align}
    \omega_n' \tau = 0 \quad \text{or} \quad \omega_n' \tau = \pi.
    \nonumber
\end{align}
For the system evolving to the orthogonal state $S(\tau)=0$, we have:
\begin{align}
    \text{Re}[S(\tau)] &= \frac{1}{\mathcal{K} (\tau)}\sum_{n=0}^{N} e^{-\gamma_n' \tau} \cos(\omega_n' \tau) |c_n|^2 = 0, \nonumber\\
    \text{Im}[S(\tau)] &= \frac{1}{\mathcal{K} (\tau)}\sum_{n=0}^{N} e^{-\gamma_n' \tau} \sin(\omega_n' \tau) |c_n|^2 = 0.
    \nonumber
\end{align}

Assuming $\tau < \pi/\omega_{N}'$ (with $\omega_{0}' \leqslant \omega_{1}' \leqslant \dots \leqslant \omega_{N}'$), we get $\sin(\omega_n' \tau) > 0$ for all $n$, which contradicts $S(\tau)=0$. Thus the minimum evolution time satisfies:
\begin{align}
    \tau_{\text{min}} = \frac{\pi}{\omega_{N}'}.
    \nonumber
\end{align}

For an initial state that is a superposition of three or more eigenstates, we have:
\begin{align}
    \text{Im}\left[S\left(\frac{\pi}{\omega_{N}'}\right)\right]
    &= \frac{1}{\mathcal{K} (\pi/\omega_{N}')} \bigg[
    e^{-\gamma_0' \pi/\omega_{N}'} \sin(0) |c_0|^2 \nonumber\\
    &\quad + e^{-\gamma_1' \pi/\omega_{N}'} \sin\left(\frac{\omega_{1}'\pi}{\omega_{N}'}\right) |c_1|^2 \nonumber\\
    &\quad + \dots + e^{-\gamma_{N}' \pi/\omega_{N}'} \sin(\pi) |c_N|^2
    \bigg] \geqslant 0,
    \nonumber
\end{align}
which implies the fastest initial state (FIS) must be a superposition of only two eigenstates.

For $\tau_{\text{min}}= \pi/\omega_{N}'$, substituting the two-state superposition into $S(\tau)=0$, we get:
\begin{align}
    \text{Im}\left[S\left(\frac{\pi}{\omega_{N}'}\right)\right] &= 0, \nonumber\\
    \text{Re}\left[S\left(\frac{\pi}{\omega_{N}'}\right)\right] &= \frac{e^{-\gamma_0' \pi/\omega_{N}'} |c_0|^2 - e^{-\gamma_1' \pi/\omega_{N}'} |c_1|^2}{\mathcal{K} (\pi/\omega_{N}')} = 0,
    \nonumber
\end{align}
which requires:
\begin{align}
    \frac{|c_0|^2}{|c_1|^2} = e^{(\gamma_0' - \gamma_1') \pi/\omega_{N}'}.
    \nonumber
\end{align}
Therefore, the fastest initial state (FIS) corresponding to the shortest evolution time given by the ML-type bound is
\begin{align}\label{mls}
    |\psi(0)\rangle = \frac{1}{\sqrt{\mathcal{N} }} (\tilde{c}_0|\psi_0\rangle + \tilde{c}_1|\psi_1\rangle),
\end{align}
where $|\psi_n\rangle$ satisfies $\hat{H}'|\psi_n\rangle =  ({\omega}_n'+i{\gamma}_n')|\psi_n\rangle,$ $n=0,1.$
$
    \tilde{c}_n = \text{exp}( \frac{\pi \gamma_n'}{2\omega_1'}),
$
and the normalization constant $\mathcal{N}$ is
$
    \mathcal{N} = \text{exp}( \frac{\pi {\gamma}_0'}{{\omega}_1'}) + \text{exp}( \frac{\pi {\gamma}_1'}{{\omega}_1'}).
$
When the Hamiltonian is Hermitian, namely $\gamma_n=0$, the corresponding result is $|\psi(0)\rangle=(|\psi_0\rangle + |\psi_1\rangle)/\sqrt{2}$,
 {which is exactly  the result in Ref.~\cite{Margolus1998}.}
In Appendix \ref{appendix A1}, we prove that the shortest time of the system is $\pi/\omega_N$, if and only if the initial state of the system is Eq.(\ref{mls}).
In other words, for any initial state of a quantum system, the shortest time $\tau$ that the system evolves to a quantum state orthogonal
to the initial one satisfies $\tau \geqslant \pi/\omega_N$.

\subsection{MT-type Quantum Speed Limit Bound}

Next, we derive a MT-type bounds related to the variance of the non-Hermitian Hamiltonian. We will use a method similar to that  in Ref.\cite{Levitin2009}. The variance of $\hat{\Omega}$ in the initial state is defined as:
\begin{align}
    2\Delta\Omega^2(0)
    &= \langle \psi(0)|\hat{\Omega}^{\dagger}\hat{\Omega}|\psi(0)\rangle + \langle \psi(0)|\hat{\Omega}\hat{\Omega}^{\dagger}|\psi(0)\rangle \nonumber\\
    &\quad - 2\langle \psi(0)|\hat{\Omega}^{\dagger}|\psi(0)\rangle \langle \psi(0)|\hat{\Omega}|\psi(0)\rangle \nonumber\\
    &= \sum_{n=0}^{N} \omega_n'^2 |c_n|^2 + \sum_{m=0}^{N} \omega_m'^2 |c_m|^2 \nonumber\\
    &\quad - 2\sum_{n,m} \omega_n' \omega_m' |c_n|^2 |c_m|^2 \nonumber\\
    &= \sum_{n,m} (\omega_n' - \omega_m')^2 |c_n|^2 |c_m|^2
    \nonumber\\
    &= \sum_{n,m} (\omega_n' - \omega_m')^2 C_{n,m},
    \label{variance_def}
\end{align}
where we used the normalization condition $\sum_{n=0}^{N}|c_n|^2=1$, and $C_{n,m}$ is defined as $C_{n,m}=|c_n|^2 |c_m|^2$.

The  inner product of the initial state and  final state $S(\tau)$ reads,
\begin{align}
    |S(\tau)|^2 = \frac{1}{\mathcal{K}^2 (\tau)}\sum_{n,m=0}^{N} e^{-(\gamma_n' + \gamma_m')\tau} e^{i(\omega_n' - \omega_m')\tau} C_{n,m}. \nonumber
\end{align}
Using the symmetry of the summation, we simplify it to:
\begin{align}
     |S(\tau)|^2 = \frac{1}{\mathcal{K}^2 (\tau)}\sum_{n,m} e^{-(\gamma_n' + \gamma_m')t} \cos\left[(\omega_n' - \omega_m')\tau\right] C_{n,m}.
    \nonumber
\end{align}
 {Applying the inequality \cite{Levitin2009} for any nonnegative number $x$:
\begin{align*}
    \cos x \geqslant 1 - \frac{4}{\pi^2} x \sin x - \frac{2}{\pi^2} x^2,
\end{align*}
where the equality holds if and only if $x = 0$ or $x = \pm \pi$.  Considering  $e^{-(\gamma_n' + \gamma_m')\tau}\geqslant 0$ and} $C_{n,m}\geqslant 0$, we get:
\begin{align}\label{mt_ineq_step}
    |S(\tau)|^2
    &\geqslant \frac{1}{\mathcal{K}^2 (t)}\bigg\{\sum_{n,m} e^{-(\gamma_n' + \gamma_m')\tau} C_{n,m} \nonumber\\
    &\quad - \frac{4}{\pi^2} \sum_{n,m} e^{-(\gamma_n' + \gamma_m')\tau} (\omega_n' - \omega_m') \tau \nonumber\\
    &\qquad \times \sin\left[(\omega_n' - \omega_m')\tau\right] C_{n,m} \nonumber\\
    &\quad - \frac{2}{\pi^2} \sum_{n,m} e^{-(\gamma_n' + \gamma_m')\tau} (\omega_n' - \omega_m')^2 \tau^2 C_{n,m}\bigg\}.
\end{align}

For the second term, we use the fact that $d(|S(t)|^2·\mathcal{K}^2 (t))/dt = 0$ when $|S(\tau)|^2=0$, where the derivative is:
\begin{align}
    \frac{d(|S(t)|^2·\mathcal{K}^2 (t))}{dt}
    = -&\sum_{n,m} (\gamma_n' + \gamma_m') e^{-(\gamma_n' + \gamma_m')t} \nonumber\\
    &\times\cos\left[(\omega_n' - \omega_m')t\right] C_{n,m} \nonumber\\
    \quad - &\sum_{n,m} (\omega_n' - \omega_m') e^{-(\gamma_n' + \gamma_m')t} \nonumber\\
    &\times\sin\left[(\omega_n' - \omega_m')t\right] C_{n,m}.
    \nonumber\\
    = &0\nonumber\\
\end{align}

We define the auxiliary quantity:
\begin{align}
    \mathcal{R}(\tau)
    &= 4 \sum_{n=0}^{N} \mathrm{Tr}\left[\hat{\Gamma } e^{-2\hat{\Gamma } \tau}\mathcal{X}_{n,n}\rho(0)\right] \mathrm{Tr}\left[\mathcal{X}_{n,n}\rho(0)\right] \nonumber\\
    &\quad - 2\mathrm{Tr}\left[\hat{\Gamma } e^{-\hat{\Gamma } \tau}\rho(0)\right]\mathrm{Tr}\left[e^{-\hat{\Gamma } \tau}\rho(0)\right],
    \label{R_def}
\end{align}
where $\mathcal{X}_{n,n} =  |\psi_n\rangle \langle \widetilde{\psi}_n|$ is the projection operator.
For the third term in Eq.(\ref{mt_ineq_step}), using the variance definition Eq.(\ref{variance_def}) and the sorted imaginary parts $\gamma_{\text{min}} = \gamma_{\phi(0)}' \leqslant \gamma_{\phi(1)}' \leqslant \dots \leqslant \gamma_{\phi(N)}' = \gamma_{\text{max}}$, we get:
\begin{align}
      \sum_{n,m} e^{-(\gamma_n' + \gamma_m')\tau} (\omega_n' - \omega_m')^2 \tau^2 C_{n,m}
    \leqslant  2 \Delta\Omega^2(0) \tau^2 e^{-\gamma_{\phi(1)}' \tau}.
    \nonumber
\end{align}

we find the time $\tau$ for the initial state $\rho(0)$ to evolve into a  state orthogonal to it,
\begin{align}\label{mt2}
    {
\mathcal{F}_{\text{MT}} (\tau) =
\frac
{\sqrt{
        \mathcal{R} (\tau)\tau
    +[{      \Delta \hat{\Omega } ^2(0) \tau^2 }]e^{-\gamma_{\phi (1)}'\tau}
}}
{
    \Tr[e^{-\hat{\Gamma } \tau} \rho(0)]
}
 \geqslant \frac{\pi}{2},
    }
\end{align}
where $ \mathcal{R}(\tau) $=$4\sum_{n}{\Tr[\hat{\Gamma } e^{-2\hat{\Gamma } \tau}\mathcal{X}_{n,n}\rho(0)] \Tr[\mathcal{X}_{n,n}\rho(0)]}
-2\Tr[\hat{\Gamma } e^{-\hat{\Gamma } \tau}\rho(0)]\Tr[e^{-\hat{\Gamma } \tau}\rho(0)],
$
with  $\mathcal{X}_{n,n} =  |\psi_n\rangle \langle \widetilde{\psi}_n|$,
$\gamma_{\text{min}}'=
\gamma_{\phi(0)}' \leqslant \gamma_{\phi(1)}' \leqslant \cdots
\leqslant \gamma_{\phi(N)}'= \gamma_{\text{max}}'$
 are the  imaginary parts of the eigenvalue of $\hat{H}$ in increasing order.
This is the bound of MT-type QSL given in this work. For details of the derivation, we  refer to Appendix \ref{appendix MT}.
$\mathcal{F}_{\text{MT}} (\tau)$ defined in Eq. (\ref{mt2}) plays the role of $\Delta E \cdot \tau$ in Hermitian systems, and it returns to the Hermitian one with $\hat{\Gamma}=0.$

\subsection{Saturation Conditions for MT-type Bound}
The results show that, similar to ML- and MT-type bounds of Hermitian system, QSL bounds for non-Hermitian systems are also restricted by mean and variance of the Hamiltonian.
The  the bound of MT-type QSL is saturated with
\begin{align*}
     (\omega_n' - \omega_m')\tau = 0 \quad \text{or} \quad  (\omega_n' - \omega_m')\tau = \pm \pi.
\end{align*}
The initial state that can satisfy the constraint condition is also Eq.(\ref{mls}), which means that under the condition of Eq.(\ref{mls}),
 {the shortest evolution time given by ML- and MT-type bounds is the same.}
For  Hermitian systems, namely $\gamma_n=0$, the bound of MT-type in this work is {consistent with} Ref.~\cite{Mandelstam1945}.

\subsection{The derivation of weak ML and MT-type bounds}\label{appendix E}

Now we discuss in details the ML-type bounds. We start from the ML-type bound given in Eq.(\ref{ml2}):
\begin{align}
    \mathcal{F}_{\text{ML}} (\tau) = \frac{\Tr [\hat{\Omega } e^{-\hat{\Gamma } \tau} \rho(0)] }{\Tr [e^{-\hat{\Gamma } \tau} \rho(0)]} \cdot \tau\geqslant \frac{\pi}{2}.
    \nonumber
\end{align}

For $\Tr [\hat{\Omega } e^{-\hat{\Gamma } \tau} \rho(0)] \tau$, we have:
\begin{align}
    \Tr [\hat{\Omega } e^{-\hat{\Gamma } \tau} \rho(0)] \tau
    &= \sum_{n=0}^{N} e^{-\gamma_n'\tau} \left[ \omega_n'\tau \right] \left| c_n \right|^2
    \nonumber\\
    &\leqslant e^{-\gamma_\text{min}'\tau}\sum_{n=0}^{N}  \omega_n'\tau\left| c_n \right|^2
    \nonumber\\
    &= \langle \hat{\Omega } (0)\rangle \tau,
    \nonumber
\end{align}
where the non-Hermitian biorthogonal normalization relation $\sum_{n=0}^{N}  \left| c_n \right|^2=1$ and $\gamma_{\text{min}}'=0$ is used. Here $\langle \hat{\Omega } (0)\rangle$ denotes the mean value of $\hat{\Omega } =\sum_{n=0}^{N} \omega_n'| \psi_n \rangle \langle \widetilde{\psi}_n |$ in the initial state Eq.(\ref{int}).

For $\Tr [e^{-\hat{\Gamma } \tau} \rho(0)]$, we have:
\begin{align}
    \Tr [e^{-\hat{\Gamma } \tau} \rho(0)] = \sum_{n=0}^{N} e^{-\gamma_n'\tau} \left| c_n \right|^2 \geqslant  e^{-\gamma_\text{max}'\tau},
    \nonumber
\end{align}
where $\gamma_{\text{max}}'$ denotes the maximum imaginary parts of the eigenvalues.

Considering $\gamma_{\text{max}}'\geq \gamma_n'$ for all $n$, we can simplify the ML-type bound to a weak ML-type bound:
\begin{align}\label{wml}
    \mathcal{F}_{\text{wML}} (\tau) = \frac{\langle \hat{\Omega } (0)\rangle }{e^{-\gamma_{\text{max}}'\tau}}\cdot \tau\geqslant \frac{\pi}{2}.
\end{align}

Collecting all these together, we have:
\begin{align}
    \mathcal{F}_{\text{wML}} (\tau) \geqslant \mathcal{F}_{\text{ML}} (\tau) \geqslant \frac{\pi}{2}.
    \nonumber
\end{align}
The equality in Eq.(\ref{wml}) holds if and only if $\gamma_{\text{min}}' = \gamma_{\text{max}}'$, i.e., $\gamma_n = 0$ for all $n$, and the initial state is $|\psi(0)\rangle = \frac{1}{\sqrt{2}} (|\psi_0\rangle + |\psi_1\rangle)$, where $|\psi_n\rangle$ is the eigenstate of the Hamiltonian $\hat{H}$ and satisfies $\hat{H}|\psi_n\rangle =nE_1|\psi_n\rangle$, $n=0,1$.

Similar to the weak ML-type bound, we now derive the weak MT-type bound. We start from the MT-type bound given in Eq.(\ref{mt2}):
\begin{align}
    \mathcal{F}_{\text{MT}} (\tau) =
    \frac
    {\sqrt{
            \mathcal{R} (\tau)\tau
        +[{      \Delta \hat{\Omega } ^2(0) \tau^2 }]e^{-\gamma_{\phi (1)}'\tau}
    }}
    {
        \Tr[e^{-\hat{\Gamma } \tau} \rho(0)]
    }
     \geqslant \frac{\pi}{2},
    \nonumber
\end{align}
where
\begin{align}
    \mathcal{R}(\tau)
    &= 4 \sum_{n=0}^{N} \mathrm{Tr}\left[\hat{\Gamma } e^{-2\hat{\Gamma } \tau}\mathcal{X}_{n,n}\rho(0)\right] \mathrm{Tr}\left[\mathcal{X}_{n,n}\rho(0)\right] \nonumber\\
    &\quad - 2\mathrm{Tr}\left[\hat{\Gamma } e^{-\hat{\Gamma } \tau}\rho(0)\right]\mathrm{Tr}\left[e^{-\hat{\Gamma } \tau}\rho(0)\right],
    \nonumber
\end{align}
with $\mathcal{X}_{n,n} =  |\psi_n\rangle \langle \widetilde{\psi}_n|$ being the projection operator.

For $\mathcal{R}(\tau)$, we have:
\begin{align*}
    \mathcal{R}(\tau)
    =& 4  \sum_{n}{\Tr[\hat{\Gamma } e^{-2\hat{\Gamma } \tau}\mathcal{X}_{n,n}\rho(0)] \Tr[\mathcal{X}_{n,n}\rho(0)]}\\
     &- 2\Tr[\hat{\Gamma } e^{-\hat{\Gamma } \tau}\rho(0)]\Tr[ e^{-\hat{\Gamma } \tau}\rho(0)]   \nonumber\\
    =& \frac{16}{\pi^2} \sum_{n=0}^{N} \gamma_n'  e^{-2\gamma_n'\tau} c_n^4 \\
    &- \frac{8}{\pi^2} \sum_{\substack{ n, m = 0}}^{N} \gamma_{\text{n}}' e^{-(\gamma_{\text{n}}'+\gamma_{\text{m}}')\tau} C_{n,m} \nonumber\\
    \geqslant &-\frac{8}{\pi^2} \gamma_{\text{max}}'.
    \nonumber
\end{align*}

For the third term in the MT-type bound, using the sorted imaginary parts $\gamma_{\text{min}}' = \gamma_{\phi(0)}' \leqslant \gamma_{\phi(1)}' \leqslant \dots \leqslant \gamma_{\phi(N)}' = \gamma_{\text{max}}'$, we have:
\begin{align}
    \sum_{n,m} e^{-(\gamma_n' + \gamma_m')\tau} (\omega_n' - \omega_m')^2 \tau^2 C_{n,m}
    \leqslant  2 \Delta\Omega^2(0) \tau^2 e^{-\gamma_{\phi(1)}' \tau}.
    \nonumber
\end{align}

Considering $\gamma_{\text{max}}'\geq \gamma_n'$ for all $n$, the MT-type bound in Eq.(\ref{mt2}) reduces to:
\begin{align}\label{WMTI}
    \mathcal{F}_{\text{wMT}} (\tau) =\frac{\sqrt{\Delta \hat{\Omega }^{2}(0)
    \tau^2 e^{-\gamma_{\phi (1)}'\tau}- 2\gamma_{\text{max}}'\tau}}
    {e^{-\gamma_{\text{max}}'\tau}}  \geqslant \frac{\pi}{2}.
\end{align}

The relation with the bound of MT-type QSL is:
\begin{align}
    \mathcal{F}_{\text{wMT}} (\tau) \geqslant \mathcal{F}_{\text{MT}} (\tau) \geqslant \frac{\pi}{2}.
    \nonumber
\end{align}
The equality in Eq.(\ref{WMTI}) holds for the same condition as that of Eq.(\ref{wml}). Details can be found in  Appendix \ref{appendix E}.


\section{Approaching to the bounds $\mathcal{F}_{\text{ML}} (\tau) / \mathcal{F}_{\text{MT}} (\tau) $}
So far, we have derived and discussed  the ML- and MT-type
bounds in non-Hermitian systems.
Now we focus on  the other widely discussed issue in QSL studies \cite{Margolus1998,Sun2021b,Sun2019}:
how $\tau$ approaches the combined bound defined by $\tau_{\text{comb}}=\max\{\tau_{\text{ML}},\tau_{\text{MT}}\}$
for different $\tau_{\text{ML}}/\tau_{\text{MT}}$ in  non-Hermitian systems.
As proved,  $\tau_{\text{ML}}/\tau_{\text{MT}}=1$ is attainable
with initial state given in Eq.(\ref{mls}). In this case,
$
    \mathcal{F}_{\text{ML}} (\tau) = \mathcal{F}_{\text{MT}} (\tau) = \frac{\pi}{2}.
$
Namely, the evolution time $\tau$ is $\tau_{\text{comb}}$  with this initial state.
For the other initial states of the two-level system  with eigenvalues $E_0 = 0$ and
$E_1 = \omega +i\gamma$,  the evolution time $\tau$ should satisfy
$
    \text{Im}[\text{S}(\tau)] = e^{-\gamma \tau} \sin (\omega \tau) |c_1|^2 = 0
$,
such that the system can evolve from an initial state to its orthogonal states.
Here $S(\tau) = \langle \psi(0)|\psi(\tau)\rangle$. It is easy to find that,
$
    \tau = \frac{k\pi}{\omega} , \quad k \in \mathbb{Z}_{+}. \nonumber
$
Since these times are not continuous, any two-level system is not the system of which   $\tau_{\text{comb}}$ is very close to the bound given in the last section. Hence, the simplest system to study how an initial state approaching the FIS  is a non-Hermitian three-level system.


For Hermitian systems, Ref. \cite{Levitin2009} introduced a ratio $\alpha=\frac{\Delta E}{E}$ to characterize the attainability of the bound. For non-Hermitian systems, the ratio defined in this way no longer works, instead we define $\alpha = \mathcal{F}_{\text{ML}} (\tau) / \mathcal{F}_{\text{MT}} (\tau)$. For $\alpha < 1$, simple calculation yields
$
\text{Im}[S(\tau)] = p_1 e^{y_1} \sin x_1 + p_2 e^{y_2} \sin x_2 = 0
$ and $
\text{Re}[S(\tau)] = p_0 e^{y_0} + p_1 e^{y_1} \cos x_1 + p_2 e^{y_2} \cos x_2 = 0,
$
where $x_n=\omega_n \tau$ and $y_n=\gamma_n \tau$, $\tau$ denotes the evolution time for the system from an initial state to its orthogonal states. Without loss of generality,  we assume   $\gamma_0 > \gamma_2 >\gamma_1$,
and $x_2 = \pi + x_1 - \delta \sin x_1$.
For an general  initial state $|\psi(0)\rangle = c_0 |E_0\rangle
 + c_1|E_1\rangle + c_2|E_2\rangle$  with  $p_n=|c_n|^2$ standing for the population,
 $|E_n\rangle$  the eigenstates of the Hamiltonian $\hat{H}$,
 $\omega_n$ and $\gamma_n$   the real and imaginary parts of eigenvalues $\lambda_n = \omega_n + i\gamma_n$,
we obtain
$
    p_0 = \delta \mathcal{L}  e^{y_2 - y_0} + \mathcal{O}(\delta^2) ,
$ $
    p_1 = \mathcal{L}{e^{y_2}} - \delta \mathcal{L}^2 {e^{y_2} (\cos x_1 + e^{y_2 - y_0})} + \mathcal{O}(\delta^2),
$ and $
    p_2 = \mathcal{L}{e^{y_2}} - \delta \mathcal{L}^2 {e^{y_2 - y_0} - e^{y_2} \cos x_1} + \mathcal{O}(\delta^2),
$
with $\mathcal{L} ^{-1}={1 + e^{y_2}}$.
   $\alpha = \mathcal{F}_{\text{ML}} (\tau)/\mathcal{F}_{\text{MT}} (\tau)<1$ leads to
$
    \alpha = \frac{\pi}{\pi + 2x_1} < 1.
$
This means $x_1 = \pi(1 - \alpha)/(2\alpha)$. Hence, we have,
\begin{align*}
    \mathcal{F}_{\text{MT}}(\tau) = \frac{\pi}{2} + \mathcal{B}(\tau)\cdot\delta + \mathcal{O}(\delta^2).
\end{align*}
Notice the MT-type bound can only be attainable  with Eq.(\ref{mls}) and

$\mathcal{B}(\tau)>0$. Here $\mathcal{B}(\tau)$ is defined as

$
\mathcal{B}(\tau) = \frac{1}{2(1 - a(\tau))} \Biggl\{
a(\tau) \Biggl[ -\frac{(\gamma_0 - \gamma_2) \tau e_2(\tau)}{\pi} $$
             - \frac{(\gamma_0 + \gamma_2) \tau e_0(\tau)}{\pi} $$
             + \frac{\pi e_0(\tau)}{4} K - S \Biggr] $$
+ b(\tau) \Biggl[ -\frac{2 \gamma_2 \tau [e_2(\tau)]^2 e_0(\tau)}{\pi}$$
              - \pi e_2(\tau) e_0(\tau) $$
              - \frac{\pi (1 - e_2(\tau)) C}{2} \Biggr] \Biggr\},
$
with
$a(\tau) = (1 + e^{\gamma_2 \tau})^{-1}$,
$b(\tau) = [a(\tau)]^2 $,
$e_0(\tau) = e^{-\gamma_0 \tau},$
$e_2(\tau) = e^{\gamma_2 \tau} $,
$\theta = {\pi (1 - \alpha)}/{2\alpha}, $
$C = \cos \theta,$
$S = \sin \theta, $ and
$K = 1 + {\alpha^{-2}}.$

For any attainable  $\tau$, it satisfies $\mathcal{F}_{\text{MT}}(\tau) \geqslant \frac{\pi}{2}$. Here the equality holds if and only if  the initial state is given by Eq.(\ref{mls}).
For  $\gamma_0, \gamma_2 > 0$ and finite $\tau$, $\mathcal{B}(\tau) > 0$. Defining  $\mathcal{B} \equiv \inf_\tau \mathcal{B}(\tau) > 0$ and   choosing $\delta < \frac{\pi}{2} \epsilon \mathcal{B}^{-1}$ for any $\epsilon > 0$, we arrive at
\[
\frac{\pi}{2} < \mathcal{F}_{\text{MT}}(\tau) \leqslant \frac{\pi}{2}(1 + \epsilon).
\]
Details of derivation can be found in Appendix \ref{appendix f}.
This result suggests  that for $\mathcal{F}_{\text{MT}}(\tau) > \mathcal{F}_{\text{ML}}(\tau)$,
the MT-type bound can be infinitely close to $\frac{\pi}{2}$ with certain initial states. In other words, the MT-type bound is tight though it in this case is not attainable: the system can approach   arbitrarily close the bound when the parameter in the initial stats approaches its limit.

For $\alpha > 1$, we consider  initial states $|\psi(0)\rangle = c_0 |E_0\rangle
 + c_1|E_1\rangle + c_{2k+1}|E_{2k+1}\rangle$, which might meet the requirement that
 $S(\tau)=\langle \widetilde{\psi}(0)|\psi(t)\rangle=0$.  Simple algebra yields,
$
\text{Im}[S(\tau)] = p_1 e^{y_1} \sin x_1 + p_{2k+1} e^{y_{2k+1}} \sin x_{2k+1} = 0  $ and $
\text{Re}[S(\tau)] = p_0 e^{y_0} + p_1 e^{y_1} \cos x_1 + p_{2k+1} e^{y_{2k+1}} \cos x_{2k+1} = 0,
$
with   $y_n = \gamma_n \tau$,  $0 = \gamma_0 < \gamma_1 < \gamma_{2k+1}$ and the normalization $p_0 + p_1 + p_{2k+1} = 1$.
Setting $x_0 = 0$, $x_1 = \pi$, and $x_{2k+1} = (2k+1)\pi$ could not change the dynamics of the system, we find from $\text{Im}[S(\tau)] = 0$ that
$
p_0 = \mathcal{N} ^{-1} \left[ \left(1 - ({\beta}/{k^2})\right) e^{\gamma_1 \tau} + ({\beta}/{k^2}) e^{\gamma_{2k+1} \tau} \right],$ $
 p_1 = \mathcal{N} ^{-1} \left(1 - ({\beta}/{k^2})\right),$ and $
 p_{2k+1} = \mathcal{N} ^{-1} \left( ({\beta}/{k^2}) \right).
$
Here $\mathcal{N}  = \left(1 - ({\beta}/{k^2})\right) (e^{\gamma_1 \tau} + 1) + ({\beta}/{k^2}) (e^{\gamma_{2k+1} \tau} + 1)$ is the normalization constant, and  the imaginary part $y_1 = \gamma_1 \tau$, $y_{2k+1} = \gamma_{2k+1} \tau$ with $0 = \gamma_0 < \gamma_1 < \gamma_{2k+1}$. Straightforward calculation  yields,
$
\alpha = { \sqrt{1 + 4\beta (1 + e^{-\gamma_1 \tau})} }{( 1 + ({2\beta}/{k}) e^{(\gamma_{2k+1} - \gamma_1) \tau} )^{-1}}.
$
Solving the above equations, we find $\beta,$ namely,
$
    \beta = \frac{1}{4}{(\alpha^2 - 1)}{\left(1 + e^{-\gamma_1 \tau} - ({\alpha^2}/{k}) e^{(\gamma_{2k+1} - \gamma_1) \tau}\right)^{-1}} + \mathcal{O}\left({1}/{k^2}\right).
$
Finally,  we find $\mathcal{F}_{\mathrm{ML}}(\tau)$
\begin{align*}
\mathcal{F}_{\mathrm{ML}}(\tau) = \dfrac{\pi}{2} + \dfrac{\pi (\alpha^2 - 1)}{4k (1 + e^{-\gamma_1 \tau})} e^{(\gamma_{2k+1} - \gamma_1) \tau} + \mathcal{O}\left({1}/{k^2}\right).
\end{align*}
Neglecting $\mathcal{O}\left({1}/{k^2}\right)$ terms, the ML-type bounds can be written as
$
\mathcal{F}_{\mathrm{ML}}(\tau) = \frac{\pi}{2} + \frac{1}{k} \mathcal{B}(\tau) + \mathcal{O}\left(\frac{1}{k^2}\right),
$
where $\mathcal{B}(\tau) = {\pi (\alpha^2 - 1) e^{(\gamma_{2k+1} - \gamma_1) \tau}}/{4 (1 + e^{-\gamma_1 \tau})} $ and $\mathcal{B} \equiv \inf_{\tau} \mathcal{B}(\tau) > 0$ is a positive constant independent of time. For any $\epsilon > 0$ and  choosing  $k > 2(\pi\epsilon)^{-1} \mathcal{B}$, we arrive at
\begin{align*}
\frac{\pi}{2} < \mathcal{F}_{\mathrm{ML}}(\tau) \leqslant \frac{\pi}{2}(1+\epsilon).
\end{align*}
The lower bound follows directly from $\mathcal{B} > 0$ and its asymptotic expansion in powers of $k$, while the upper bound is guaranteed by the exponential decay of $\mathcal{B}(\tau)$ and the choice of $k$. Details can be found in Appendix \ref{appendix G}.

We thus claim that for $\alpha = 1$,
there is an initial state $|\psi(0)\rangle$ such that
$\mathcal{F}_{\mathrm{ML}}(\tau)$ and $\mathcal{F}_{\mathrm{MT}}(\tau)$
are equal. But for $\alpha \neq 1$,  although there is no initial state
which can make the bound attainable, the QSL can approach arbitrarily close to the bound. For $\alpha < 1$ and
any $\epsilon >0$, the MT-type bound is tighter than that of the ML-type,
and there is an initial state $|\psi(0)\rangle$ such
that $\frac{\pi}{2} < \mathcal{F}_{\mathrm{MT}}(\tau)
\leqslant \frac{\pi}{2}(1+\epsilon)$. Whereas for $\alpha > 1$
and any $\epsilon >0$, the ML-type bound is tight more than the MT-type,
and there is an initial state $|\psi(0)\rangle$ such
that $\frac{\pi}{2} < \mathcal{F}_{\mathrm{ML}}(\tau)
\leqslant \frac{\pi}{2}(1+\epsilon)$.
These results suggest that while no initial state achieves $\tau_{\text{comb}}$,
one can always find an initial state making $\tau$ arbitrarily close to $\tau_{\text{comb}}$.

\section{Comparison of Different  Bounds}
We will employ a two-level system to compare our QSL bounds with that in the earlier publications.
In Appendix \ref{appendix H}, the  general two-level non-Hermitian system can be described by,
\begin{align}
    \hat{H} = \xi \begin{pmatrix}
        \cos\vartheta  e^{i\varrho} & \sin\varrho \sin \beta \\
        \sin\varrho \cos \beta & -\cos\vartheta e^{i\varrho}
     \end{pmatrix}   \nonumber,
\end{align}
where $\xi\in(0, +\infty )$,
parameter $\varrho , \beta\in[0 , 2 \pi]$, and $\vartheta\in[0 , \pi]$.
By shifting  the Hamiltonian $\hat{H}$ to $\hat{H}_{t}$ by $\hat{H}_{t} = \hat{H} - \kappa \mathcal{I} $ with $\kappa = - \xi   \sqrt{ \cos^2 \vartheta  e^{i 2 \varrho} + \frac{1}{2} \sin^2 \varrho \sin 2\beta },$
its eigenvalues can be reexpressed as $\lambda_{-} = 0$, $\lambda_{+} = -2\kappa = \mu +i \nu$.
Here $\mu = \text{Re}[\lambda_+]$, $\nu = \text{Im}[\lambda_+]$.
We consider a general initial state
$|\psi(0) \rangle = \cos \alpha|\lambda_{-} \rangle +\sin \alpha e^{i\phi}|\lambda_{+} \rangle $,
with $\alpha \in (0 , \frac{\pi}{2}]$,  $\phi \in (0 , 2\pi]$, and denote
 $\Omega = \mu |\lambda_+\rangle \langle\lambda_+ |$ and $\Gamma = \nu |\lambda_+\rangle \langle\lambda_+ |$.
After simple algebra,  Eq.(\ref{ml2}) follows,
\begin{align}
    \mathcal{F}_{\text{ML}} (t) = \frac{\tan^2 \alpha \mu t}{e^{\nu t}+ \tan^2 \alpha}.
    \nonumber
\end{align}
The QSL bound determined by $\mathcal{F}_{\text{ML}} (\tau_{\text{ML}}) = \frac{\pi}{2}$.
The  variance $\Delta \hat{\Omega } = \mu$,
$\mathcal{R} (t)$ defined in Eq.(\ref{mt2}) can be expressed as
$\mathcal{R} (t) =2 \nu t  e^{-2\nu t} \sin^4 \alpha
                -2 \nu t  e^{-\nu t} \sin^2 \alpha  \cos^2 \alpha$.

For the MT-type bound, the left-hand side of Eq.(\ref{mt2}) for this model becomes
\begin{align}
    \mathcal{F}_{\text{MT}} (t)= \frac
    {\sqrt{
            -2 \nu t  \tan^4 \alpha
            +(2 \nu t
        +\mu^2 t^2   )e^{\nu t}\tan^2 \alpha
    }}
    {
        e^{\nu t} +  \tan^2 \alpha
    }.\nonumber
\end{align}
The MT-type bound $\tau_{\text{MT}}$ is determined by  $\mathcal{F}_{\text{MT}} (\tau_{\text{MT}})=\frac{1}{2}\pi.$ The  two inequalities, $\mathcal{F}_{\text{MT}}(\tau_{\text{MT}})\geq \frac{\pi}{2}$ and $\mathcal{F}_{\text{ML}}(\tau_{\text{ML}})\geq \frac{\pi}{2}$  together  give the lower bound of the time for the system to evolve
to  states orthogonal to the initial ones. The final bound should  satisfy $\tau \geqslant \text{max}\{ \tau_{\text{ML}} , \tau_{\text{MT}}\} $.
The ML-type bound $\tau_{\text{ML}}$ and the MT-type bound $\tau_{\text{MT}}$ are often
different for different parameters and initial states. In the following,  we analyze the optimal
initial state leading to the fastest evolution of the system and explore the feature for the initial states nearby the optimal one.
For the initial state leading to the fastest evolution speed of the system,
the following conditions need to satisfy (see Eq.(\ref{mls})),
$
|c_0|^2/ |c_1|^2 = \tan^2 \alpha = e^{\nu \tau},
$
where $\tau = \frac{\pi}{\mu}$ is the shortest evolution time.
With this initial state, $\mathcal{F}_{\text{ML}} (t)$ and $\mathcal{F}_{\text{MT}} (t)$ satisfy,
$
    \mathcal{F}_{\text{ML}} (\tau) = \mathcal{F}_{\text{MT}} (\tau) = \frac{\pi}{2}.\nonumber
$
This means that $\mathcal{F}_{\text{ML}} (t)$ and $\mathcal{F}_{\text{MT}} (t)$ can
simultaneously obtain the shortest evolution time with the same fastest
initial state.

In the last section, we focus on the analysis of
the relation between $\tau_{\text{ML}}$ and
$\tau_{\text{MT}}$ near the fastest  initial
state. It is ready  to  compare the
 the combination bound
$\tau_{\text{comb}} = \text{max} \{ \tau_{\text{ML}},\tau_{\text{MT}}\} $
with that in earlier publications, for example in
Refs.~\cite{Sun2021b,Sun2019}, where the author gives
a special quantum velocity limit via
the geometric phase $\varphi _{g}.$
Solving $\varphi _{\text{dyn}} = 0$,
the QSL bound given in Refs.~\cite{Sun2021b,Sun2019} follows,
$
\tau_{\text{G}} \geqslant \frac{\pi}{2\frac{1}{\tau}\int_{0}^{\tau} |\frac{d\phi}{dt}| \,dt }.
$
For a general two-level non-Hermitian Hamiltonian, the
 QSL bound can be expressed as,
\begin{align}    \label{eq7}
    \mathcal{F}_{\text{G}} (t)=\frac{\sqrt{\mu^2+\nu^2} }{\nu }
    \left(\arctan \left( e^{\nu t} \tan \alpha \right)-\alpha  \right),  
\end{align}
where $\tau_{\text{G}}$ is the solution of
$\mathcal{F}_{\text{G}} (t) =\frac{\pi}{2}$.
Details can be found in Appendix \ref{appendix I}.

\begin{figure}[h]
    \centering
    \includegraphics[width=\linewidth]{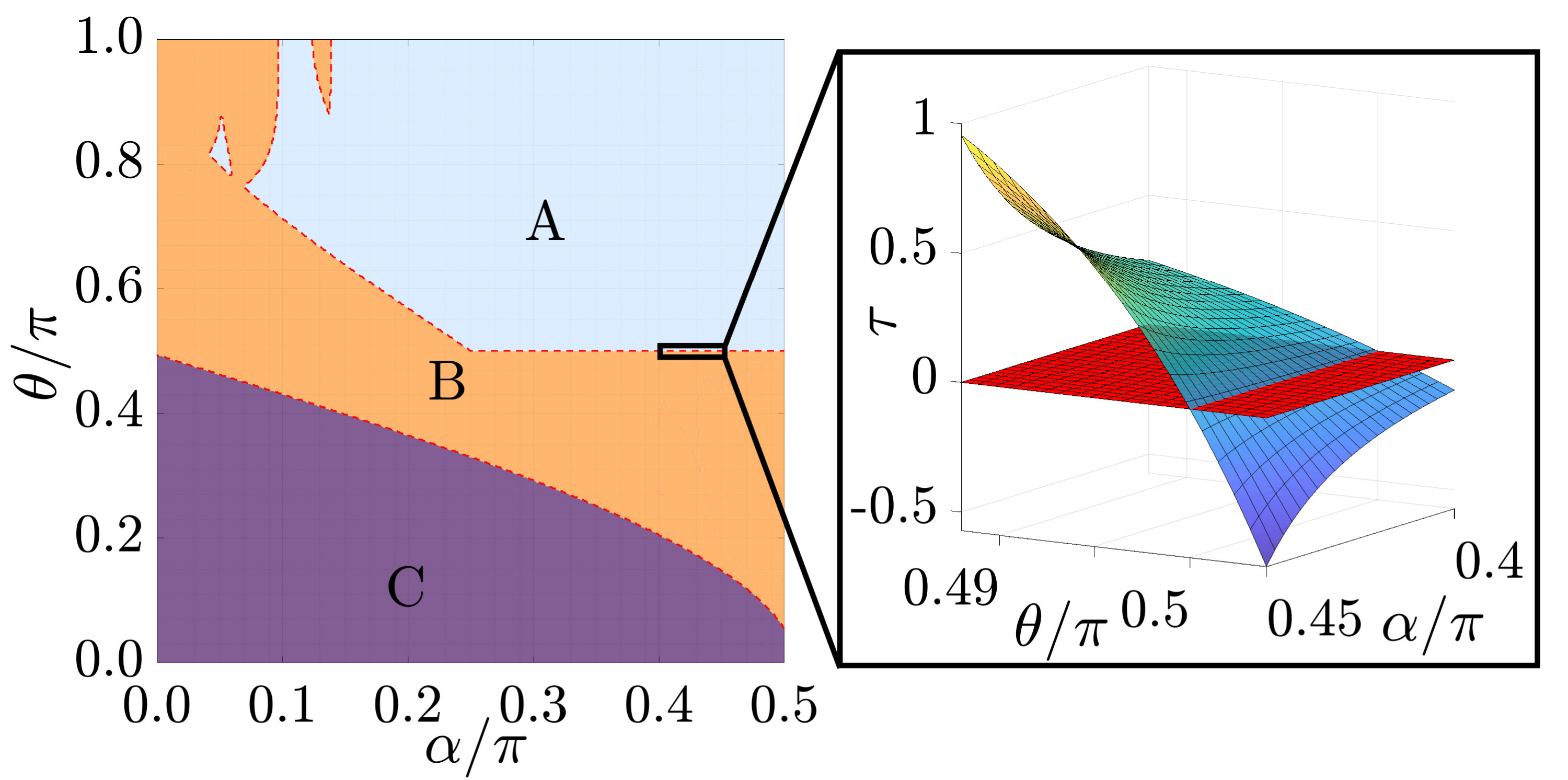}
    \caption{Left panel: $\Delta \tau' = \tau_{\text{comb}}' - \tau_{\text{G}}'$ versus $\theta$ and $\alpha$.
    Region A (blue) is for $\tau_{\text{comb}}' < \tau_{\text{G}}'$,
    region B (orange) is for $\tau_{\text{comb}}' > \tau_{\text{G}}'$,
    and region C (purple) is for $\tau_{\text{comb}}'$ and $\tau_{\text{G}}'$ having no solution. Right panel: 
    $\Delta \tau'$ with $\theta \in [0.485\pi, 0.505\pi]$ and $\alpha \in [0.4\pi, 0.45\pi]$. The
    red region is plotted for reference, where $\Delta \tau' = 0$.}
    \label{npic2}
\end{figure}


For $\hat{H}_{t} = \hat{H} - \kappa \mathcal{I} $, its eigenvalue are
$\lambda_- = 0$ and $\lambda_{+} = 2\kappa = 2\xi   \sqrt{ \cos^2 \vartheta   e^{i 2 \varrho } + \frac{1}{2} \sin^2 \varrho  \sin 2\beta} = 2\xi r e^{i\theta}
$, where $\theta \in (0,\pi)$. Therefore
$\mu = \text{Re}[\lambda_{1}] = 2\xi r \cos(\theta) = \zeta  \mu'$ and $\nu = \text{Im}[\lambda_{1}] = 2\xi r \sin(\theta) = \zeta  \nu'$, where $\zeta = 2\xi r$.
Thus the three bounds for system $\hat{H}_{t}$ read
$
    \mathcal{F}_{\text{ML}} (\mu, \nu, t) = \mathcal{F}_{\text{ML}} (\mu', \nu', t'),
$
where $t'=\zeta t$.
We define $\Delta \tau' = \tau_{\text{comb}}' - \tau_{\text{G}}'$
to quantify the deference between our bound and the Geometric bound denoted by $\tau_{\text{G}}$ given in Eq. (\ref{eq7}) that in this case reduces to analyse the dependence of $\Delta \tau'$ on $\mu' = \cos\theta \in (0,1)$
and $\nu' = \sin\theta \in (-1,1)$.
Scaling $\mu = \zeta\mu'$ and $\nu = \zeta\nu'$, the corresponding calculation result
does not change the sign of $\Delta \tau'$. Details can be found in  Appendix \ref{appendix J}.

We numerically calculate $\Delta \tau'$ and show the results in
Fig.\ref{npic2}. From Fig.\ref{npic2}, we can get that when $\nu' = \sin\theta$ is large, $\tau_{\text{comb}}$ and $\tau_{\text{G}}$ will have no solution.
This is due to the fact that for ML-type bound and MT-type bound, when $\nu' $ is large, a grows faster with $e^{\nu' t}$, which makes there is no $t$ such
that $\mathcal{F}_{\text{ML}} (\mu', \nu', t')=\pi/2$ and $\mathcal{F}_{\text{MT}} (\mu', \nu', t')=\pi/2$. For G-type bound, its solution can be expressed as $\tau_{\text{G}}={\nu}^{-1}\mathrm{ln}\left(\tan\left(\alpha+\frac{\pi\nu}{2\sqrt{\mu^2+\nu^2}}\right)\tan^{-1}\alpha\right)$, and there is no solution for
$\tau_{\text{G}}$ when $\alpha+\frac{\pi\nu}{2\sqrt{\mu^2+\nu^2}} > \pi/2$. When $\nu' < 0$ and $\alpha$ is large, $\tau_{\text{G}}$ is larger than $\tau_{\text{comb}} $,
but no matter what $\mu'$ and $\nu'$ parameter conditions, there exists an initial state which makes $\tau_{\text{comb}} > \tau_{\text{G}}$.

\begin{figure}[h]
    \centering
    \includegraphics[width=\linewidth]{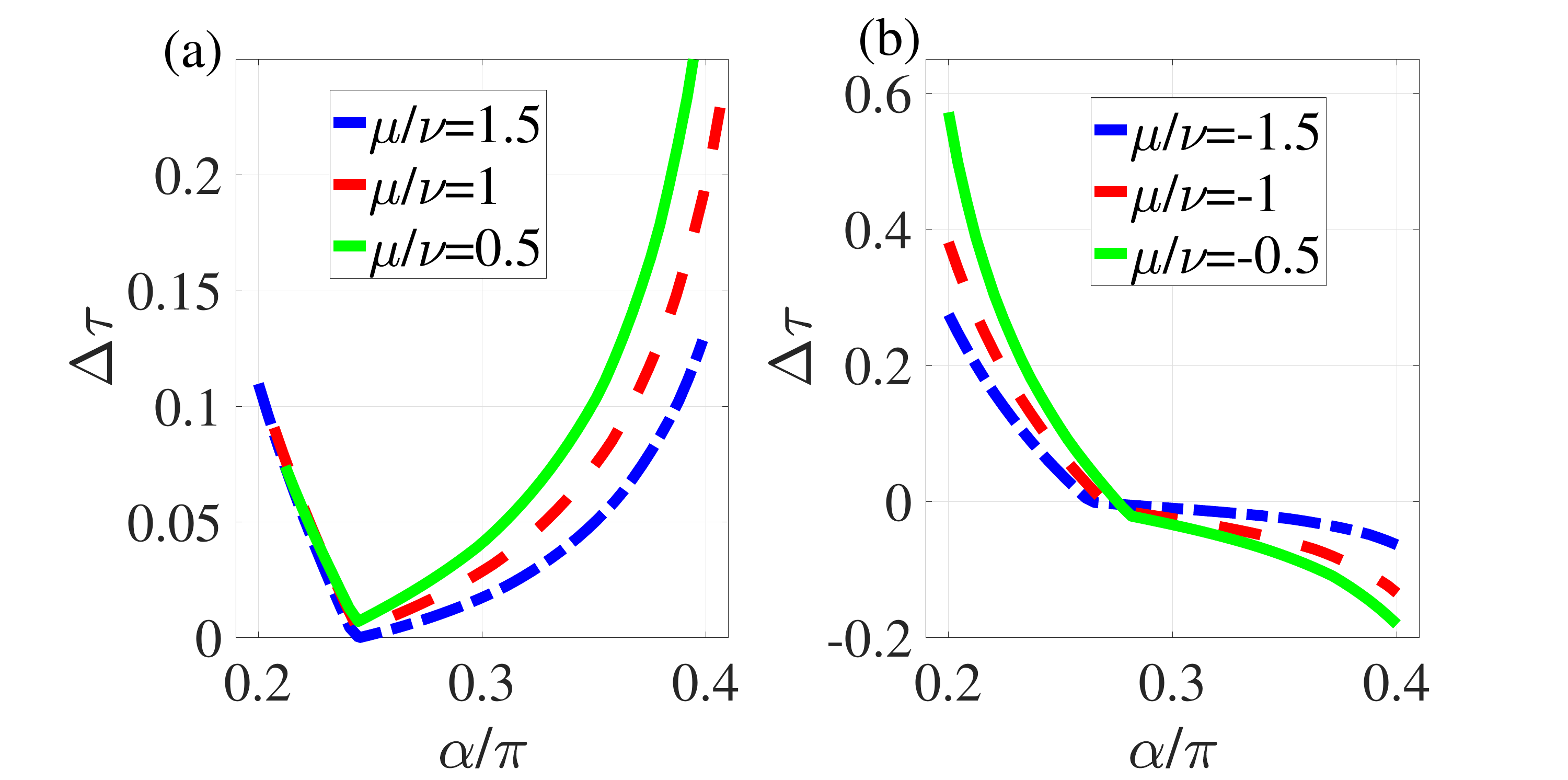}
    \caption{
    {(a), $\Delta \tau$ versus $\alpha / \pi$ for different $\mu/\nu$. }
    Green solid line is for $\mu/\nu = 0.5$,
    red dotted line  $\mu/\nu = 1$,
    and blue dashed dot line   $\mu/\nu = 1.5$. 
    (b) The same as in Figure (a) but with $-\mu/\nu$ with
    respect to that in Figure (a).
    }
    \label{npic3}
\end{figure}

We show the dependence of the QSL bounds $\tau_{\text{comb}}$ and $\tau_{\text{G}}$ on the real part of the eigenvalue for different values of $\mu/\nu$.
For the non-Hermitian two-level system in Fig.\ref{npic3}, with $|\mu/\nu|$ increasing, the difference between the two type QSL bounds  decreases.
This is because no matter what kind of QSL type bound,
it would increase with the increase of $|\mu/\nu|$. Namely,
regardless of the sign of $\nu$,
$\mathcal{F}_{\text{A}} (\tau_{\text{A}, 0.5}) \leqslant
\mathcal{F}_{\text{A}} (\tau_{\text{A}, 1}) \leqslant
\mathcal{F}_{\text{A}} (\tau_{\text{A}, 1.5}) (\text{A = ML, MT, G})$, where
$\mathcal{F}_{\text{A}} (\tau_{\text{A}, |\mu/\nu|}) (|\mu/\nu| = 0.5, 1, 1.5)$ is the corresponding QSL
bound when $\mu/\nu$ is equal to $0.5, 1, 1.5$ or $-0.5, -1, -1.5$.
Because of $\mathcal{F}_{\text{ML}} (\tau_{\text{ML} })
= \mathcal{F}_{\text{MT}} (\tau_{\text{MT} })
= \mathcal{F}_{\text{G}} (\tau_{\text{G} })$,
it corresponds to the QSL bound $\tau_{\text{ML}}$, $\tau_{\text{MT}}$, $\tau_{\text{G} }$.
This  lead to an increase in the QSL bound
as $|\mu / \nu|$ increases, resulting in a reduction in the difference $|\Delta \tau|$.

\section{Example}
The results presented so far give and discuss the bound on QSL. We are now going to  simulate the dynamics and find the  relation between the QSL bound $\tau_{\text{comb}}$ and the  time $\tau$. Remind that $\tau$ denotes the time for the system to evolve from an initial state to its orthogonal states.
As shown, the evolution time of a two-level non-Hermitian  system is
$
    \tau = \frac{k\pi}{\mu} , \quad k \in \mathbb{Z}_{+}.
$
Since $\tau$ here is not continuous, we can not get any insight on  how the evolution time approaches  to the QSL bound
through a two-level system. This motivates us to  apply  a three-level $\mathcal{P} \mathcal{T} $-system\cite{Wu2022,Hao2023} in the numerical simulation.
The $\mathcal{P} \mathcal{T} $-system can be realized by a three-coil wireless power transfer(WPT) system, which  describes a non-Hermitian system consisting of
three resonators with gain and loss. The Hamilonian of such a system can be written as
\begin{align}
    \hat{H}_{\text{WPT}} = \begin{bmatrix}
       \varsigma + i\eta & \kappa & 0 \\
       \kappa & \varsigma & \kappa \\
       0 & \kappa & \varsigma - i\eta
    \end{bmatrix}, \nonumber
\end{align}
where $\varsigma$ stands for
the resonant frequency of the three resonators, $\eta$ is
the dissipation and gain rate of the system, and $\kappa$ represents
the interaction rate between the resonators.
Its eigenvalues are,
$
    \lambda_{0,2} = \varsigma \mp \sqrt{2\kappa^2 - \eta^2}, \quad \lambda_{1} = \varsigma.
$
For \(2\kappa^2 > \eta^2\), the system exhibits $\mathcal{P} \mathcal{T}$-symmetry\cite{Bender1998,Dorey2001,Berry2004,Heiss2012,Bender2003}. When \(2\kappa^2 = \eta^2\) is the exceptional point (EP) of the Hamiltonian. When \(2\kappa^2 < \eta^2\), $\mathcal{P} \mathcal{T}$-symmetry is broken.

For the case of $\mathcal{P} \mathcal{T} $-symmetry,
the imaginary part $\gamma_0 = \gamma_1 = \gamma_2 = 0$.
With  $|\psi(0)\rangle = \sum_{n=0}^{2} c_{n} |\psi_n\rangle$ as the initial state, where $|\psi_n\rangle$ is the eigenstate of $\hat{H}_{\text{WPT}}$,
The bound of ML-type QSL for
the WPT system is
\begin{align}
    {
        \mathcal{F}_{\text{ML}} (\tau) = {(\langle \hat{H}_{\text{WPT}} (0)\rangle + \sqrt{2\kappa^2 - \eta^2} - \varsigma ) \tau}\geqslant \frac{\pi}{2},
    } \nonumber
\end{align}
and the bound of MT-type QSL is
\begin{align}
    \mathcal{F}_{\text{MT}} (\tau) ={\Delta \hat{H}_{\text{WPT}}(0)
    \tau }
      \geqslant \frac{\pi}{2}. \nonumber
\end{align}

\begin{figure}[h]
    \centering
    \includegraphics[width=\linewidth]{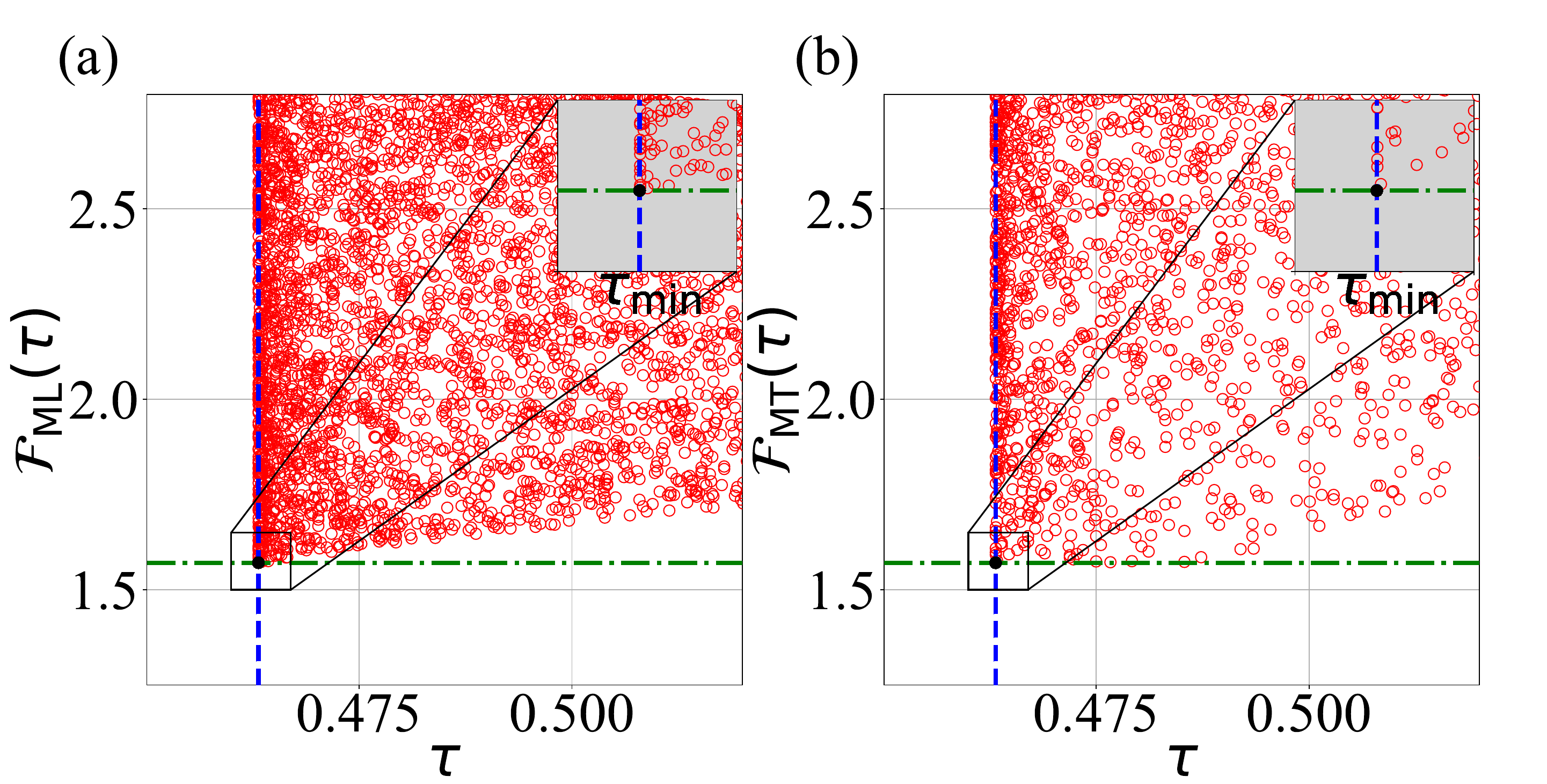}
    \caption{
     {
    $\mathcal{F}_{\text{ML}} (\tau)$ (a) and $\mathcal{F}_{\text{MT}} (\tau)$ (b) versus time $\tau$. }
    $\tau$ is the time for the quantum system to evolve from an initial state to its orthogonal ones.  To plot this figure, we choose $\kappa = 2.5$ and $\eta, \varsigma = 1.$
    The red points are for  random initial states,
    the blue line corresponds to the shortest evolution time
    $\tau_{\text{min}}=\frac{\pi}{2\sqrt{2\kappa^2 - \eta^2}}$,
    the green line is for  $\frac{\pi}{2}$,
    and the black point is a point generated for the initial state of
    $|\psi(0)\rangle = \frac{1}{\sqrt{2}}(|\psi_0\rangle + |\psi_2\rangle)$.
    }
    \label{mtaml}
\end{figure}
Fig.\ref{mtaml} shows the relation between $\mathcal{F}_{\text{ML}} (\tau)$
and $\mathcal{F}_{\text{MT}} (\tau)$ and the time $\tau$.
From the figure we can see all the red points are on
the right side of the blue line, above the green line.
 {This means that the evolution
time $\tau$ for all initial states are satisfied the inequality given in Eq.(\ref{ml2}) and Eq.(\ref{mt2}).
Here, we zoom in on the intersection point of the
green and blue lines. The black point at this cross
point corresponds to the FIS,
which demonstrates that the evolution time for this
initial state is the shortest and  that the inequalities
in Eq.(\ref{ml2}) and Eq.(\ref{mt2}) take equal sign in this case.}

\begin{figure}[h]
    \centering
    \includegraphics[width=\linewidth]{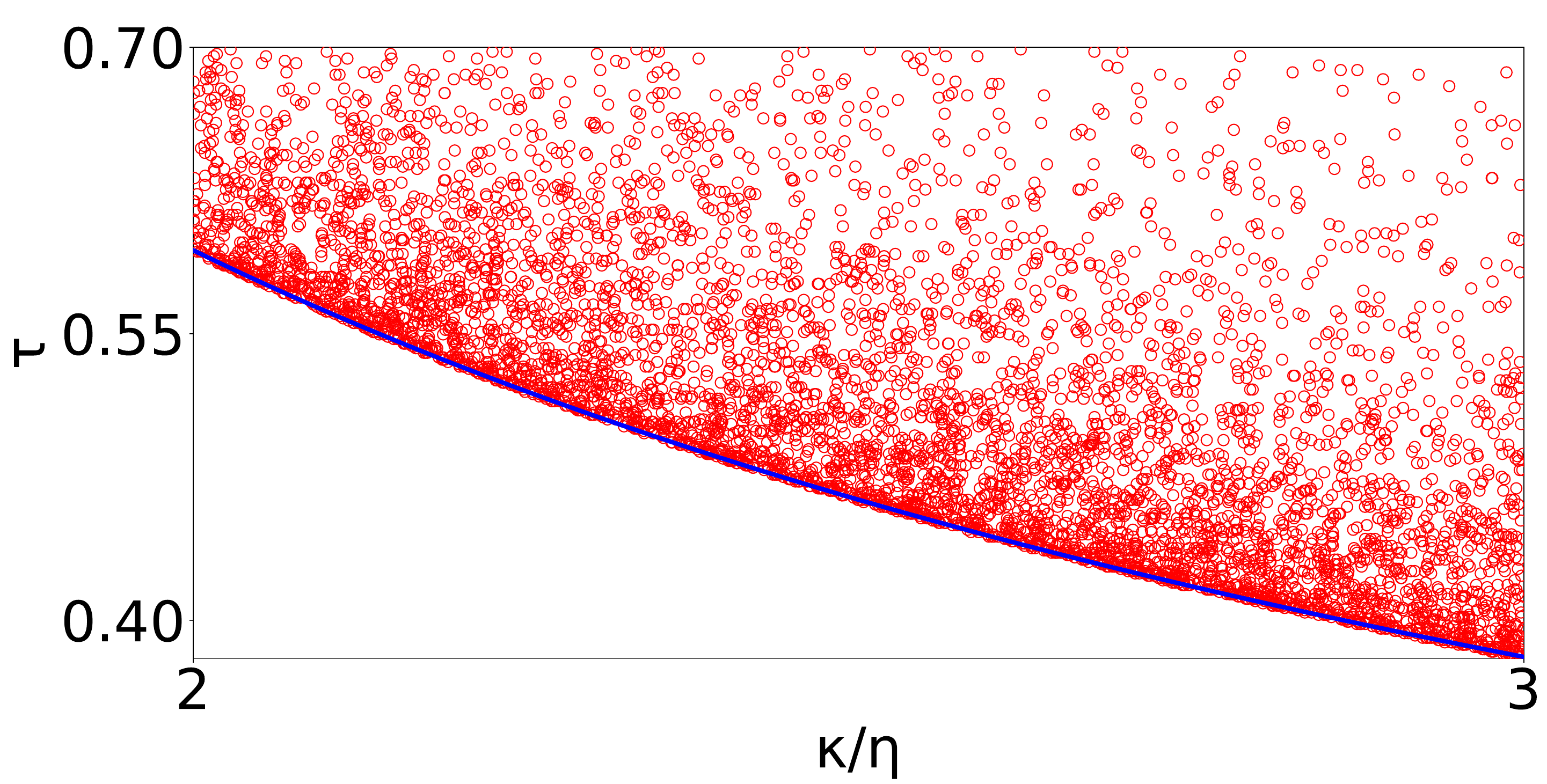}
    \caption{Upper and lower bound of  the evolution time needed for the system from an initial states to its orthogonal states as a function of $\kappa/\eta$ ($\varsigma =1$).
        The red points in the figure are  obtained
        for  randomly generated  initial states.
        The blue solid line bounds the shortest evolution
        time $\tau_{\text{min}}$.
    }
    \label{mtmlm}
\end{figure}

 {In addition to the shortest evolution time,
there is a maximum evolution time when
the imaginary parts of the all
eigenvalues are the same.}
Fig.\ref{mtmlm}  shows the dependence of the lower and upper bounds of
QSL on parameter $\kappa/\eta$. The blue solid  line is for the
bound $\tau_{\text{min}}=\frac{\pi}{2\sqrt{2\kappa^2 - \eta^2}}$ of QSL. {The red point line represents  the shortest time
for a randomly generated initial state
to evolve to its orthogonal states.
With the range of parameters shown in Fig.\ref{mtmlm},
the evolution time for an arbitrary initial state
should satisfy $\tau
\geqslant \tau_{\text{min}}$.}

For the case of $\mathcal{P} \mathcal{T} $-symmetry  broken, the denominator of ML-type function $\mathcal{F}_{\text{ML}} (\tau)$ and MT-type function $\mathcal{F}_{\text{ML}} (\tau)$ is $0$, which means that the time for any initial state evolves to its orthogonal state is infinite.
This is due to the evolution of the Hamiltonian for each quantum state
$
    |\psi(t)\rangle = e^{-i\hat{H}' t}|\psi_n\rangle=e^{-i\omega'_{n}t}e^{-\gamma'_{n}t}|\psi_n\rangle=e^{-\gamma'_{n}t}|\psi_n\rangle,
$
where $\hat{H}' = \hat{H} -(\varsigma - \sqrt{\eta^2 - 2\kappa^2}\mathbb{I} )$, this result shows that
the evolution governing by the Hamiltonian can only change the length of the base vector, not allows the base vector to evolve to the others. Therefore, it can never evolve into a quantum state orthogonal to it.
 {This can be verified} by the fidelity between the initial state and the final state
$S(t)$,
\begin{align}
    S(t) = \frac{1}{\mathcal{K} (t)} \sum_{n=0}^{N} e^{-\gamma_n' t} \left| c_n \right|^2 \geqslant e^{-\gamma_{\text{max}}t} >0, \nonumber
\end{align}
where $\mathcal{K} (t) =\sum_{n}^{N} |c_n|^2 e^{ -  2\gamma_n' t} \leqslant 1$ and $e^{-\gamma_n' t} \geqslant e^{-\gamma_{\text{max}}t}$ are applied.
This result implies that  $S(t)=0$ leads to $t \to \infty  $.
This verifies the conclusion given by Eq.(\ref{ml2})
and Eq.(\ref{mt2}).

\section{Conclusion}
In this work, we extend the ML-type  and the MT-type bounds from
Hermitian  to  non-Hermitian systems. The extention is not trivial in the sense that the method and results are different.  We found the FIS
that leads to the fastest evolution of the non-Hermitian system.
We analyze and calculate the QSL of non-Hermitian systems by utilizing
the inner product between the initial and final states  with the
non-Hermitian biorthogonal basis.
Under different ML and MT type bounds, we analyzes
the proximity between the evolution time and the QSL bound.
We show that when $\alpha = \mathcal{F}_{\text{ML}} (\tau) / \mathcal{F}_{\text{MT}} (\tau) = 1$,
the combination bound $\tau_{\text{comb}}$ of ML and MT
type bounds coincides with the evolution time
$\tau$. For $\alpha \neq 1$, though the
QSL bound $\tau_{\text{comb}}$ does not coincide with
the evolution time $\tau$, there exists
an initial state that that renders
$\tau$ arbitrarily close to $\tau_{\text{comb}}$.
To verify and illustrate our theory, we have derived  the QSL   and the shortest evolution times with different parameters for a
non-Hermitian three-level system, and  compare our results with
the earlier proposed ones. In addition, we analyze the ML-type bound and the MT-type bound near
the fastest evolving initial state, and compare them with the QSL
bound in the earlier publications. As an illustration, we use a three-level
$\mathcal{P} \mathcal{T} $-symmetric system  to show how
the evolution time $\tau$ approaches the
combined QSL bound $\tau_{\text{comb}}$.

\section{Acknowledgments} 
This work is supported by National Natural Science Foundation of
China (NSFC) under Nos. 12575010 and 12175033.


\appendix

\begin{widetext}


\section{model of QSL}\label{appendix 0}
We consider a general non-Hermitian Hamiltonian $ \hat{H} $, its time-independent Schrodinger equation reads ,
    \begin{align}
        \hat{H} | \psi_n\rangle = E_n | \psi_n\rangle = (\omega_{n} - i \gamma_{n}) | \psi_n\rangle.
                \nonumber
    \end{align}
    Here, $E_n$ is the eigenvalue of $\hat{H}$,
    $| \psi_n\rangle$ is the eigenstate of $\hat{H}$, i.e.,  the right eigenstate of the system.
    The eigenequation of its adjoint Hamiltonian $ \hat{H}^\dagger $ can be expressed as
    \begin{align}
        \hat{H}^\dagger | \widetilde{\psi}_n\rangle = E_n^* | \widetilde{\psi}_n\rangle = (\omega_{n} + i \gamma_{n}) | \widetilde{\psi}_n\rangle,
            \nonumber
    \end{align}
    where $ E_n^* $ and $ | \widetilde{\psi}_n\rangle $ denote the eigenvalues and eigenstates of $ \hat{H}^\dagger $, respectively. For  non-degenerate states, it satisfies the biorthogonal relation,
    $
        \langle \widetilde{\psi}_n | \psi_m\rangle = \delta_{nm} , \langle \psi_m | \widetilde{\psi}_n\rangle = \delta_{nm}.
    $
   But neither
    $
        \langle \psi_n | \psi_m\rangle \neq \delta_{nm},$ nor $ \langle \widetilde{\psi}_n | \widetilde{\psi}_m\rangle \neq \delta_{nm}
    $.

    We denote the minimum value of the real part of the non-Hermitian eigenvalue as $ \omega_{\text{\text{min}}} $, and the \text{min}imum value of the imaginary part as $ \gamma_{\text{min}} $. Define
    $
        \hat{H}' = \hat{H} - (\omega_{\text{min}} + i \gamma_{\text{min}}) \mathbb{I},
    $
    the eigenstate of $ \hat{H}' $ is same as $\hat{H}$,
    \begin{align}
        \hat{H}' | \psi_n\rangle = (\omega'_{n} - i \gamma'_{n}) | \psi_n\rangle, \nonumber
    \end{align}
    where
    $
        \omega'_{n} = \omega_{n} - \omega_{\text{min}} , \gamma'_{n} = \gamma_{n} - \gamma_{\text{min}}.
    $
    Here $ \omega'_{n} $ and $ \gamma'_{n} $ are nonnegative.
    The purpose of this transformation is to make the formulation clear and easy.

    In this work, we choose a superposition of the eigenstates of the Hamiltonian $H$ as the initial state,
    $
        |\psi(0)\rangle = \sum_{n}^{N} c_n  | \psi_n\rangle,
    $
    where $ c_n $ is a complex number.
    The initial state satisfies the biorthogonal normalization relation,
    $
        \mathcal{K} (0) =\langle\widetilde{\psi}(0)|\psi(0)\rangle = \sum_{n}^{N} |c_n|^2 = 1,
    $
    where $K(0)$ is the initial normalization coefficient.
    The state at time $t$ reads,
    \begin{align}
        |\psi(t)\rangle = \frac{1}{\mathcal{K} (t)}\sum_{n}^{N} e^{-i(\omega_n' - i \gamma_n') t} c_n  | \psi_n\rangle,
            \nonumber
    \end{align}
    where $\mathcal{K} (t)$ is the normalized coefficient at $t$, $\mathcal{K}^2 (t) =\langle\widetilde{\psi}(t)|\psi(t)\rangle = \sum_{n}^{N} |c_n|^2 e^{-2\gamma_n' t}$.
     {Here $| \tilde{\psi}(0) \rangle ,| \tilde{\psi}(t) \rangle \in \widetilde{\mathcal{V}} = \text{Span}(\{|\widetilde{\psi}_n\rangle\}_{n=0}^{N})$,
    and $| \psi(0) \rangle ,| \psi(t) \rangle \in \mathcal{V} = \text{Span}(\{|\psi_n\rangle\}_{n=0}^{N})$.}

    \section{Derivation of ML-type bound}\label{appendix A}
    For a non-Hermitian Hamiltonian, we can write it in terms of its complete biorthonormal set of basis,
    \begin{align}
        \hat{H} &= \sum_{n=0}^{N} (\omega_n'-i\gamma_n' )| \psi_n \rangle \langle \widetilde{\psi}_n | \nonumber\\
        &=\sum_{n=0}^{N} \omega_n'| \psi_n \rangle \langle \widetilde{\psi}_n |-i\sum_{n=0}^{N} \gamma_n' | \psi_n \rangle \langle \widetilde{\psi}_n | \nonumber\\
        &\equiv \hat{\Omega } - i\hat{\Gamma }. \nonumber
    \end{align}
    The inner product of the initial state and the final state $|\psi(t)\rangle$ is,
    \begin{align}
        S(t) &= \langle \tilde{\psi}(0) | \psi(t) \rangle = \frac{1}{\mathcal{K} (t)} \sum_{n=0}^{N} e^{-i(\omega_n' - i\gamma_n')t} \left| c_n \right|^2,
        \nonumber
    \end{align}
     {where $\mathcal{K}^2 (t) = \langle \tilde{\psi}(t) | \psi(t) \rangle$ is the normalized coefficient of $| \psi(t) \rangle$.
    $| \tilde{\psi}(0) \rangle ,| \tilde{\psi}(t) \rangle \in \widetilde{\mathcal{V}} = \text{Span}(\{|\widetilde{\psi}_n\rangle\}_{n=0}^{N})$,
    and $| \psi(0) \rangle ,| \psi(t) \rangle \in \mathcal{V} = \text{Span}(\{|\psi_n\rangle\}_{n=0}^{N})$.}
    The real part and imaginary part of $S(t)$ are respectively,
    \begin{align}
        \text{Re}[S(t)] &= \frac{1}{\mathcal{K} (t)} \sum_{n=0}^{N} e^{-\gamma_n't} \cos(\omega_n't) \left| c_n \right|^2  \nonumber\\
        \text{Im}[S(t)] &= \frac{1}{\mathcal{K} (t)} \sum_{n=0}^{N} e^{-\gamma_n't} \sin(\omega_n't) \left| c_n \right|^2.  \nonumber
    \end{align}
    At  time $t=\tau$, $S(\tau) = 0$ follows,
    \begin{align}
        \text{Re}[S(\tau)] = 0, \quad \text{Im}[S(\tau)] = 0. \nonumber
    \end{align}
Next, we apply the inequality,
    \begin{align}
        \cos x \geqslant 1 - \frac{2}{\pi}(x + \sin x). \nonumber
    \end{align}
This inequality holds for $x \geqslant 0$, and the equality holds for $x = 0$ or $x = \pi$.
Because $e^{-\gamma_n'\tau} \geqslant 0$, $|c_n|^2 \geqslant 0$ and $\omega_n' \geqslant 0$, we have
    \begin{align}
        {\mathcal{K} (\tau)}\text{Re}[S(\tau)] &= \sum_{n=0}^{N} e^{-\gamma_n'\tau} \cos(\omega_n'\tau) \left| c_n \right|^2 \nonumber \\
        &\geqslant \sum_{n=0}^{N} e^{-\gamma_n'\tau} \left| c_n \right|^2 - \sum_{n=0}^{N} e^{-\gamma_n'\tau} \left[ \frac{2}{\pi} (\omega_n'\tau + \sin(\omega_n'\tau)) \right] \left| c_n \right|^2.
        \nonumber
    \end{align}
    The first term on the right side of the equation is,
    \begin{align}
        \sum_{n=0}^{N} e^{-\gamma_n'\tau} \left| c_n \right|^2 {=\Tr [e^{-\hat{\Gamma } \tau} \rho(0)] } .
        \nonumber
    \end{align}
    The second term satisfies the equation,
    \begin{align}
        -\sum_{n=0}^{N} e^{-\gamma_n'\tau} \left[ \frac{2}{\pi} \omega_n'\tau \right] \left| c_n \right|^2
        {= -\frac{2}{\pi}\Tr [e^{-\hat{\Gamma } \tau} \hat{\Omega }  \rho(0)]  \tau}.
        \nonumber
    \end{align}
    For the ML type bound,
    the last term is related to $\text{Im}[S(\tau)]$,
    \begin{align}
        {
            \mathcal{K} (\tau)\text{Re}[S(\tau)] \geqslant \Tr [e^{-\hat{\Gamma } \tau} \rho(0)] -\frac{2}{\pi}\Tr [e^{-\hat{\Gamma } \tau} \hat{\Omega }  \rho(0)] \tau- \frac{2}{\pi} \mathcal{K} (\tau)\text{Im}[S(\tau)].
        }   \nonumber
    \end{align}
    Therefore the bound of ML type is,
    \begin{align}
        {
            \tau \geqslant \frac{\pi}{2}\frac{ \Tr [e^{-\hat{\Gamma } \tau} \rho(0)]}{  \Tr [e^{-\hat{\Gamma } \tau} \hat{\Omega }  \rho(0)]}.
        }   \nonumber
    \end{align}
    The ML-type bound can be rewritten as,
    \begin{align}
            {
                \mathcal{F}_{\text{ML}} (\tau) = \frac{\Tr [e^{-\hat{\Gamma } \tau} \hat{\Omega }  \rho(0)] \tau}{\Tr [e^{-\hat{\Gamma } \tau} \rho(0)]} \geqslant \frac{\pi}{2}.
            }   \nonumber
    \end{align}

    \section{Derivation of the upper and lower limits of the evolution time independent of the initial state}\label{appendix A1}
    For the system evolving to its orthogonal state $S(\tau)=0$, we have
    \begin{align}
        \text{Re}[S(\tau)] &= \frac{1}{\mathcal{K} (\tau)}\sum_{n=0}^{N} e^{-\gamma_n' \tau} \cos(\omega_n' \tau) |c_n|^2 = 0, \nonumber\\
        \text{Im}[S(\tau)] &= \frac{1}{\mathcal{K} (\tau)}\sum_{n=0}^{N} e^{-\gamma_n' \tau} \sin(\omega_n' \tau) |c_n|^2 = 0.  \nonumber
    \end{align}
     {For the shortest evolution time $\tau$(from an initial state to its orthogonal states) of the system, if we assume $\tau < \frac{\pi}{\omega_{N}'}$, then
    \begin{align}
         \sin(\omega_n' t) > 0. \nonumber
    \end{align}
    This contradicts with $\text{Re}[S(\tau)]=\text{Im}[S(\tau)]=0$. Therefore, the time to finish the shortest evolution should satisfy,
    \begin{align}
        \tau_{\text{min}} = \frac{\pi}{\omega_{N}'}.    \nonumber
    \end{align}
    Here $\omega_{0}' \leqslant \omega_{1}' \leqslant \cdots \leqslant \omega_{N}'$.
    If the initial state is a superposition of  three or more eigenstates $|\psi_n\rangle$,
    then for the shortest evolution time $\tau_{\text{min}}$,
    \begin{align}
        \text{Im}[S(\frac{\pi}{\omega_{N}'})]
            = \frac{1}{\mathcal{K} (\frac{\pi}{\omega_{N}'})}
            (e^{-\gamma_0' \frac{\pi}{\omega_{N}'}} \sin(0) |c_0|^2
            + e^{-\gamma_1' \frac{\pi}{\omega_{N}'}} \sin(\omega_{1}'\frac{\pi}{\omega_{N}'}) |c_1|^2
            + \cdots
            + e^{-\gamma_{N}' \frac{\pi}{\omega_{N}'}} \sin(\pi) |c_N|^2)
            \geqslant 0.    \nonumber
    \end{align}
    This suggests  that the initial state leading to the shortest evolution time  is a superposition of two eigenstates.}
    For  time $\tau_{\text{min}}= \frac{\pi}{\omega_{N}'}$ to finish the evolution with   maximum speed, we have
    \begin{align}
        \text{Im}[S(\frac{\pi}{\omega_{N}'})]   &= \frac{1}{\mathcal{K} (\frac{\pi}{\omega_{N}'})}(e^{-\gamma_0' \frac{\pi}{\omega_{N}'}} \sin(0) |c_0|^2 + e^{-\gamma_1' \frac{\pi}{\omega_{N}'}} \sin(\pi) |c_1|^2) \nonumber\\
                                                &= 0.\nonumber\\
        \text{Re}[S(\frac{\pi}{\omega_{N}'})]   &= \frac{1}{\mathcal{K} (\frac{\pi}{\omega_{N}'})}(e^{-\gamma_0' \frac{\pi}{\omega_{N}'}} \cos(0) |c_0|^2 + e^{-\gamma_1' \frac{\pi}{\omega_{N}'}} \cos(\pi) |c_1|^2) \nonumber\\
                                                &= \frac{1}{\mathcal{K} (\frac{\pi}{\omega_{N}'})}(e^{-\gamma_0' \frac{\pi}{\omega_{N}'}} |c_0|^2 - e^{-\gamma_1' \frac{\pi}{\omega_{N}'} } |c_1|^2) \nonumber\\
                                                &= 0.   \nonumber
    \end{align}
    In other words, the system can reach the maximum evolution speed if and only if the initial state is,
    \begin{align} \label{eq,state}
        \frac{|c_0|^2}{|c_1|^2} = e^{(\gamma_0 - \gamma_1) \frac{\pi}{\omega_{N}'}},
    \end{align}
When the system has the same imaginary part of the eigenvalue,
    \begin{align}
        \text{Re}[S(\tau)] &= \frac{e^{-\gamma_n' \tau}}{\mathcal{K} (\tau)}\sum_{n=0}^{N}  \cos(\omega_n' \tau) |c_n|^2 = 0,   \nonumber\\
        \text{Im}[S(\tau)] &= \frac{e^{-\gamma_n' \tau}}{\mathcal{K} (\tau)}\sum_{n=0}^{N}  \sin(\omega_n' \tau) |c_n|^2 = 0.   \nonumber
    \end{align}
    I.e.,
    \begin{align}
        \sum_{n=0}^{N}  \cos(\omega_n' \tau) |c_n|^2 = 0, \nonumber\\
        \sum_{n=0}^{N}  \sin(\omega_n' \tau) |c_n|^2 = 0.  \nonumber
    \end{align}
    This means that if $\tau$ satisfies $\omega_n' \tau \geqslant 2\pi$,  $\omega_n' \tau - 2\pi$ also satisfies $\text{Re}[S(\tau)]
    =\text{Im}[S(\tau)] = 0$. Then the evolution
    time $\tau$ satisfies,
    \begin{align}
        \tau \leqslant \frac{2\pi}{\omega_{N}'}.    \nonumber
    \end{align}

    \section{The derivation of weak ML and MT-type bounds}\label{appendix E}
    Now we discuss in details the ML-type bounds,
    \begin{align}
        {
            \mathcal{F}_{\text{ML}} (\tau) = \frac{\Tr [e^{-\hat{\Gamma } \tau} \hat{\Omega }  \rho(0)] \tau}{\Tr [e^{-\hat{\Gamma } \tau} \rho(0)]} \geqslant \frac{\pi}{2}.
        }   \nonumber
    \end{align}
     {For $\Tr [e^{-\hat{\Gamma } \tau} \hat{\Omega }  \rho(0)] \tau$,} we have,
    \begin{align}
        \Tr [e^{-\hat{\Gamma } \tau} \hat{\Omega }  \rho(0)] \tau
        =\sum_{n=0}^{N} e^{-\gamma_n'\tau} \left[ \omega_n'\tau \right] \left| c_n \right|^2
        \leqslant e^{-\gamma_\text{min}'\tau}\sum_{n=0}^{N}  \omega_n'\tau\left| c_n \right|^2
        =\langle \hat{\Omega } (0)\rangle \tau. \nonumber
    \end{align}
    Here the non-Hermitian biorthogonal normalization relation $\sum_{n=0}^{N}  \left| c_n \right|^2=1$ and $\gamma_{\text{min}}'=0$ is used.
     {For $\Tr [e^{-\hat{\Gamma } \tau} \rho(0)]$,} we have,
    \begin{align}
        {\Tr [e^{-\hat{\Gamma } \tau} \rho(0)]} = \sum_{n=0}^{N} e^{-\gamma_n'\tau} \left| c_n \right|^2 \geqslant  e^{-\gamma_\text{max}'\tau}.
        \nonumber
    \end{align}
    Therefore, the weak ML-type bound reads,
    \begin{align}
        {
            \mathcal{F}_{\text{WML}} (\tau) = \frac{\langle \hat{\Omega } (0)\rangle \tau}{e^{-\gamma_{\text{max}}'\tau}}\geqslant \frac{\pi}{2}.
        }   \nonumber
    \end{align}
    Collecting all these together, we have,
    \begin{align}
        \mathcal{F}_{\text{WML}} (\tau) \geqslant \mathcal{F}_{\text{ML}} (\tau) \geqslant
        \frac{\pi}{2}.  \nonumber
    \end{align}
    The equality holds if and only if,
    \begin{align}
        \gamma_{\text{min}}' = \gamma_{\text{max}}'.    \nonumber
    \end{align}
    For MT-type bound,
    \begin{align}
        {
    \mathcal{F}_{\text{MT}} (\tau) =
    \frac
    {\sqrt{
            -\mathcal{R} (\tau)\tau
        +[{      \Delta \hat{\Omega } ^2 \tau^2 }]e^{-\gamma_{\phi (1)}' \tau}
    }}
    {
        \Tr[e^{-\hat{\Gamma } \tau} \rho(0)]
    }
     \geqslant \frac{\pi}{2},
        }   \nonumber
    \end{align}
    where,
    \begin{align}
        \mathcal{R}(\tau)  =& 4  \sum_{n}{\Tr[\hat{\Gamma } e^{-2\hat{\Gamma }
        \tau}\mathcal{X}_{n,n}\rho(0)] \Tr[\mathcal{X}_{n,n}\rho(0)]} -
        2\Tr[\hat{\Gamma } e^{-\hat{\Gamma } \tau}\rho(0)]\Tr[ e^{-\hat{\Gamma }
        \tau}\rho(0)]   \nonumber\\
        =&\frac{16}{\pi^2} \sum_{n=0}^{N} \gamma_n'  e^{-2\gamma_n'\tau} c_n^4
            - \frac{8}{\pi^2} \sum_{\substack{ n, m = 0}}^{N} \gamma_{\text{n}}'
            e^{-(\gamma_{\text{n}}'+\gamma_{\text{m}}')\tau} |c_n|^2 |c_m|^2
            \nonumber\\
        \geqslant&
        -\frac{8}{\pi^2} \gamma_{\text{max}}'.  \nonumber
    \end{align}
    The weak MT-type bound is,
    \begin{align}
            \mathcal{F}_{\text{WMT}} (\tau) =\frac{\sqrt{\Delta \hat{\Omega }^{2} \tau^2 e^{-\gamma_{\phi (1)}'\tau}- 2\gamma_{\text{max}}'\tau}}{e^{-\gamma_{\text{max}}'\tau}}  \geqslant \frac{\pi}{2}.
            \nonumber
        \end{align}
    The relation with the bound of ML-type QSL is ,
    \begin{align}
        \mathcal{F}_{\text{WMT}} (\tau) \geqslant \mathcal{F}_{\text{MT}} (\tau) \geqslant
        \frac{\pi}{2}.  \nonumber
    \end{align}
    The equality holds if and only if,
    \begin{align}
        \gamma_{\text{min}}' = \gamma_{\text{max}}'.    \nonumber
    \end{align}

    \section{Derivation of MT-type bound}\label{appendix MT}
    For $\hat{\Omega }$, we have
    \begin{align}
        \hat{\Omega } = \sum_{n=0}^{N} \omega_n' | \psi_n \rangle \langle \widetilde{\psi}_n |.   \nonumber
    \end{align}
    The variance of $\hat{\Omega }$ in the initial state is defined as,
    \begin{align}
        2\Delta \hat{\Omega }^{2}&= \langle \psi(0)|\hat{\Omega }^{\dagger}\hat{\Omega }|\psi(0)\rangle + \langle \psi(0)|\hat{\Omega }^{\dagger}\hat{\Omega }|\psi(0)\rangle - 2\langle \psi(0)|\hat{\Omega }^{\dagger}|\psi(0)\rangle \langle \psi(0)|\hat{\Omega }|\psi(0)\rangle\nonumber\\
        &= \sum_{n=0}^{N} \omega_n'^2 |c_n|^2 + \sum_{m=0}^{N} \omega_m'^2 |c_m|^2 - 2\sum_{n=0}^{N} \sum_{m=0}^{N} \omega_n' \omega_m' |c_n|^2 |c_m|^2 \nonumber\\
        &= \sum_{n=0}^{N} \sum_{m=0}^{N} (\omega_n' - \omega_m')^2 |c_n|^2 |c_m|^2. \nonumber
    \end{align}
     {In the above derivation,  $\sum_{n=0}^{N}|c_n|^2=\sum_{m=0}^{N} |c_m|^2=1$ was applied.}
    The modular square considering the inner product of the initial state and the final state is,
    \begin{align}
        \mathcal{K}^2 (t)|S(t)|^2 = \sum_{n=0}^{N} \sum_{m=0}^{N} e^{-(\gamma_n' + \gamma_m')t} e^{i(\omega_n' - \omega_m')t} |c_n|^2 |c_m|^2.
        \nonumber
    \end{align}
     Since for any two cases $n = k, m = j$ and $n = j, m = k$,
    \begin{align}
        \gamma_k' + \gamma_j' &= \gamma_j' + \gamma_k', \quad \omega_j' - \omega_k' = -(\omega_k' - \omega_j').
        \nonumber
    \end{align}
    That is,
    \begin{align}
        \mathcal{K}^2 (t) |S(t)|^2 = \sum_{n=0}^{N} \sum_{m=0}^{N} e^{-(\gamma_n' + \gamma_m')t} \cos[(\omega_n' - \omega_m')t] |c_n|^2 |c_m|^2.
        \nonumber
    \end{align}
    Using the inequality,
    \begin{align}
        \cos x \geqslant 1 - \frac{4}{\pi^2} x \sin x - \frac{2}{\pi^2} x^2,
        \nonumber
    \end{align}
     which holds for any real number $x$, and the equality holds if and only if for $x = 0$ or $x = \pm \pi$. With this inequality, simple algebra follows,
    \begin{align}
        \mathcal{K}^2 (t)|S(t)|^2 \geqslant & \sum_{n=0}^{N} \sum_{m=0}^{N} e^{-(\gamma_n' + \gamma_m')t}
        |c_n|^2 |c_m|^2
            -  \frac{4}{\pi^2} \sum_{n=0}^{N} \sum_{m=0}^{N} e^{-(\gamma_n' + \gamma_m')t} (\omega_n' - \omega_m') t \sin[(\omega_n' - \omega_m')t] |c_n|^2 |c_m|^2 \nonumber\\
            &- \frac{2}{\pi^2} \sum_{m=0}^{N} \sum_{m=0}^{N} e^{-(\gamma_n' + \gamma_m')t} (\omega_n' - \omega_m')^2 t^2 |c_n|^2 |c_m|^2.
            \nonumber
        \end{align}
    The first term satisfies,
    \begin{align}
        \sum_{n=0}^{N} \sum_{m=0}^{N} e^{-(\gamma_n' + \gamma_m')t} |c_n|^2 |c_m|^2
        {= (\Tr[e^{-\hat{\Gamma } t} \rho(0)])^2}.
        \nonumber
    \end{align}
    Since $|S(t)|^2 \geqslant 0$, and $|S(t)|^2$ is continuous, then if $|S(t)|^2 = 0$, $\frac{d|S(t)|^2}{dt} = 0$ must hold,
    \begin{align}
        \frac{d|S(t)|^2}{dt} = & -\sum_{n=0}^{N} \sum_{m=0}^{N} (\gamma_n' + \gamma_m') e^{-(\gamma_n' + \gamma_m')t} \cos[(\omega_n' - \omega_m')t] |c_n|^2 |c_m|^2 \nonumber\\
        & - \sum_{n=0}^{N} \sum_{m=0}^{N} (\omega_n' - \omega_m') e^{-(\gamma_n' + \gamma_m')t} \sin[(\omega_n' - \omega_m')t] |c_n|^2 |c_m|^2.
        \nonumber
    \end{align}
    Therefore, the second term can be written as,
    \begin{align}
        \frac{4}{\pi^2} \sum_{n=0}^{N} \sum_{m=0}^{N} (\gamma_n' + \gamma_m') t e^{-(\gamma_n' + \gamma_m')t} \cos[(\omega_n' - \omega_m')t] |c_n|^2 |c_m|^2
            =& \frac{4}{\pi^2} \sum_{\substack{n = m \\ n, m = 0}}^{N} (\gamma_n' + \gamma_m') t e^{-(\gamma_n' + \gamma_m')t} \cos[(\omega_n' - \omega_m')t] |c_n|^2 |c_m|^2 \nonumber\\
            & + \frac{4}{\pi^2} \sum_{\substack{n \neq m \\ m, n = 0}}^{N} (\gamma_n' + \gamma_m') t e^{-(\gamma_n' + \gamma_m')t} \cos[(\omega_n' - \omega_m')t] c_i^2 c_j^2 \nonumber\\
            \geqslant & \frac{4}{\pi^2} \sum_{n=0}^{N} 2\gamma_n' t e^{-2\gamma_n't} c_n^4\nonumber\\
            &- \frac{4}{\pi^2} \sum_{\substack{n \neq m \\ n, m = 0}}^{N} (\gamma_{\text{n}}'+\gamma_{\text{m}}') t e^{-(\gamma_{\text{n}}'+\gamma_{\text{m}}')t} |c_n|^2 |c_m|^2 \nonumber\\
            =&\frac{16}{\pi^2} \sum_{n=0}^{N} \gamma_n' t e^{-2\gamma_n't} c_n^4\nonumber\\
            &- \frac{8}{\pi^2} \sum_{\substack{ n, m = 0}}^{N} \gamma_{\text{n}}' t e^{-(\gamma_{\text{n}}'+\gamma_{\text{m}}')t} |c_n|^2 |c_m|^2
            \nonumber
    \end{align}
    \begin{align}
        \mathcal{R}(t)  = 4  \sum_{n}{\Tr[\hat{\Gamma } e^{-2\hat{\Gamma } t}\mathcal{X}_{n,n}\rho(0)] \Tr[\mathcal{X}_{n,n}\rho(0)]} - 2\Tr[\hat{\Gamma } e^{-\hat{\Gamma } t}\rho(0)]\Tr[ e^{-\hat{\Gamma } t}\rho(0)],
        \nonumber
    \end{align}
    where,
    \begin{align}
        \mathcal{X}_{n,n} =  |\psi_n\rangle \langle \widetilde{\psi}_n|.
        \nonumber
    \end{align}
    The last item is related to the variance, and $\gamma_{\text{min}}'=0$,
    \begin{align}
        {- \frac{2}{\pi^2} \sum_{n,m=0}^{N}  e^{-(\gamma_n' + \gamma_m')t} (\omega_n' - \omega_m')^2 t^2 |c_n|^2 |c_m|^2}
        {=} & {- \frac{2}{\pi^2} \sum_{\substack{n \neq m \\ m, n = 0}}^{N}  e^{-(\gamma_n' + \gamma_m')t} (\omega_n' - \omega_m')^2 t^2 |c_n|^2 |c_m|^2} \nonumber\\
        {\geqslant} & - \frac{4}{\pi^2} e^{- \gamma_{\phi(1)}' t}    \Delta \hat{\Omega } ^2 t^2 ,
        \nonumber
    \end{align}
    where $\gamma_{\text{min}}'=
    \gamma_{\phi(0)}' \leqslant \gamma_{\phi(1)}' \leqslant \cdots
    \leqslant \gamma_{\phi(N)}'= \gamma_{\text{max}}'$ and $\gamma_{\phi(n)}'
    $ is
    the ordering of the imaginary part $\gamma_{n}'$ of the eigenvalue of $\hat{H}$.
    Therefore, due to $|S(\tau)|^2 =0$, the bound of the MT-type is,
    \begin{align}
            {
         -\mathcal{R} (\tau)\tau
        +{      \Delta \hat{\Omega } ^2 e^{-\gamma_{\phi(1)}' \tau}\tau^2 }
        \geqslant \frac{\pi^2}{4}  (\Tr[e^{-\hat{\Gamma } \tau} \rho(0)])^2 >0.
            }\nonumber
    \end{align}
    The bound of {MT-type} is,
    \begin{align}
            {
        \mathcal{F}_{\text{MT}} (\tau) =
        \frac
        {\sqrt{
                -\mathcal{R} (\tau)\tau
            +[{      \Delta \hat{\Omega } ^2 \tau^2 }]e^{-\gamma_{\phi(1)}' \tau}
        }}
        {
            \Tr[e^{-\hat{\Gamma } \tau} \rho(0)]
        }
        \geqslant \frac{\pi}{2}.
            }\nonumber
    \end{align}

\section{a general two-level non-hermitian system} \label{appendix f}

For a general two-level non-Hermitian Hamiltonian,
\begin{align}
       \hat{H} = \begin{pmatrix}
            \omega_0 + i\gamma_0 & s \\
            t & \omega_1 + i\gamma_1
        \end{pmatrix},
        \nonumber
\end{align}
For simplicity, we define
\begin{align}
    \hat{H}' = \hat{H} - (\frac{\omega_0 + \omega_1}{2}
    + i\frac{\gamma_0 + \gamma_1}{2})\hat{I}
    = \begin{pmatrix}
        \frac{\omega_0 - \omega_1}{2}
        + i\frac{\gamma_0 - \gamma_1}{2} & s \\
         t & -\frac{\omega_0 - \omega_1}{2}
         - i\frac{\gamma_0 - \gamma_1}{2}
     \end{pmatrix}.
     \nonumber
\end{align}
Clearly, $\hat{H}'$ and $\hat{H}$ lead to the same dynamics of the system.
$\hat{H}'$ can be rewritten to be,
\begin{align}
    \hat{H}'  &= \begin{pmatrix}
        re^{i\theta} & s \\
         t & -re^{i\theta}
     \end{pmatrix}\nonumber\\
     &= \sqrt{r^2+s^2+t^2}\begin{pmatrix}
        \frac{r}{\sqrt{r^2+s^2+t^2}}e^{i\theta} & \frac{s}{\sqrt{r^2+s^2+t^2}} \\
        \frac{t}{\sqrt{r^2+s^2+t^2}} & -\frac{r}{\sqrt{r^2+s^2+t^2}}e^{i\theta}
     \end{pmatrix},
     \nonumber
\end{align}
where $r = \sqrt{(\frac{\omega_0 - \omega_1}{2})^2 + (\frac{\gamma_0 - \gamma_1}{2})^2}$,
and $\theta$ is the angle of the complex number $\frac{\omega_0 - \omega_1}{2} + i\frac{\gamma_0 - \gamma_1}{2}$.
Here we have defined,
\begin{align}
    \frac{r}{\sqrt{r^2+s^2+t^2}} &= \cos\alpha, \nonumber\\
    \frac{s}{\sqrt{r^2+s^2+t^2}} &= \sin\theta \sin \beta, \nonumber\\
    \frac{t}{\sqrt{r^2+s^2+t^2}} &= \sin\theta \cos \beta.
    \nonumber
\end{align}
The Hamiltonian then becomes,
\begin{align} \label{eq,general two-level}
    \hat{H}' = \xi \begin{pmatrix}
        \cos\alpha e^{i\theta} & \sin\theta \sin \beta \\
        \sin\theta \cos \beta & -\cos\alpha e^{i\theta}
     \end{pmatrix},
\end{align}
where $\xi = \sqrt{r^2+s^2+t^2} \in (0, +\infty )$,
and  $\theta , \beta \in [0 , 2 \pi]$, $\alpha \in [0 , \pi]$. Its eigenvalue is,
\begin{align}
    \lambda_{\pm} = \pm  \xi   \sqrt{ \cos^2 \alpha  e^{i 2 \theta} + \frac{1}{2} \sin^2 \theta \sin 2\beta }.\nonumber
\end{align}
The eigenstate is,
\begin{align}
    |\lambda_{\pm}\rangle = \begin{pmatrix} \sin \theta \sin \beta \\ \xi^{-1}\lambda_{\pm} - \cos \alpha e^{i\theta} \end{pmatrix}
    = \begin{pmatrix} \sin \theta \sin \beta \\ \pm \sqrt{ \cos^2 \alpha  e^{i 2 \theta} + \frac{1}{2} \sin^2 \theta \sin 2\beta } - \cos \alpha e^{i\theta}
    \end{pmatrix}   \nonumber
\end{align}

\section{
Derivation of $\mathcal{F}_{\text{ML}} (t)$,
$\mathcal{F}_{\text{MT}} (t)$ and $\mathcal{F}_{\text{G}} (t)$
in non-Hermitian two-levels
} \label{appendix G}

For a two-level non-Hermitian Hamiltonian, we can always choose the zeroth
eigenvalue $\lambda_0$ to be  zero, so that its eigenvalue can be reexpressed as,
\begin{align}
    \lambda_0 = 0 \quad \text{and} \quad
    \lambda_1 = \mu + i\nu . \nonumber
\end{align}
The initial state of the system can be chosen to be
$|\psi(0)\rangle = \cos\alpha |\lambda_0\rangle+ e^{i\phi} \sin\alpha |\lambda_1\rangle$,
the ML-type bound for the non-Hermitian two-level can be written as,

\begin{align}
        \mathcal{F}_{\text{ML}} (t) = \frac{\Tr [e^{-\hat{\Gamma } t} \hat{\Omega }
        \rho(0)] t}{\Tr [e^{-\hat{\Gamma } t} \rho(0)]}
        = \frac{
            \sum_{n=0}^{1} e^{-\gamma_n't}  \omega_n't  \left| c_n \right|^2
        } {
            \sum_{n=0}^{1} e^{-\gamma_n't} \left| c_n \right|^2
        }
        =\frac{\sin^2 \alpha \mu t}{e^{-\nu t}\cos^2 \alpha+ \sin^2 \alpha} = \frac{\tan^2 \alpha \mu t}{e^{-\nu t}+ \tan^2 \alpha}.\nonumber
\end{align}
For the MT-type bound, where $\mathcal{R}(t)$ can be expressed as,
\begin{align}
    \mathcal{R}(t)  =& 4  \sum_{n=0}^{1}{\Tr[\hat{\Gamma } e^{-2\hat{\Gamma } t}\mathcal{X}_{n,n}\rho(0)] \Tr[\mathcal{X}_{n,n}\rho(0)]} - 2\Tr[\hat{\Gamma } e^{-\hat{\Gamma } t}\rho(0)]\Tr[ e^{-\hat{\Gamma } t}\rho(0)]
        \nonumber\\
        =&4 \sum_{n=0}^{1} \gamma_n' t e^{-2\gamma_n't} c_n^4
            - 2 \sum_{\substack{ n, m = 0}}^{1} \gamma_{\text{n}}' t e^{-(\gamma_{\text{n}}'+\gamma_{\text{m}}')t} |c_n|^2 |c_m|^2
            \nonumber\\
        =&  -2 \nu   e^{2\nu t} \sin^4 \alpha
            +2 \nu   e^{\nu t} \sin^2 \alpha  \cos^2 \alpha  ,\nonumber
\end{align}
$\Delta \hat{\Omega } ^2$ can be expressed as,
\begin{align}
    \Delta \hat{\Omega } ^2 = \frac{1}{2}\sum_{n=0}^{1} \sum_{m=0}^{1} (\omega_n' - \omega_m')^2 |c_n|^2 |c_m|^2
    = \mu^2  \sin^2 \alpha \cos^2 \alpha, \nonumber
\end{align}
the MT-type bound then reads,
\begin{align}
    \mathcal{F}_{\text{MT}} (t) &=
        \frac
        {\sqrt{
                -\mathcal{R} (t)t
            +[{      \Delta \hat{\Omega } ^2 t^2 }]e^{-\gamma_{\phi(1)}' t}
        }}
        {
            \Tr[e^{-\hat{\Gamma } t} \rho(0)]
        } \nonumber\\
        &= \frac
        {\sqrt{
                 2 \nu t  e^{2\nu t} \sin^4 \alpha
                -2 \nu t  e^{\nu t} \sin^2 \alpha  \cos^2 \alpha
            +\mu^2 t^2 \sin^2 \alpha \cos^2 \alpha  e^{\nu t}
        }}
        {
             \cos^2 \alpha + e^{\nu t} \sin^2 \alpha
        }. \nonumber\\
        &= \frac
        {\sqrt{
                2 \nu t  \tan^4 \alpha
                -2 \nu t  e^{-\nu t} \tan^2 \alpha
            +\mu^2 t^2 \tan^2 \alpha  e^{-\nu t}
        }}
        {
            e^{-\nu t} +  \tan^2 \alpha
        }. \nonumber
\end{align}
For the G-type bound, we can set
$\langle \hat{H}(t)\rangle $ to be zero, which means that
\begin{align}
    \hat{H}_{\text{G}} = \hat{H} - \langle \hat{H}(t)\rangle \hat{I}, \nonumber
\end{align}
$\hat{H}_{\text{G}}$ satisfies,
\begin{align}
    \mathcal{A} _{\text{GP}} = \langle \widetilde{\psi}(t)   | \frac{d}{dt} \psi(t)\rangle =0. \nonumber
\end{align}
Then $|\dot{\varphi _g} |^2$ can be expressed as,
\begin{align}
    |\dot{\varphi _g} |^2 &= \langle \hat{H}_{\text{G}}^\dagger \hat{H}_{\text{G}}\rangle - \langle \hat{H}_{G}^\dagger \rangle \langle  \hat{H}_{\text{G}}\rangle    \nonumber \\
        &= \frac{(\mu^2+\nu^2)\sin^2 \alpha e^{2\nu t}}{\cos^2 \alpha + e^{2\nu t} \sin^2 \alpha} - \big|\frac{(\mu+i\nu)\sin^2 \alpha e^{2\nu t}}{(\cos^2 \alpha + e^{2\nu t} \sin^2 \alpha)^2} \big|^2  \nonumber\\
        &= (\mu^2+\nu^2)\frac{ e^{2\nu t} \sin^2 \alpha \cos^2 \alpha}{( \cos^2 \alpha +e^{2\nu t} \sin^2 \alpha)^2}, \nonumber
\end{align}
where $\mathcal{F}_{\text{G}} (t)$ takes,
\begin{align}
    \mathcal{F}_{\text{G}} (t) &= \int_{0}^{t} |\dot{\varphi _g} | \,dt \nonumber\\
        &= \int_{0}^{t} \sqrt{(\mu^2+\nu^2)\frac{ e^{2\nu t} \sin^2 \alpha \cos^2 \alpha}{( \cos^2 \alpha +e^{2\nu t} \sin^2 \alpha)^2}} dt  \nonumber\\
        &= \sqrt{\mu^2+\nu^2} \sin \alpha \cos \alpha   \int_{0}^{t}
        \frac{e^{\nu t}}{ \cos^2 \alpha +e^{2\nu t} \sin^2 \alpha} dt . \nonumber
\end{align}
Since $\alpha \in [0, \pi/2]$, we have $\sin \alpha > 0$ and $\cos \alpha > 0$, so $|\sin \alpha \cos \alpha| = \sin \alpha \cos \alpha$. Thus the expression simplifies to,
\begin{align*}
\sqrt{\mu^2 + \nu^2}  \sin \alpha \cos \alpha \int_{0}^{t} \frac{e^{\nu t}}{\cos^2 \alpha + e^{2\nu t} \sin^2 \alpha}  dt.
\end{align*}
First, consider the integral,
\begin{align*}
\mathcal{I}  = \int_{0}^{t} \frac{e^{\nu t}}{\cos^2 \alpha + e^{2\nu t} \sin^2 \alpha}  dt.
\end{align*}
Apply the substitution $u = e^{\nu t}$. Then $du = \nu e^{\nu t}  dt$, so $dt = \dfrac{du}{\nu u}$. The limits change as follows, when $t = 0$, $u = 1$; when $t = t$, $u = e^{\nu t}$. Simple algebra gives,
\begin{align*}
\mathcal{I}  = \int_{1}^{e^{\nu t}} \frac{u}{\cos^2 \alpha + u^2 \sin^2 \alpha} \cdot \frac{du}{\nu u} = \frac{1}{\nu} \int_{1}^{e^{\nu t}} \frac{du}{\cos^2 \alpha + u^2 \sin^2 \alpha}.
\end{align*}
 $\cos^2 \alpha$ from the denominator is,
\begin{align*}
\cos^2 \alpha + u^2 \sin^2 \alpha = \cos^2 \alpha (1 + u^2 \tan^2 \alpha).
\end{align*}
So,
\begin{align*}
\mathcal{I}  = \frac{1}{\nu} \int_{1}^{e^{\nu t}} \frac{du}{\cos^2 \alpha (1 + u^2 \tan^2 \alpha)} = \frac{1}{\nu \cos^2 \alpha} \int_{1}^{e^{\nu t}} \frac{du}{1 + (u \tan \alpha)^2}.
\end{align*}
Now apply the substitution $w = u \tan \alpha$. Then $dw = \tan \alpha  du$, so $du = \dfrac{dw}{\tan \alpha}$. The limits change as follows, when $u = 1$, $w = \tan \alpha$; when $u = e^{\nu t}$, $w = e^{\nu t} \tan \alpha$. Simple derivation gives,
\begin{align*}
\mathcal{I}  = \frac{1}{\nu \cos^2 \alpha} \int_{\tan \alpha}^{e^{\nu t} \tan \alpha} \frac{1}{1 + w^2} \cdot \frac{dw}{\tan \alpha} = \frac{1}{\nu \cos^2 \alpha \tan \alpha} \int_{\tan \alpha}^{e^{\nu t} \tan \alpha} \frac{dw}{1 + w^2}.
\end{align*}
Simplify the coefficient,
\begin{align*}
\frac{1}{\cos^2 \alpha \tan \alpha} = \frac{1}{\cos^2 \alpha} \cdot \frac{\cos \alpha}{\sin \alpha} = \frac{1}{\sin \alpha \cos \alpha}.
\end{align*}
The integral simplifies to,
\begin{align*}
\int \frac{dw}{1 + w^2} = \arctan w.
\end{align*}
So,
\begin{align*}
\mathcal{I}  = \frac{1}{\nu \sin \alpha \cos \alpha} \left[ \arctan w \right]_{\tan \alpha}^{e^{\nu t} \tan \alpha} = \frac{1}{\nu \sin \alpha \cos \alpha} \left( \arctan(e^{\nu t} \tan \alpha) - \arctan(\tan \alpha) \right).
\end{align*}
Since $\alpha \in (0, \pi/2)$, we have $\arctan(\tan \alpha) = \alpha$, so,
\begin{align*}
\mathcal{I}  = \frac{1}{\nu \sin \alpha \cos \alpha} \left( \arctan(e^{\nu t} \tan \alpha) - \alpha \right).
\end{align*}
Substituting  this equation  back into the original expression, we find
\begin{align*}
\sqrt{\mu^2 + \nu^2}  \sin \alpha \cos \alpha \cdot \mathcal{I}  = \sqrt{\mu^2 + \nu^2}  \sin \alpha \cos \alpha \cdot \frac{1}{\nu \sin \alpha \cos \alpha} \left( \arctan(e^{\nu t} \tan \alpha) - \alpha \right).
\end{align*}
The $\sin \alpha \cos \alpha$ terms cancel,
\begin{align*}
\sqrt{\mu^2 + \nu^2}  \sin \alpha \cos \alpha \cdot \mathcal{I}  = \frac{\sqrt{\mu^2 + \nu^2}}{\nu} \left( \arctan(e^{\nu t} \tan \alpha) - \alpha \right).
\end{align*}
Therefore, the integral reduces  to,
\begin{align*}
    \mathcal{F}_{\text{G}} (t) =\frac{\sqrt{\mu^{2} + \nu^{2}}} {\nu} \left( \arctan\left(e^{\nu t} \tan \alpha\right) - \alpha \right).
\end{align*}


\section{Near the shortest evolution time given by the MT-type bound}\label{appendix H}

In Appendix ~\ref{appendix A1}, Eq(\ref{eq,state}) gives the condition that the initial state meets the shortest evolution requirement  should satisfy,
\begin{align}
    \tan^2 \alpha = \frac{|c_1|^2}{|c_0|^2} = e^{(\gamma_1 - \gamma_0) \tau} = e^{ - \nu \tau}, \nonumber
\end{align}
where $\tau = \frac{\pi}{\mu}$ is the shortest evolution time of the system.
Applying the two conditions into the ML-type bound, we have
\begin{align}
    \mathcal{F}_{\text{ML}}(\tau)=\frac{\tan^2\alpha\cdot\mu \tau}{e^{-\nu \tau}+\tan^2\alpha}=\frac{e^{-\nu\pi/\mu}\cdot\mu\cdot\frac{\pi}{\mu}}{e^{-\nu\cdot\pi/\mu}+e^{-\nu\pi/\mu}}=\frac{\pi e^{-\nu\pi/\mu}}{2e^{-\nu\pi/\mu}}=\frac{\pi}{2}.\nonumber
\end{align}
Similary, for the MT-type bound, we have
\begin{align}
    \mathcal{F}_{\text{MT}} (\tau)&=\frac
        {\sqrt{
                2 \nu \tau  \tan^4 \alpha
                -2 \nu \tau  e^{-\nu \tau} \tan^2 \alpha
            +\mu^2 \tau^2 \tan^2 \alpha  e^{-\nu \tau}
        }}
        {
            e^{-\nu \tau} +  \tan^2 \alpha
        }\nonumber\\
    & = \frac
    {\sqrt{
            2 \nu \tau  e^{-2\nu \tau}
            -2 \nu \tau  e^{-\nu \tau} e^{-\nu \tau}
        +\mu^2 (\frac{\pi}{\mu})^2 e^{-\nu \tau}  e^{-\nu \tau}
    }}
    {
        e^{-\nu \tau} +  e^{-\nu \tau}
    }\nonumber\\
    & = \frac{\pi}{2}.  \nonumber
\end{align}
 This result  shows that the ML-type bound and the MT-type bound are equal at the shortest evolution time $\tau = \frac{\pi}{\mu}$,
which is consistent with the observation  in Appendix ~\ref{appendix A1}.

To simplify the calculation, we define
\begin{align}
    \mathcal{F}_{\text{ML}} &= \frac{\mathcal{F}_{\text{\{ML ,UP\}}}}{\mathcal{F}_{\text{\{ML ,DOWN\}}}} =\frac{\Tr [e^{-\hat{\Gamma } \tau} \hat{\Omega }  \rho(0)] \tau}{\Tr [e^{-\hat{\Gamma } \tau} \rho(0)]} \nonumber\\
    \mathcal{F}_{\text{MT}} &= \frac{\mathcal{F}_{\text{\{MT ,UP\}}}}{\mathcal{F}_{\text{\{MT ,DOWN\}}}} =\frac
    {\sqrt{
            -\mathcal{R} (\tau)\tau
        +[{      \Delta \hat{\Omega } ^2 \tau^2 }]e^{-\gamma_{\phi (1)}' \tau}
    }}
    {
        \Tr[e^{-\hat{\Gamma } \tau} \rho(0)]   \nonumber
    }
\end{align}
Consider a set of initial states $|\psi(0)\rangle = c_0|0\rangle + c_1|E_1\rangle + c_2|E_2\rangle$. The probabilities $p_i = |c_i|^2$ and phase angles $x_i = 2\pi E_i \tau / \hbar$ ($i=1,2$) at time $\tau$ satisfy,
\begin{align}
p_1 e^{y_1} \sin x_1 + p_2 e^{y_2} \sin x_2 &= 0, \label{eq,mod20} \\
p_0 e^{y_0} + p_1 e^{y_1} \cos x_1 + p_2 e^{y_2} \cos x_2 &= 0, \label{eq,mod21}
\end{align}
with $y_n = \gamma_n \tau$, $\gamma_1 = 0 < \gamma_2 < \gamma_0$, and the normalization $p_0 + p_1 + p_2 = 1$.

Set $x_2 = \pi + x_1 - \delta \sin x_1$, where $0<\delta \ll 1$ is a  parameter. Substituting these equation into \eqref{eq,mod20} and using $e^{y_1} = e^0 = 1$, we have
\begin{align*}
p_1 \sin x_1 + p_2 e^{y_2} \sin(\pi + x_1 - \delta \sin x_1) &= 0, \\
p_1 \sin x_1 - p_2 e^{y_2} \sin(x_1 - \delta \sin x_1) &= 0.
\end{align*}
Expanding $\sin(x_1 - \delta \sin x_1) \approx \sin x_1 - \delta \sin x_1 \cos x_1$ for small $\delta$ yields,
\begin{align}
p_1 \sin x_1 - p_2 e^{y_2} (\sin x_1 - \delta \sin x_1 \cos x_1) &= 0, \nonumber \\
\sin x_1 (p_1 - p_2 e^{y_2} + \delta p_2 e^{y_2} \cos x_1) &= 0. \label{eq,simplified}
\end{align}
For $\sin x_1 \neq 0$, Eq. \eqref{eq,simplified} reduces to,
\begin{equation}
p_1 = p_2 e^{y_2} (1 - \delta \cos x_1) + \mathcal{O}(\delta^2). \label{eq,p1_relation}
\end{equation}
Substitute $x_2$ and Eq. \eqref{eq,p1_relation} into Eq. \eqref{eq,mod21} and  expand $\cos x_2$,
\begin{align*}
\cos x_2 &= \cos(\pi + x_1 - \delta \sin x_1) \\
&= -\cos(x_1 - \delta \sin x_1) \\
&\approx -(\cos x_1 + \delta \sin^2 x_1).
\end{align*}
Equation \eqref{eq,mod21} becomes,
\begin{align*}
p_0 e^{y_0} + p_1 \cos x_1 + p_2 e^{y_2} (-\cos x_1 - \delta \sin^2 x_1) &= 0.
\end{align*}
Considering  Eq. \eqref{eq,p1_relation}, we have
\begin{align*}
p_0 e^{y_0} + \left[p_2 e^{y_2} (1 - \delta \cos x_1)\right] \cos x_1 - p_2 e^{y_2} (\cos x_1 + \delta \sin^2 x_1) &= 0.
\end{align*}
Simplifying the trigonometric terms,
\begin{align*}
(1 - \delta \cos x_1)\cos x_1 - \cos x_1 - \delta \sin^2 x_1 &= -\delta (\cos^2 x_1 + \sin^2 x_1) = -\delta,
\end{align*}
we find,
\begin{equation}
p_0 e^{y_0} = \delta p_2 e^{y_2} + \mathcal{O}(\delta^2). \label{eq,p0_relation}
\end{equation}
The normalization $p_0 + p_1 + p_2 = 1$ is met  using \eqref{eq,p1_relation} and \eqref{eq,p0_relation}. Define $C = e^{y_2 - y_1} = e^{y_2}$ and $K = 1 + C$,
\begin{align*}
\delta p_2 e^{y_2 - y_0} + p_2 e^{y_2} (1 - \delta \cos x_1) + p_2 &= 1, \\
p_2 \left[ \delta e^{y_2 - y_0} + e^{y_2} (1 - \delta \cos x_1) + 1 \right] &= 1.
\end{align*}
Expanding $p_2$ to the first order in $\delta$, we have
\begin{equation}
p_2 = \frac{1}{K} - \delta \frac{e^{y_2 - y_0} - e^{y_2} \cos x_1}{K^2} + \mathcal{O}(\delta^2). \label{eq,p2_final}
\end{equation}
Substituting Eq. \eqref{eq,p2_final} into Eq. \eqref{eq,p1_relation} and Eq. \eqref{eq,p0_relation}, we find
\begin{align}
p_1 &= \frac{e^{y_2}}{K} - \delta \frac{e^{y_2} (\cos x_1 + e^{y_2 - y_0})}{K^2} + \mathcal{O}(\delta^2),  \nonumber\\
p_0 &= \delta \frac{e^{y_2 - y_0}}{K} + \mathcal{O}(\delta^2). \nonumber
\end{align}

The expression of $\mathcal{F}_{\text{\{ML ,UP\}}}(\tau) = x_1 e^{y_1} p_1 + x_2 e^{y_2} p_2$ into the first order in $\delta$ yields,
\begin{align}
x_2 &= \pi + x_1 - \delta \sin x_1, \label{eq,x2_perturb} \\
p_1 &= \frac{e^{y_2}}{1 + e^{y_2}} - \delta \frac{e^{y_2} (\cos x_1 + e^{y_2 - y_0})}{(1 + e^{y_2})^2} + \mathcal{O}(\delta^2), \label{eq,p1_perturb} \\
p_2 &= \frac{1}{1 + e^{y_2}} - \delta \frac{e^{y_2 - y_0} - e^{y_2} \cos x_1}{(1 + e^{y_2})^2} + \mathcal{O}(\delta^2), \label{eq,p2_perturb}
\end{align}
where $y_n = \gamma_n \tau$, $\gamma_1 = 0$, $0 < \gamma_2 < \gamma_0$, and $K = 1 + e^{y_2}$. Using Eqs \eqref{eq,x2_perturb}-\eqref{eq,p2_perturb}, we find
\noindent Zeroth-order term ($\delta^0$),
\begin{align*}
\mathcal{F}_{\text{\{ML,UP\}}}^{(0)} &= (2x_1 + \pi) \frac{e^{y_2}}{1 + e^{y_2}} \nonumber \\
&= (2x_1 + \pi) \frac{\exp(\gamma_2 \tau)}{1 + \exp(\gamma_2 \tau)},
\end{align*}
and
\noindent first-order coefficient ($\delta^1$),
\begin{align}
\mathcal{F}_{\text{\{ML ,UP\}}}^{(1)} &= -\frac{e^{y_2}}{(1 + e^{y_2})^2} \left[ e^{y_2 - y_0} (2x_1 + \pi) + \cos x_1 \left( x_1 (1 - e^{y_2}) - \pi e^{y_2} \right) \right] \nonumber \\
&\quad - \sin x_1 \frac{e^{y_2}}{1 + e^{y_2}}, \label{eq,fml_first_raw}
\end{align}
which simplifies to,
\begin{align}
\mathcal{F}_{\text{\{ML ,UP\}}}^{(1)} &= -\frac{\exp(\gamma_2 \tau)}{(1 + \exp(\gamma_2 \tau))^2} \left[ \exp((\gamma_2 - \gamma_0) \tau) (2x_1 + \pi) \right. \nonumber \\
&\quad \left. + \cos x_1 \left( x_1 (1 - \exp(\gamma_2 \tau)) - \pi \exp(\gamma_2 \tau) \right) \right] \nonumber \\
&\quad - \sin x_1 \frac{\exp(\gamma_2 \tau)}{1 + \exp(\gamma_2 \tau)}. \label{eq,fml_first_simplified}
\end{align}

\noindent Complete first-order expansion,
\begin{align}
\mathcal{F}_{\text{\{ML ,UP\}}}(\tau) &= \mathcal{F}_{\text{\{ML ,UP\}}}^{(0)} + \delta \mathcal{F}_{\text{\{ML ,UP\}}}^{(1)} + \mathcal{O}(\delta^2) \nonumber \\
&= (2x_1 + \pi) \frac{\exp(\gamma_2 \tau)}{1 + \exp(\gamma_2 \tau)} \nonumber \\
&\quad - \delta \left\{ \frac{\exp(\gamma_2 \tau)}{(1 + \exp(\gamma_2 \tau))^2} \left[ \exp((\gamma_2 - \gamma_0) \tau) (2x_1 + \pi) \right. \right. \nonumber \\
&\quad \left. \left. + \cos x_1 \left( x_1 (1 - \exp(\gamma_2 \tau)) - \pi \exp(\gamma_2 \tau) \right) \right] + \sin x_1 \frac{\exp(\gamma_2 \tau)}{1 + \exp(\gamma_2 \tau)} \right\} \nonumber \\
&\quad + \mathcal{O}(\delta^2). \label{eq,fml_full}
\end{align}
The parameter $\delta$ is related to the ground-state occupation via $p_0 = \delta \frac{\exp((\gamma_2 - \gamma_0)\tau)}{1 + \exp(\gamma_2 \tau)} + \mathcal{O}(\delta^2)$. For $\gamma_n \to 0$, Eq. \eqref{eq,fml_full} reduces to the case of unitary evolution.

 The expression for $\mathcal{F}_{\{\text{MT, UP}\}}(\tau)^2 = -\mathcal{R}(\tau)\tau + \Delta^2 e^{-\gamma_{\phi (1)}' \tau}$, where,
\begin{align*}
\mathcal{R}(\tau)\tau &= -2y_0 e^{2y_0} p_0^2 - 2y_2 e^{2y_2} p_2^2 + 2\left[ y_0 e^{y_0} p_0 p_1 + (y_0 + y_2) e^{y_0 + y_2} p_0 p_2 + y_2 e^{y_2} p_2 p_1 \right], \\
\Delta^2 e^{-\gamma_{\phi (1)}' \tau}&= e^{y_2} \left[ x_1^2 p_0 p_1 + x_2^2 p_0 p_1 + (x_1 - x_2)^2 p_1 p_2 \right],
\end{align*}
with $y_n = \gamma_n \tau$ and $\gamma_1 = 0$. The perturbative expansions are given by,
\begin{align*}
p_0 &= \delta \frac{e^{y_2 - y_0}}{1 + e^{y_2}} + \mathcal{O}(\delta^2), \\
p_1 &= \frac{e^{y_2}}{1 + e^{y_2}} - \delta \frac{e^{y_2} (\cos x_1 + e^{y_2 - y_0})}{(1 + e^{y_2})^2} + \mathcal{O}(\delta^2), \\
p_2 &= \frac{1}{1 + e^{y_2}} - \delta \frac{e^{y_2 - y_0} - e^{y_2} \cos x_1}{(1 + e^{y_2})^2} + \mathcal{O}(\delta^2), \\
x_2 &= \pi + x_1 - \delta \sin x_1 + \mathcal{O}(\delta^2).
\end{align*}

Substituting $\delta = 0$ into the expansions,
\begin{align}
p_0^{(0)} &= 0, \quad p_1^{(0)} = \frac{e^{y_2}}{1 + e^{y_2}}, \quad p_2^{(0)} = \frac{1}{1 + e^{y_2}}, \quad x_2^{(0)} = \pi + x_1.
\end{align}
Substituting these into $\mathcal{R}(\tau)$, we have
\begin{align*}
\mathcal{R}^{(0)}(\tau)\tau &= -2y_0 e^{2y_0} (0)^2 - 2y_2 e^{2y_2} \left(\frac{1}{1 + e^{y_2}}\right)^2 \nonumber \\
&\quad + 2\left[ y_0 e^{y_0} (0) \left(\frac{e^{y_2}}{1 + e^{y_2}}\right) + (y_0 + y_2) e^{y_0 + y_2} (0) \left(\frac{1}{1 + e^{y_2}}\right) + y_2 e^{y_2} \left(\frac{1}{1 + e^{y_2}}\right) \left(\frac{e^{y_2}}{1 + e^{y_2}}\right) \right] \nonumber \\
&= 0 - 2y_2 e^{2y_2} \frac{1}{(1 + e^{y_2})^2} + 2y_2 e^{y_2} \frac{e^{y_2}}{(1 + e^{y_2})^2} \nonumber \\
&= -\frac{2y_2 e^{2y_2}}{(1 + e^{y_2})^2} + \frac{2y_2 e^{2y_2}}{(1 + e^{y_2})^2} = 0.
\end{align*}
Similarly, for $\Delta^2$,
\begin{align*}
[\Delta^2 e^{-\gamma_{\phi (1)}' \tau}]^{(0)} &= e^{y_2} \left[ x_1^2 (0) \left(\frac{e^{y_2}}{1 + e^{y_2}}\right) + (\pi + x_1)^2 (0) \left(\frac{e^{y_2}}{1 + e^{y_2}}\right) + (x_1 - (\pi + x_1))^2 \left(\frac{e^{y_2}}{1 + e^{y_2}}\right) \left(\frac{1}{1 + e^{y_2}}\right) \right] \nonumber \\
&= e^{y_2} \left[ 0 + 0 + (-\pi)^2 \frac{e^{y_2}}{(1 + e^{y_2})^2} \right] = \pi^2 \frac{e^{2y_2}}{(1 + e^{y_2})^2}.
\end{align*}
Thus,
\begin{align*}
\mathcal{F}_{\{\text{MT, UP}\}}^{(0)}(\tau)^2 &= -\mathcal{R}^{(0)}(\tau) + \Delta^{2(0)} = \pi^2 \frac{e^{2y_2}}{(1 + e^{y_2})^2}, \\
\mathcal{F}_{\{\text{MT, UP}\}}^{(0)}(\tau) &= \pi \frac{e^{y_2}}{1 + e^{y_2}}.
\end{align*}
The first-order correction $\mathcal{F}_{\{\text{MT, UP}\}}^{(1)}(\tau)$ is derived from the terms with $\delta$ in $\mathcal{F}_{\{\text{MT, UP}\}}(\tau)^2$. Substituting the expansions into $\mathcal{R}(\tau)$, we arrive at,
\begin{align*}
\mathcal{R}^{(1)}(\tau)\tau &= -4y_2 e^{2y_2} p_2^{(0)} p_2^{(1)} + 2y_0 e^{y_0} p_0^{(1)} p_1^{(0)} + 2(y_0 + y_2) e^{y_0 + y_2} p_0^{(1)} p_2^{(0)} \nonumber \\
&\quad + 2y_2 e^{y_2} \left( p_2^{(0)} p_1^{(1)} + p_2^{(1)} p_1^{(0)} \right)
\end{align*}
with coefficients,
\begin{align*}
p_0^{(1)} &= \frac{e^{y_2 - y_0}}{1 + e^{y_2}}, \quad p_1^{(1)} = -\frac{e^{y_2} (\cos x_1 + e^{y_2 - y_0})}{(1 + e^{y_2})^2}, \quad p_2^{(1)} = -\frac{e^{y_2 - y_0} - e^{y_2} \cos x_1}{(1 + e^{y_2})^2}.
\end{align*}
Substituting these into the expression for $\mathcal{R}^{(1)}(\tau)$,
\begin{align*}
\mathcal{R}^{(1)}(\tau) &= -4y_2 e^{2y_2} \left(\frac{1}{1 + e^{y_2}}\right) \left(-\frac{e^{y_2 - y_0} - e^{y_2} \cos x_1}{(1 + e^{y_2})^2}\right) \nonumber \\
&\quad + 2y_0 e^{y_0} \left(\frac{e^{y_2 - y_0}}{1 + e^{y_2}}\right) \left(\frac{e^{y_2}}{1 + e^{y_2}}\right) \nonumber \\
&\quad + 2(y_0 + y_2) e^{y_0 + y_2} \left(\frac{e^{y_2 - y_0}}{1 + e^{y_2}}\right) \left(\frac{1}{1 + e^{y_2}}\right) \nonumber \\
&\quad + 2y_2 e^{y_2} \left[ \left(\frac{1}{1 + e^{y_2}}\right) \left(-\frac{e^{y_2} (\cos x_1 + e^{y_2 - y_0})}{(1 + e^{y_2})^2}\right) + \left(-\frac{e^{y_2 - y_0} - e^{y_2} \cos x_1}{(1 + e^{y_2})^2}\right) \left(\frac{e^{y_2}}{1 + e^{y_2}}\right) \right].
\end{align*}
Similarly for $[\Delta^2 e^{-\gamma_{\phi (1)}' \tau}]^2$,
\begin{align*}
[\Delta^2 e^{-\gamma_{\phi (1)}' \tau}]^{(1)} &= e^{y_2} \Big[ p_0^{(1)} p_1^{(0)} \left( x_1^2 + (x_2^{(0)})^2 \right) + \pi^2 \left( p_1^{(0)} p_2^{(1)} + p_1^{(1)} p_2^{(0)} \right) - 2\pi \sin x_1  p_1^{(0)} p_2^{(0)} \Big]
\end{align*}
with $x_2^{(0)} = \pi + x_1$ and $x_2^{(1)} = -\sin x_1$. Then
\begin{align*}
[\Delta^2 e^{-\gamma_{\phi (1)}' \tau}]^{(1)} &= e^{y_2} \Bigg[ \left(\frac{e^{y_2 - y_0}}{1 + e^{y_2}}\right) \left(\frac{e^{y_2}}{1 + e^{y_2}}\right) \left( x_1^2 + (\pi + x_1)^2 \right) \nonumber \\
&\quad + \pi^2 \left[ \left(\frac{e^{y_2}}{1 + e^{y_2}}\right) \left(-\frac{e^{y_2 - y_0} - e^{y_2} \cos x_1}{(1 + e^{y_2})^2}\right) + \left(-\frac{e^{y_2} (\cos x_1 + e^{y_2 - y_0})}{(1 + e^{y_2})^2}\right) \left(\frac{1}{1 + e^{y_2}}\right) \right] \nonumber \\
&\quad - 2\pi \sin x_1 \left(\frac{e^{y_2}}{1 + e^{y_2}}\right) \left(\frac{1}{1 + e^{y_2}}\right) \Bigg].
\end{align*}
Simplify these expressions, the first-order correction to $\mathcal{F}_{\{\text{MT, UP}\}}(\tau)^2$ takes,
\begin{align*}
\mathcal{F}_{\{\text{MT, UP}\}}^2{}^{(1)}(\tau) &= -2 (y_0 - y_2) \frac{e^{2y_2}}{(1+e^{y_2})^2} - 4 y_2 \frac{e^{2y_2} e^{y_2 - y_0}}{(1+e^{y_2})^3} - 2 (y_0 + y_2) \frac{e^{y_2 - y_0}}{(1+e^{y_2})^2} \nonumber \\
&\quad + e^{y_2 - y_0} \frac{e^{y_2} (2x_1^2 + 2\pi x_1 + \pi^2)}{(1+e^{y_2})^2} - \pi^2 e^{y_2} \frac{2 e^{y_2 - y_0} + (1 - e^{y_2}) \cos x_1}{(1+e^{y_2})^3} \nonumber \\
&\quad - 2\pi \frac{e^{y_2} \sin x_1}{(1+e^{y_2})^2}.
\end{align*}
The first-order correction to $\mathcal{F}_{\{\text{MT, UP}\}}(\tau)$ is then,
\begin{align*}
\mathcal{F}_{\{\text{MT, UP}\}}^{(1)}(\tau) &= \frac{\mathcal{F}_{\{\text{MT, UP}\}}^2{}^{(1)}(\tau)}{2 \mathcal{F}_{\{\text{MT, UP}\}}^{(0)}(\tau)} = \frac{\mathcal{F}_{\{\text{MT, UP}\}}^2{}^{(1)}(\tau) (1 + e^{y_2})}{2 \pi e^{y_2}}.
\end{align*}
Substituting and simplifying term by term,
\begin{align*}
\mathcal{F}_{\{\text{MT, UP}\}}^{(1)}(\tau) &=
  -\frac{(y_0 - y_2) e^{y_2}}{\pi (1+e^{y_2})}
  -\frac{2 y_2 e^{2y_2 - y_0}}{\pi (1+e^{y_2})^2}
  -\frac{(y_0 + y_2) e^{-y_0}}{\pi (1+e^{y_2})} \nonumber \\
&\quad + \frac{e^{-y_0} (2x_1^2 + 2\pi x_1 + \pi^2)}{2\pi (1+e^{y_2})}
  - \frac{\pi e^{y_2 - y_0}}{(1+e^{y_2})^2} \nonumber \\
&\quad - \frac{\pi (1 - e^{y_2}) \cos x_1}{2 (1+e^{y_2})^2}
  - \frac{\sin x_1}{1+e^{y_2}}.
\end{align*}

\[
\mathcal{F}_{\text{MT}}(\tau) = \frac{\mathcal{F}_{\{\text{MT, UP}\}}(\tau)}{\mathcal{F}_{\{\text{MT, DOWN}\}}(\tau)}
\]

The denominator is given by,
\[
\mathcal{F}_{\{\text{MT, DOWN}\}}(\tau) = p_0 e^{\gamma_0 \tau} + p_1 + p_2 e^{\gamma_2 \tau}
\]
with coefficients,
\begin{align*}
p_0 &= \delta \frac{e^{\gamma_2 \tau - \gamma_0 \tau}}{1 + e^{\gamma_2 \tau}} + \mathcal{O}(\delta^2) \\
p_1 &= \frac{e^{\gamma_2 \tau}}{1 + e^{\gamma_2 \tau}} - \delta \frac{e^{\gamma_2 \tau} (\cos x_1 + e^{\gamma_2 \tau - \gamma_0 \tau})}{(1 + e^{\gamma_2 \tau})^2} + \mathcal{O}(\delta^2) \\
p_2 &= \frac{1}{1 + e^{\gamma_2 \tau}} - \delta \frac{e^{\gamma_2 \tau - \gamma_0 \tau} - e^{\gamma_2 \tau} \cos x_1}{(1 + e^{\gamma_2 \tau})^2} + \mathcal{O}(\delta^2)
\end{align*}

The dominant term (zeroth-order in $\delta$) is,
\begin{align*}
\mathcal{F}_{\{\text{MT, DOWN}\}}^{(0)}(\tau) &= \left. \mathcal{F}_{\{\text{MT, DOWN}\}}(\tau) \right|_{\delta=0} \\
&= \frac{e^{\gamma_2 \tau}}{1 + e^{\gamma_2 \tau}} + \frac{1}{1 + e^{\gamma_2 \tau}} e^{\gamma_2 \tau} \\
&= \frac{e^{\gamma_2 \tau} + e^{\gamma_2 \tau}}{1 + e^{\gamma_2 \tau}} \\
&= \frac{2e^{\gamma_2 \tau}}{1 + e^{\gamma_2 \tau}}
\end{align*}

The numerator with $x_1 = \pi(1 - \alpha)/(2\alpha)$ is,
\begin{align*}
\mathcal{F}_{\{\text{MT, UP}\}}(\tau) =& \pi \frac{e^{\gamma_2 \tau}}{1 + e^{\gamma_2 \tau}} \\
& + \delta \left[ -\frac{(\gamma_0 - \gamma_2) \tau e^{\gamma_2 \tau}}{\pi (1 + e^{\gamma_2 \tau})} - \frac{2 \gamma_2 \tau e^{(2\gamma_2 - \gamma_0) \tau}}{\pi (1 + e^{\gamma_2 \tau})^2} - \frac{(\gamma_0 + \gamma_2) \tau e^{-\gamma_0 \tau}}{\pi (1 + e^{\gamma_2 \tau})} \right. \\
& \left. + \frac{\pi e^{-\gamma_0 \tau}}{4 (1 + e^{\gamma_2 \tau})} \left(1 + \frac{1}{\alpha^{2}}\right) - \frac{\pi e^{(\gamma_2 - \gamma_0) \tau}}{(1 + e^{\gamma_2 \tau})^2} \right. \\
& \left. - \frac{\pi (1 - e^{\gamma_2 \tau}) \cos \left( \dfrac{\pi (1 - \alpha)}{2\alpha} \right)}{2 (1 + e^{\gamma_2 \tau})^2} - \frac{\sin \left( \dfrac{\pi (1 - \alpha)}{2\alpha} \right)}{1 + e^{\gamma_2 \tau}} \right] \\
& + \mathcal{O}(\delta^{2})
\end{align*}

Combining numerator and dominant denominator,
\begin{align*}
\mathcal{F}_{\text{MT}}(\tau) &= \frac{\mathcal{F}_{\{\text{MT, UP}\}}(\tau)}{\mathcal{F}_{\{\text{MT, DOWN}\}}^{(0)}(\tau)} = \frac{\mathcal{F}_{\{\text{MT, UP}\}}(\tau)}{\dfrac{2e^{\gamma_2 \tau}}{1 + e^{\gamma_2 \tau}}} \\
&= \frac{1 + e^{\gamma_2 \tau}}{2e^{\gamma_2 \tau}} \mathcal{F}_{\{\text{MT, UP}\}}(\tau)
\end{align*}

Substituting the numerator expression,
\begin{align*}
\mathcal{F}_{\text{MT}}(\tau) =& \frac{1 + e^{\gamma_2 \tau}}{2e^{\gamma_2 \tau}} \left[ \pi \frac{e^{\gamma_2 \tau}}{1 + e^{\gamma_2 \tau}} \right. \\
& + \delta \left( -\frac{(\gamma_0 - \gamma_2) \tau e^{\gamma_2 \tau}}{\pi (1 + e^{\gamma_2 \tau})} - \frac{2 \gamma_2 \tau e^{(2\gamma_2 - \gamma_0) \tau}}{\pi (1 + e^{\gamma_2 \tau})^2} - \frac{(\gamma_0 + \gamma_2) \tau e^{-\gamma_0 \tau}}{\pi (1 + e^{\gamma_2 \tau})} \right. \\
& + \frac{\pi e^{-\gamma_0 \tau}}{4 (1 + e^{\gamma_2 \tau})} \left(1 + \frac{1}{\alpha^{2}}\right) - \frac{\pi e^{(\gamma_2 - \gamma_0) \tau}}{(1 + e^{\gamma_2 \tau})^2} \\
& \left. \left. - \frac{\pi (1 - e^{\gamma_2 \tau}) \cos \left( \dfrac{\pi (1 - \alpha)}{2\alpha} \right)}{2 (1 + e^{\gamma_2 \tau})^2} - \frac{\sin \left( \dfrac{\pi (1 - \alpha)}{2\alpha} \right)}{1 + e^{\gamma_2 \tau}} \right) \right] + \mathcal{O}(\delta^{2})
\end{align*}

The zeroth-order term simplifies to,
\begin{align*}
\frac{1 + e^{\gamma_2 \tau}}{2e^{\gamma_2 \tau}} \cdot \pi \frac{e^{\gamma_2 \tau}}{1 + e^{\gamma_2 \tau}} = \frac{\pi}{2}
\end{align*}

Thus $\mathcal{F}_{\text{MT}}(\tau)$ has the structure,
\[
\mathcal{F}_{\text{MT}}(\tau) = \frac{\pi}{2} + \delta \mathcal{B}(\tau) + \mathcal{O}(\delta^2)
\]
where the first-order coefficient $\mathcal{B}(\tau)$ is,
\begin{align*}
\mathcal{B}(\tau) =& \frac{1 + e^{\gamma_2 \tau}}{2e^{\gamma_2 \tau}} \left[ -\frac{(\gamma_0 - \gamma_2) \tau e^{\gamma_2 \tau}}{\pi (1 + e^{\gamma_2 \tau})} - \frac{2 \gamma_2 \tau e^{(2\gamma_2 - \gamma_0) \tau}}{\pi (1 + e^{\gamma_2 \tau})^2} - \frac{(\gamma_0 + \gamma_2) \tau e^{-\gamma_0 \tau}}{\pi (1 + e^{\gamma_2 \tau})} \right. \\
& + \frac{\pi e^{-\gamma_0 \tau}}{4 (1 + e^{\gamma_2 \tau})} \left(1 + \frac{1}{\alpha^{2}}\right) - \frac{\pi e^{(\gamma_2 - \gamma_0) \tau}}{(1 + e^{\gamma_2 \tau})^2} \\
& \left. - \frac{\pi (1 - e^{\gamma_2 \tau}) \cos \left( \dfrac{\pi (1 - \alpha)}{2\alpha} \right)}{2 (1 + e^{\gamma_2 \tau})^2} - \frac{\sin \left( \dfrac{\pi (1 - \alpha)}{2\alpha} \right)}{1 + e^{\gamma_2 \tau}} \right]
\end{align*}

For time $\tau$ that  satisfies $\mathcal{F}_{\text{MT}}(\tau) \geqslant \frac{\pi}{2}$, where the equality holds if and only if $\tau = \frac{\pi}{\mu_2-\mu_1} (|\psi(0)\rangle = e^{-\gamma_1 \tau}|E_{0}\rangle + e^{-\gamma_0 \tau}|E_{1}\rangle)$,
 we consider parameters ($\gamma_0, \gamma_2 > 0$ and finite $\tau$) without loss of generality, $\mathcal{B}(\tau) > 0$. Let $\mathcal{B} \equiv \inf_t \mathcal{B}(\tau) > 0$. Then for any $\epsilon > 0$, choosing $\delta < \frac{\pi}{2} \epsilon \mathcal{B}^{-1}$ guarantees,
\[
\frac{\pi}{2} < \mathcal{F}_{\text{MT}}(\tau) \leqslant \frac{\pi}{2}(1 + \epsilon)
\]

\section{Derivation of the case near the shortest evolution time for ML-type bound}    \label{appendix I}

Similar to the discussion for  MT type bound, we consider  a set of initial states $|\psi(0)\rangle = c_0|0\rangle + c_1|E_1\rangle + c_{2k+1}|E_{2k+1}\rangle$. The probabilities $p_i = |c_i|^2$ and phase angles $x_i = 2\pi E_i \tau / \hbar$ (for $i = 1, 2k+1$) satisfy,
\begin{align*}
p_1 e^{y_1} \sin x_1 + p_{2k+1} e^{y_{2k+1}} \sin x_{2k+1} &= 0,  \\
p_0 e^{y_0} + p_1 e^{y_1} \cos x_1 + p_{2k+1} e^{y_{2k+1}} \cos x_{2k+1} &= 0,
\end{align*}
with $y_n = \gamma_n \tau$ satisfying $0 = \gamma_0 < \gamma_1 < \gamma_{2k+1}$, and the normalization condition $p_0 + p_1 + p_{2k+1} = 1$. The phase angles are specifically chosen as $x_0 = 0$, $x_1 = \pi$, and $x_{2k+1} = (2k+1)\pi$.
Substituting the chosen phase angles into the trigonometric functions yields,
\begin{align*}
\sin x_1 &= \sin \pi = 0, \\
\sin x_{2k+1} &= \sin((2k+1)\pi) = 0, \\
\cos x_1 &= \cos \pi = -1, \\
\cos x_{2k+1} &= \cos((2k+1)\pi) = -1, \\
e^{y_0} &= e^{\gamma_0 \tau} = e^0 = 1.
\end{align*}
The first line of equation simplifies to,
\[
p_1 e^{y_1} \cdot 0 + p_{2k+1} e^{y_{2k+1}} \cdot 0 = 0,
\]
and the second line of equation reduces to,
\[
p_0 \cdot 1 + p_1 e^{y_1} \cdot (-1) + p_{2k+1} e^{y_{2k+1}} \cdot (-1) = 0,
\]
namely,
\begin{equation}
p_0 - p_1 e^{y_1} - p_{2k+1} e^{y_{2k+1}} = 0.
\label{eq,ortho_constraint}
\end{equation}
This equation should be solved together with the normalization condition,
\begin{equation}
p_0 + p_1 + p_{2k+1} = 1.
\label{eq,norm_constraint}
\end{equation}

Solving these equations, we find from Equation (\ref{eq,ortho_constraint}),
\[
p_0 = p_1 e^{y_1} + p_{2k+1} e^{y_{2k+1}}.
\]
Substituting this into Equation (\ref{eq,norm_constraint}) gives,
\[
(p_1 e^{y_1} + p_{2k+1} e^{y_{2k+1}}) + p_1 + p_{2k+1} = 1,
\]
or,
\begin{equation}
p_1 (e^{y_1} + 1) + p_{2k+1} (e^{y_{2k+1}} + 1) = 1.
\label{eq,combined}
\end{equation}
To make the discussion clear, we introduce a ratio $\beta$ that is defined by,
\[
\frac{p_{2k+1}}{p_1} = \frac{\beta/k^2}{1 - \beta/k^2},
\]
which allows us to rewrite,
\[
p_{2k+1} = \left( \frac{\beta/k^2}{1 - \beta/k^2} \right) p_1.
\]
Substituting this into Equation (\ref{eq,combined}) yields,
\[
p_1 (e^{y_1} + 1) + \left( \frac{\beta/k^2}{1 - \beta/k^2} \right) p_1 (e^{y_{2k+1}} + 1) = 1.
\]
Factoring $p_1$,
\[
p_1 \left[ (e^{y_1} + 1) + \frac{\beta/k^2}{1 - \beta/k^2} (e^{y_{2k+1}} + 1) \right] = 1,
\]
and solving for $p_1$, we have
\[
p_1 = \frac{1}{(e^{y_1} + 1) + \dfrac{\beta/k^2}{1 - \beta/k^2} (e^{y_{2k+1}} + 1)}.
\]
Multiplying numerator and denominator by $1 - \beta/k^2$ simplifies this to,
\[
p_1 = \frac{1 - \beta/k^2}{(1 - \beta/k^2)(e^{y_1} + 1) + (\beta/k^2) (e^{y_{2k+1}} + 1)}.
\]
The expression for $p_{2k+1}$ follows,
\[
p_{2k+1} = \frac{\beta/k^2}{(1 - \beta/k^2)(e^{y_1} + 1) + (\beta/k^2) (e^{y_{2k+1}} + 1)}.
\]
Finally, substituting this into the expression for $p_0$,
\[
p_0 = p_1 e^{y_1} + p_{2k+1} e^{y_{2k+1}} = \frac{(1 - \beta/k^2) e^{y_1} + (\beta/k^2) e^{y_{2k+1}}}{(1 - \beta/k^2)(e^{y_1} + 1) + (\beta/k^2) (e^{y_{2k+1}} + 1)}.
\]
The denominator is the normalization factor $\mathcal{N} $,
\[
\mathcal{N}  = \left(1 - \frac{\beta}{k^2}\right) (e^{\gamma_1 \tau} + 1) + \frac{\beta}{k^2} (e^{\gamma_{2k+1} \tau} + 1).
\]
Collecting these together, we find the probability  given by,
\begin{align*}
p_0 &= \mathcal{N} ^{-1} \left[ \left(1 - \frac{\beta}{k^2}\right) e^{\gamma_1 \tau} + \frac{\beta}{k^2} e^{\gamma_{2k+1} \tau} \right], \\
p_1 &= \mathcal{N} ^{-1} \left(1 - \frac{\beta}{k^2}\right), \\
p_{2k+1} &= \mathcal{N} ^{-1} \left( \frac{\beta}{k^2} \right),
\end{align*}
where $\mathcal{N} $ is the  normalization constant.

The ratio $\dfrac{\mathcal{F}_{\{\mathrm{ML},\ \mathrm{UP}\}}(\tau)}{p_0 + p_1 e^{y_1} + p_{2k+1} e^{y_{2k+1}}}$ is evaluated using the exact expressions for the probabilities and the function $\mathcal{F}$. The probabilities are given by,
\begin{align*}
p_0 &= \mathcal{N} ^{-1} \left[ \left(1 - \frac{\beta}{k^2}\right) e^{\gamma_1 \tau} + \frac{\beta}{k^2} e^{\gamma_{2k+1} \tau} \right], \\
p_1 &= \mathcal{N} ^{-1} \left(1 - \frac{\beta}{k^2}\right), \\
p_{2k+1} &= \mathcal{N} ^{-1} \left( \frac{\beta}{k^2} \right),
\end{align*}
where $\mathcal{N}  = \left(1 - \frac{\beta}{k^2}\right) (e^{\gamma_1 \tau} + 1) + \frac{\beta}{k^2} (e^{\gamma_{2k+1} \tau} + 1)$ is the normalization factor, and $y_1 = \gamma_1 \tau$, $y_{2k+1} = \gamma_{2k+1} \tau$ with $0 = \gamma_0 < \gamma_1 < \gamma_{2k+1}$. The function $\mathcal{F}$ is defined as,
\[
\mathcal{F}_{\{\mathrm{ML},\ \mathrm{UP}\}}(\tau) = \pi e^{\gamma_1 \tau} p_1 + \pi(2k+1) e^{\gamma_{2k+1} \tau} p_{2k+1}.
\]

The denominator simplifies to,
\begin{align*}
\mathcal{F}_{\{\mathrm{ML},\ \mathrm{DOWN}\}}(\tau) =
p_0 + p_1 e^{y_1} + p_{2k+1} e^{y_{2k+1}} &= \mathcal{N} ^{-1} \left[ \left(1 - \frac{\beta}{k^2}\right) e^{\gamma_1 \tau} + \frac{\beta}{k^2} e^{\gamma_{2k+1} \tau} \right] \nonumber \\
&\quad + \mathcal{N} ^{-1} \left(1 - \frac{\beta}{k^2}\right) e^{\gamma_1 \tau} + \mathcal{N} ^{-1} \left( \frac{\beta}{k^2} \right) e^{\gamma_{2k+1} \tau} \nonumber \\
&= \mathcal{N} ^{-1} \left[ 2\left(1 - \frac{\beta}{k^2}\right) e^{\gamma_1 \tau} + 2\frac{\beta}{k^2} e^{\gamma_{2k+1} \tau} \right] \nonumber \\
&= 2p_0.
\end{align*}
The ratio is then expressed as,
\begin{align*}
\mathcal{F}_{\mathrm{ML}}(\tau) =\dfrac{\mathcal{F}_{\{\mathrm{ML},\ \mathrm{UP}\}}(\tau)}{p_0 + p_1 e^{y_1} + p_{2k+1} e^{y_{2k+1}}} &= \frac{\pi e^{\gamma_1 \tau} p_1 + \pi(2k+1) e^{\gamma_{2k+1} \tau} p_{2k+1}}{2p_0} \nonumber \\
&= \frac{\pi}{2} \cdot \frac{ e^{\gamma_1 \tau} p_1 + (2k+1) e^{\gamma_{2k+1} \tau} p_{2k+1} }{ p_0 }.
\end{align*}
Substituting the probability expressions and simplifying it, we have,
\begin{equation}
\mathcal{F}_{\mathrm{ML}}(\tau)  = \frac{\pi}{2} \cdot \frac{ \left(1 - \dfrac{\beta}{k^2}\right) e^{\gamma_1 \tau} + \dfrac{\beta (2k+1)}{k^2} e^{\gamma_{2k+1} \tau} }{ \left(1 - \dfrac{\beta}{k^2}\right) e^{\gamma_1 \tau} + \dfrac{\beta}{k^2} e^{\gamma_{2k+1} \tau} }.
\label{eq,exact_ratio}
\end{equation}

To obtain the asymptotic expansion up to $\mathcal{O}(1/k)$, we expand both the numerator and denominator in Equation (\ref{eq,exact_ratio}). The numerator expands as,
\begin{align*}
\mathcal{F}_{\{\mathrm{ML},\ \mathrm{UP}\}}(\tau)  &= \left(1 - \dfrac{\beta}{k^2}\right) e^{\gamma_1 \tau} + \dfrac{\beta (2k+1)}{k^2} e^{\gamma_{2k+1} \tau} \nonumber \\
&= e^{\gamma_1 \tau} - \dfrac{\beta}{k^2} e^{\gamma_1 \tau} + \dfrac{2k\beta}{k^2} e^{\gamma_{2k+1} \tau} + \dfrac{\beta}{k^2} e^{\gamma_{2k+1} \tau} \nonumber \\
&= e^{\gamma_1 \tau} + \dfrac{2\beta}{k} e^{\gamma_{2k+1} \tau} + \dfrac{\beta}{k^2} \left(e^{\gamma_{2k+1} \tau} - e^{\gamma_1 \tau}\right) \nonumber \\
&= e^{\gamma_1 \tau} + \dfrac{2\beta}{k} e^{\gamma_{2k+1} \tau} + \mathcal{O}\left(\frac{1}{k^2}\right).
\end{align*}
The denominator expands as,
\begin{align*}
\mathcal{F}_{\{\mathrm{ML},\ \mathrm{DOWN}\}}(\tau) &= \left(1 - \dfrac{\beta}{k^2}\right) e^{\gamma_1 \tau} + \dfrac{\beta}{k^2} e^{\gamma_{2k+1} \tau} \nonumber \\
&= e^{\gamma_1 \tau} - \dfrac{\beta}{k^2} e^{\gamma_1 \tau} + \dfrac{\beta}{k^2} e^{\gamma_{2k+1} \tau} \nonumber \\
&= e^{\gamma_1 \tau} + \dfrac{\beta}{k^2} \left(e^{\gamma_{2k+1} \tau} - e^{\gamma_1 \tau}\right) \nonumber \\
&= e^{\gamma_1 \tau} + \mathcal{O}\left(\frac{1}{k^2}\right).
\end{align*}
The ratio of numerator to denominator is,
\begin{align*}
\mathcal{F}_{\mathrm{ML}}(\tau) &= \frac{ e^{\gamma_1 \tau} + \dfrac{2\beta}{k} e^{\gamma_{2k+1} \tau} + \mathcal{O}(k^{-2}) }{ e^{\gamma_1 \tau} + \mathcal{O}(k^{-2}) } \nonumber \\
&= \left( 1 + \dfrac{2\beta}{k} e^{(\gamma_{2k+1} - \gamma_1) \tau} \right) \left( 1 + \mathcal{O}(k^{-2}) \right) \nonumber \\
&= 1 + \dfrac{2\beta}{k} e^{(\gamma_{2k+1} - \gamma_1) \tau} + \mathcal{O}\left(\frac{1}{k^2}\right).
\end{align*}
Multiplying by the constant factor $\pi/2$ gives the final asymptotic expression,
\begin{equation}
\mathcal{F}_{\mathrm{ML}}(\tau) = \dfrac{\pi}{2} + \dfrac{\pi\beta}{k} e^{(\gamma_{2k+1} - \gamma_1) \tau} + \mathcal{O}\left(\dfrac{1}{k^2}\right).
\label{eq,asymptotic_ratio}
\end{equation}
We derive the dominant contribution to $\mathcal{F}_{\{\text{MT}\}}(\tau)$ in the large-$k$ limit ($1/k$ expansion), retaining terms up to $\mathcal{O}(1)$ while neglecting higher-order corrections. The expression originates from the probability distributions,
\begin{align*}
p_0 &= \mathcal{N}^{-1} \left[ \left(1 - \frac{\beta}{k^2}\right) e^{\gamma_1 \tau} + \frac{\beta}{k^2} e^{\gamma_{2k+1} \tau} \right], \\
p_1 &= \mathcal{N}^{-1} \left(1 - \frac{\beta}{k^2}\right), \\
p_{2k+1} &= \mathcal{N}^{-1} \left( \frac{\beta}{k^2} \right), \\
\mathcal{N}  &= \left(1 - \frac{\beta}{k^2}\right) (e^{\gamma_1 \tau} + 1) + \frac{\beta}{k^2} (e^{\gamma_{2k+1} \tau} + 1),
\end{align*}
and,
\begin{align*}
\mathcal{F}_{\text{MT}}(\tau) = \frac{ \sqrt{ \mathcal{F}_{\{\text{MT, UP}\}}(\tau)^2 } }{ p_0 + p_1 e^{\gamma_1 \tau} + p_{2k+1} e^{\gamma_{2k+1} \tau} },
\end{align*}
where $\mathcal{F}_{\{\text{MT, UP}\}}(\tau)^2 = -\mathcal{R}(\tau)\tau + \Delta^2 e^{-\gamma_{\phi (1)}' \tau}$ with definitions,
\begin{align*}
\mathcal{R}(\tau)\tau &= -2\gamma_1 \tau e^{2\gamma_1 \tau} p_1^2 - 2\gamma_{2k+1} \tau e^{2\gamma_{2k+1} \tau} p_{2k+1}^2 \nonumber \\
&+ 2\left[ \gamma_1 \tau e^{\gamma_1 \tau} p_0 p_1 + (\gamma_1 \tau + \gamma_{2k+1} \tau) e^{(\gamma_1 + \gamma_{2k+1}) \tau} p_1 p_{2k+1} + \gamma_{2k+1} \tau e^{\gamma_{2k+1} \tau} p_{2k+1} p_0 \right], \\
\Delta^2 e^{-\gamma_{\phi (1)}' \tau} &= e^{\gamma_1 \tau} \left[ \pi^2 p_1 + ((2k+1)\pi)^2 p_{2k+1} - (\pi p_1 + (2k+1)\pi p_{2k+1})^2 \right].
\end{align*}
The spatial modes are characterized by $x_0 = 0$, $x_1 = \pi$, $x_{2k+1} = (2k+1)\pi$.

For large $k$, we define a small parameter $\epsilon \equiv \beta/k^2 \ll 1$. The rapid change $\gamma_{2k+1} \tau \to -\infty$ implies exponential suppression of terms containing $e^{\gamma_{2k+1} \tau}$. The normalization factor simplifies to,
\begin{align*}
\mathcal{N} &\approx (1 - \epsilon)(e^{\gamma_1 \tau} + 1) + \epsilon = (1 - \epsilon)e^{\gamma_1 \tau} + 1.
\end{align*}
The probabilities consequently reduce to,
\begin{align*}
p_1 &\approx \frac{1 - \epsilon}{(1 - \epsilon)e^{\gamma_1 \tau} + 1}, \\
p_{2k+1} &\approx \frac{\epsilon}{(1 - \epsilon)e^{\gamma_1 \tau} + 1}, \\
p_0 &\approx \frac{(1 - \epsilon)e^{\gamma_1 \tau}}{(1 - \epsilon)e^{\gamma_1 \tau} + 1}.
\end{align*}
Introducing the auxiliary variable $A \equiv e^{\gamma_1 \tau}$ and expanding to first order in $\epsilon$ yields,
\begin{align*}
p_1 &\approx \frac{1}{A + 1} - \epsilon \frac{1}{(A + 1)^2} + \mathcal{O}(\epsilon^2), \\
p_0 &\approx \frac{A}{A + 1} - \epsilon \frac{A}{(A + 1)^2} + \mathcal{O}(\epsilon^2), \\
p_{2k+1} &\approx \frac{\epsilon}{A + 1} + \mathcal{O}(\epsilon^2).
\end{align*}

Term $\Delta^2 e^{-\gamma_{\phi (1)}' \tau}$ requires careful expansion,
\begin{align*}
\Delta^2 e^{-\gamma_{\phi (1)}' \tau} = e^{\gamma_1 \tau} \left[ \pi^2 p_1 + ((2k+1)\pi)^2 p_{2k+1} - \pi^2 (p_1 + (2k+1)p_{2k+1})^2 \right].
\end{align*}
Expanding the quadratic term and neglecting $\mathcal{O}(p_{2k+1}^2)$ contributions (since $p_{2k+1} \sim \mathcal{O}(1/k^2)$) gives,
\begin{align*}
\Delta^2 e^{-\gamma_{\phi (1)}' \tau} &\approx e^{\gamma_1 \tau} \pi^2 \left[ p_1 + (2k+1)^2 p_{2k+1} - p_1^2 - 2(2k+1)p_1 p_{2k+1} \right].
\end{align*}
Substituting the probability expressions and retaining leading terms, we have
\begin{align*}
p_1 - p_1^2 &\approx \frac{A}{(A + 1)^2} + \mathcal{O}(\epsilon), \\
(2k+1)^2 p_{2k+1} &\approx (4k^2 + 4k + 1) \frac{\beta}{k^2} \frac{1}{A + 1} = 4\beta \left(1 + \frac{1}{k} + \frac{1}{4k^2}\right) \frac{1}{A + 1} \approx \frac{4\beta}{A + 1}, \\
2(2k+1)p_1 p_{2k+1} &\sim \mathcal{O}(1/k) \quad (\text{negligible}).
\end{align*}
Thus the considered  term simplifies to,
\begin{align*}
\Delta^2 e^{-\gamma_{\phi (1)}' \tau} &\approx e^{\gamma_1 \tau} \pi^2 \left[ \frac{A}{(A + 1)^2} + \frac{4\beta}{A + 1} \right] \nonumber \\
&= \pi^2 \left[ \frac{A^2}{(A + 1)^2} + \frac{4\beta A}{A + 1} \right] \nonumber \\
&= \pi^2 \left( u^2 + 4\beta u \right),
\end{align*}
where $u \equiv A/(A + 1) = e^{\gamma_1 \tau}/(1 + e^{\gamma_1 \tau})$. The  rate $\gamma_{\phi(1)}'$ vanishes to ensure consistent initial conditions.

The  term $\mathcal{R}(\tau)\tau$ simplifies significantly under the large-$k$ approximation. Exponentially suppressed terms and $\mathcal{O}(1/k)$ contributions are negligible,
\begin{align*}
\mathcal{R}(\tau)\tau &\approx -2\gamma_1 \tau e^{2\gamma_1 \tau} p_1^2 + 2\gamma_1 \tau e^{\gamma_1 \tau} p_0 p_1.
\end{align*}
Substituting $p_0 \approx A/(A + 1)$ and $p_1 \approx 1/(A + 1)$ yields,
\begin{align*}
\mathcal{R}(\tau)\tau &\approx 2\gamma_1 \tau e^{\gamma_1 \tau} \frac{1}{(A + 1)^2} \left[ -e^{\gamma_1 \tau} + A \right] = 0.
\end{align*}
Consequently, $\mathcal{F}_{\{\text{MT, UP}\}}(\tau)^2 \approx \Delta^2 e^{-\gamma_{\phi (1)}' \tau} = \pi^2 (u^2 + 4\beta u)$.

The denominator of $\mathcal{F}_{\{\text{MT}\}}(\tau)$ reduces to,
\begin{align*}
D &= p_0 + p_1 e^{\gamma_1 \tau} + p_{2k+1} e^{\gamma_{2k+1} \tau} \nonumber \\
&\approx p_0 + p_1 A \nonumber \\
&\approx \frac{A}{A + 1} + \frac{1}{A + 1} A = \frac{2A}{A + 1} = 2u.
\end{align*}

Combining these results provides the final expression,
\begin{align*}
\mathcal{F}_{\text{MT}}(\tau) &= \frac{ \sqrt{ \pi^2 (u^2 + 4\beta u) } }{ 2u } \nonumber \\
&= \frac{\pi}{2} \sqrt{ \frac{u^2 + 4\beta u}{u^2} } \nonumber \\
&= \frac{\pi}{2} \sqrt{ 1 + \frac{4\beta}{u} }.
\end{align*}
Substituting $u = e^{\gamma_1 \tau}/(1 + e^{\gamma_1 \tau})$ produces the explicit time-dependence,
\begin{align*}
\frac{1}{u} &= 1 + e^{-\gamma_1 \tau},
\end{align*}
resulting in a compact form,
\begin{align*}
\mathcal{F}_{\mathrm{MT}}(\tau) = \dfrac{\pi}{2} \sqrt{1 + 4\beta \left(1 + e^{-\gamma_1 \tau}\right)}
\end{align*}
This expression retains explicit $\beta$-dependence and captures the dominant $\tau$-dependent behavior for large $k$, with all subleading $1/k$ corrections systematically neglected.

Given,
\begin{align*}
\mathcal{F}_{\text{ML}}(\tau) &= \dfrac{\pi}{2} + \dfrac{\pi\beta}{k} e^{(\gamma_{2k+1} - \gamma_1) \tau} + \mathcal{O}\left(\dfrac{1}{k^2}\right), \\
\mathcal{F}_{\mathrm{MT}}(\tau) &= \dfrac{\pi}{2} \sqrt{1 + 4\beta \left(1 + e^{-\gamma_1 \tau}\right)}, \\
\alpha &= \dfrac{\mathcal{F}_{\{\mathrm{MT}\}}(\tau)}{\mathcal{F}_{\{\mathrm{ML}\}}(\tau)}.
\end{align*}
Solving for $\beta$ in terms of $\alpha$,
\begin{align*}
\alpha &= \dfrac{ \dfrac{\pi}{2} \sqrt{1 + 4\beta (1 + e^{-\gamma_1 \tau})} }{ \dfrac{\pi}{2} + \dfrac{\pi\beta}{k} e^{(\gamma_{2k+1} - \gamma_1) \tau} } \\
&= \dfrac{ \sqrt{1 + 4\beta (1 + e^{-\gamma_1 \tau})} }{ 1 + \dfrac{2\beta}{k} e^{(\gamma_{2k+1} - \gamma_1) \tau} }.
\end{align*}
Squaring both sides and rearranging,
\begin{align*}
\alpha^2 \left( 1 + \dfrac{4\beta}{k} e^{(\gamma_{2k+1} - \gamma_1) \tau} + \mathcal{O}\left(\dfrac{1}{k^2}\right) \right) &= 1 + 4\beta (1 + e^{-\gamma_1 \tau}) \\
\alpha^2 - 1 &= 4\beta (1 + e^{-\gamma_1 \tau}) - \dfrac{4\alpha^2\beta}{k} e^{(\gamma_{2k+1} - \gamma_1) \tau} + \mathcal{O}\left(\dfrac{1}{k^2}\right) \\
\beta &= \dfrac{\alpha^2 - 1}{4\left(1 + e^{-\gamma_1 \tau} - \dfrac{\alpha^2}{k} e^{(\gamma_{2k+1} - \gamma_1) \tau}\right)} + \mathcal{O}\left(\dfrac{1}{k^2}\right).
\end{align*}
Substituting into $\mathcal{F}_{\{\text{ML}\}}(\tau)$,
\begin{align*}
\mathcal{F}_{\text{ML}}(\tau) &= \dfrac{\pi}{2} + \dfrac{\pi}{k} e^{(\gamma_{2k+1} - \gamma_1) \tau} \left[ \dfrac{\alpha^2 - 1}{4\left(1 + e^{-\gamma_1 \tau} - \dfrac{\alpha^2}{k} e^{(\gamma_{2k+1} - \gamma_1) \tau}\right)} \right] + \mathcal{O}\left(\dfrac{1}{k^2}\right) \\
&= \dfrac{\pi}{2} + \dfrac{\pi (\alpha^2 - 1)}{4k} \dfrac{ e^{(\gamma_{2k+1} - \gamma_1) \tau} }{1 + e^{-\gamma_1 \tau} } \left(1 + \dfrac{\alpha^2}{k(1 + e^{-\gamma_1 \tau})} e^{(\gamma_{2k+1} - \gamma_1) \tau} + \mathcal{O}\left(\dfrac{1}{k^2}\right) \right) \\
&= \dfrac{\pi}{2} + \dfrac{\pi (\alpha^2 - 1)}{4k (1 + e^{-\gamma_1 \tau})} e^{(\gamma_{2k+1} - \gamma_1) \tau} + \mathcal{O}\left(\dfrac{1}{k^2}\right).
\end{align*}
The final expression for $\mathcal{F}_{\mathrm{ML}}(\tau)$ is,
\begin{align*}
\mathcal{F}_{\mathrm{ML}}(\tau) = \dfrac{\pi}{2} + \dfrac{\pi (\alpha^2 - 1)}{4k (1 + e^{-\gamma_1 \tau})} e^{(\gamma_{2k+1} - \gamma_1) \tau} + \mathcal{O}\left(\dfrac{1}{k^2}\right)
\end{align*}
In the large-$k$ expansion, the fidelity for the ML method admits the form,
\begin{align*}
\mathcal{F}_{\mathrm{ML}}(\tau) = \frac{\pi}{2} + \frac{1}{k} \mathcal{B}(\tau) + \mathcal{O}\left(\frac{1}{k^2}\right),
\end{align*}
where $\mathcal{B}(\tau) = \dfrac{\pi (\alpha^2 - 1)}{4 (1 + e^{-\gamma_1 \tau})} e^{(\gamma_{2k+1} - \gamma_1) \tau}$ and $\mathcal{B} \equiv \inf_{\tau} \mathcal{B}(\tau) > 0$ is a positive constant independent of time. Given any $\epsilon > 0$, selecting $k > 2\pi^{-1} \epsilon^{-1} \mathcal{B}$ ensures the bounds,
\begin{align*}
\frac{\pi}{2} < \mathcal{F}_{\mathrm{ML}}(\tau) \leqslant \frac{\pi}{2}(1+\epsilon).
\end{align*}
The lower bound follows directly from $\mathcal{B} > 0$ and the asymptotic expansion, while the upper bound is guaranteed by the exponential decay in $\mathcal{B}(\tau)$ and the choice of $k$, which controls the magnitude of the $\mathcal{O}(1/k)$ correction term.

\section{Related Derivation of Fig. 1}    \label{appendix J}
For $\hat{H}_{t} = \hat{H} - \kappa \mathcal{I} $, its eigenvalue can be expressed as
$\lambda_0 = 0$ and $\lambda_{1} = 2\kappa = 2\xi   \sqrt{ \cos^2 \vartheta   e^{i 2 \varrho } + \frac{1}{2} \sin^2 \varrho  \sin 2\beta} = 2\xi r e^{i\theta}
$, where $\theta \in (0,\pi)$. So we can get
$\mu = \text{Re}[\lambda_{1}] = 2\xi r \cos(\theta) = \zeta  \mu'$ and $\nu = \text{Im}[\lambda_{1}] = 2\xi r \sin(\theta) = \zeta  \nu'$.
For the general two-level Hamiltonian $\hat{H}_{t}$, its ML-type bound can be expressed as
\begin{align}
    \mathcal{F}_{\text{ML}} (\mu ,\nu ,t) = \frac{\tan^2 \alpha \mu t}{e^{\nu t}+ \tan^2 \alpha}=\frac{\tan^2 \alpha \mu'  \zeta t}{e^{\nu'  \zeta t}+ \tan^2 \alpha}
    = \mathcal{F}_{\text{ML}} (\mu' ,\nu' ,t'),
    \nonumber
\end{align}
where $t'=\zeta t$.Similarly, its MT-type bounds can be expressed as
\begin{align}
    \mathcal{F}_{\text{MT}} (\mu ,\nu ,t)= \frac
    {\sqrt{
            -2 \nu t  \tan^4 \alpha
            +(2 \nu t
        +\mu^2 t^2   )e^{\nu t}\tan^2 \alpha
    }}
    {
        e^{\nu t} +  \tan^2 \alpha
    }
    = \frac
    {\sqrt{
            -2 \nu' t'  \tan^4 \alpha
            +(2 \nu' t'
        +\mu'^2 t'^2   )e^{\nu' t'}\tan^2 \alpha
    }}
    {
        e^{\nu' t'} +  \tan^2 \alpha
    }
    = \mathcal{F}_{\text{MT}} (\mu' ,\nu' ,t').\nonumber
\end{align}
For G-type bounds, we can get
\begin{align}
    \mathcal{F}_{\text{G}} (\mu ,\nu ,t)=\frac{\sqrt{\mu^2+\nu^2} }{\nu }
    \left(\arctan \left( e^{\nu t} \tan \alpha \right)-\alpha  \right)
    =\frac{\zeta\sqrt{\mu'^2+\nu'^2} }{\zeta\nu' }
    \left(\arctan \left( e^{\nu' \zeta t} \tan \alpha \right)-\alpha  \right)
    = \mathcal{F}_{\text{G}} (\mu' ,\nu' ,t'),  \nonumber
\end{align}
Therefore, we only need to represent $\mu' = \cos\theta \in (0,1)$
and $\nu' = \sin\theta \in (-1,1)$ to show the change trend of $\Delta \tau'$.
For $\mu = \zeta\mu'$, $\nu = \zeta\nu'$, the corresponding calculation result
is $\zeta$ times of the result corresponding to $\Delta \tau$, but this does
not affect the positive and negative of $\Delta \tau'$.
\end{widetext}
    

\vspace*{0.5cm}
\begin{thebibliography}{10}
    \bibitem{Agnew2014} M. Agnew, R. Eliot, J. Kevin, S. Franke-Arnold, L. Sonja, Jonathan, Discriminating single-photon states unambiguously in high dimensions, Phys. Rev. Lett. \textbf{113}, 020501 (2014).

    \bibitem{Martinez2023} D. Martinez, E. S. Gmez, J. Cari$\tilde{\text{n}}$e,  L. Pereira, A. Delgado, S. P. Walborn, Certification of a non-projective qudit measurement using multiport beamsplitters, Nat. Phys. \textbf{19}, 190-195 (2023).

    \bibitem{Cai2024} W. Cai,  J. N. Zhang, Z. Hua, W. Wang, X. Pan, X. Liu, Unambiguous discrimination of general quantum operations, Sci. Adv. \textbf{10}, eadq2529 (2024).

    \bibitem{Ashhab2012} S. Ashhab, P. C. De Groot, F. Nori, Speed limits for quantum gates in multiqubit systems, Phys. Rev. A \textbf{85}, 052327 (2012).

    \bibitem{Basilewitsch2024} D. Basilewitsch, C. Dlaska, W. Lechner, Comparing planar quantum computing platforms at the quantum speed limit, Phys. Rev. Res. \textbf{6}, 023026 (2024).

    \bibitem{Bremermann1962} H. J. Bremermann, Optimization through evolution and recombination, Self-organizing systems \textbf{93}, 106 (1962).

    \bibitem{Bremermann1967} H. J. Bremermann, Quantum noise and information, in \textit{Proc. Fifth Berkeley Symp. Math. Stat. Prob.} \textbf{4}, 15-20 (1967).

    \bibitem{Caneva2009} T. Caneva, M. Murphy, T. Calarco, R. Fazio, S. Montangero, V. Giovannetti, G. E. Santoro, Optimal control at the quantum speed limit, Phys. Rev. Lett. \textbf{103}, 240501 (2009).

    \bibitem{Giovannetti2012} V. Giovannetti, L. Maccone, Sub-Heisenberg estimation strategies are ineffective, Phys. Rev. Lett. \textbf{108}, 210404 (2012).

    \bibitem{Lloyd2002} S. Lloyd, Computational capacity of the universe, Phys. Rev. Lett. \textbf{88}, 237901 (2002).

    \bibitem{Lloyd2000} S. Lloyd, Ultimate physical limits to computation, Nature \textbf{406}, 1047-1054 (2000).

    \bibitem{Deffner2017b} S. Deffner, S. Campbell, Quantum speed limits: from Heisenberg uncertainty principle to optimal quantum control, J. Phys. A: Math. Theor. \textbf{50}, 453001 (2017).

    \bibitem{Shanahan2018} B. Shanahan, A. Chenu, N. Margolus, A. del Campo, Quantum speed limits across the quantum-to-classical transition, Phys. Rev. Lett. \textbf{120}, 070401 (2018).

    \bibitem{delCampo2013} A. del Campo, I. L. Egusquiza, M. B. Plenio, S. F. Huelga, Quantum speed limits in open system dynamics, Phys. Rev. Lett. \textbf{110}, 050403 (2013).

    \bibitem{Pires2016} D. P. Pires, M. Cianciaruso, L. C. C$\acute{\text{e}}$leri, G. Adesso, D. O. Soares-Pinto, Generalized geometric quantum speed limits, Phys. Rev. X \textbf{6}, 021031 (2016).

    \bibitem{Beau2017} G. Ness, A. Alberti, Y. Sagi, Quantum Speed Limit for States with a Bounded Energy Spectrum, Phys. Rev. Lett. \textbf{129}, 140403 (2022).

    \bibitem{Uzdin2016} R. Uzdin, R. Kosloff, Speed limits in Liouville space for open quantum systems, Europhys. Lett. \textbf{115}, 40003 (2016).

    \bibitem{Hu2020} X. Hu, S. Sun, Y. Zheng, Quantum speed limit via the trajectory ensemble, Phys. Rev. A \textbf{101}, 042107 (2020).

    \bibitem{Deffner2017} S. Deffner, Geometric quantum speed limits: a case for Wigner phase space, New J. Phys. \textbf{19}, 103018 (2017).

    \bibitem{GarciaPintos2019} L. P. Garcia-Pintos, A. del Campo, Quantum speed limits under continuous quantum measurements, New J. Phys. \textbf{21}, 033012 (2019).

    \bibitem{Zhang2018} X. M. Zhang, Z. W. Cui, X. Wang, M. H. Yung, Automatic spin-chain learning to explore the quantum speed limit, Phys. Rev. A \textbf{97}, 052333 (2018).

    \bibitem{Wu2014} S. X. Wu, Y. Zhang, C. S. Yu, H. S. Yi, The initial-state dependence of the quantum speed limit, J. Phys. A: Math. Theor. \textbf{48}, 045301 (2014).

    \bibitem{Bukov2019} M. Bukov, D. Sels, A. Polkovnikov, Geometric speed limit of accessible many-body state preparation, Phys. Rev. X \textbf{9}, 011034 (2019).

    \bibitem{Marvian2015} I. Marvian, D. A. Lidar, Quantum speed limits for leakage and decoherence, Phys. Rev. Lett. \textbf{115}, 210402 (2015).

    \bibitem{Sun2019} S. Sun, Y. Zheng, Distinct bound of the quantum speed limit via the gauge invariant distance, Phys. Rev. Lett. \textbf{123}, 180403 (2019).

    \bibitem{Okuyama2018} M. Okuyama, M. Ohzeki, Quantum speed limit is not quantum, Phys. Rev. Lett. \textbf{120}, 070402 (2018).

    \bibitem{Cimmarusti2015} A. D. Cimmarusti, Z. Yan, B. D. Patterson, L. P. Corcos, P. Lefebvre, L. A. Orozco, S. Deffner, Environment-assisted speed-up of the field evolution in cavity quantum electrodynamics, Phys. Rev. Lett. \textbf{114}, 233602 (2015).

    \bibitem{Fogarty2020} T. Fogarty, S. Deffner, T. Busch, S. Campbell, Orthogonality catastrophe as a consequence of the quantum speed limit, Phys. Rev. Lett. \textbf{124}, 110601 (2020).

    \bibitem{Mukhopadhyay2018} C. Mukhopadhyay, A. Misra, S. Bhattacharya, A. K. Pati, Quantum speed limit constraints on a nanoscale autonomous refrigerator, Phys. Rev. E \textbf{97}, 062116 (2018).

    \bibitem{Deffner2020} S. Deffner, Quantum speed limits and the maximal rate of information production, Phys. Rev. Res. \textbf{2}, 013161 (2020).

    \bibitem{DelCampo2021} A. Del Campo, Probing quantum speed limits with ultracold gases, Phys. Rev. Lett. \textbf{126}, 180603 (2021).

    \bibitem{Giovannetti2003} V. Giovannetti, S. Lloyd, L. Maccone, Quantum limits to dynamical evolution, Phys. Rev. A \textbf{67}, 052109 (2003).

    \bibitem{Liu2015} C. Liu, Z. Y. Xu, S. Zhu, Quantum-speed-limit time for multiqubit open systems, Phys. Rev. A \textbf{91}, 022102 (2015).

    \bibitem{Xu2014} Z. Y. Xu, S. Luo, W. L. Yang, C. Liu, S. Zhu, Quantum speedup in a memory environment, Phys. Rev. A \textbf{89}, 012307 (2014).

    \bibitem{Deffner2010} S. Deffner, E. Lutz, Generalized Clausius inequality for nonequilibrium quantum processes, Phys. Rev. Lett. \textbf{105}, 170402 (2010).

    \bibitem{Hamma2009} A. Hamma, F. Markopoulou, I. Prmont-Schwarz, S. Severini, Lieb-Robinson Bounds and the Speed of Light from Topological Order, Phys. Rev. Lett. \textbf{102}, 017204 (2009).

    \bibitem{GarciaPintos2022} L. P. Garcia-Pintos, S. B. Nicholson, J. R. Green, A. Del Campo,  A. V. Gorshkov, Unifying quantum and classical speed limits on observables, Phys. Rev. X \textbf{12}, 011038 (2022).

    \bibitem{Carabba2022} N. Carabba, N. H$\ddot{\text{o}}$rnedal, A. del Campo, Quantum Speed Limits on Operator Flows and Correlation Functions, Quantum \textbf{6}, 884 (2022).

    \bibitem{Mohan2022} B. Mohan, A. K. Pati, Quantum speed limits for observables, Phys. Rev. A \textbf{106}, 042436 (2022).

    \bibitem{Hornedal2022} N. H$\ddot{\text{o}}$rnedal, N. Carabba, A. S. Matsoukas-Roubeas, A. del Campo, Ultimate speed limits to the growth of operator complexity, Commun. Phys. \textbf{5}, 207 (2022).

    \bibitem{Nicholson2020} S. B. Nicholson,  A. del Campo, Time-information uncertainty relations in thermodynamics, Nat. Phys. \textbf{16}, 1211-1215 (2020).

    \bibitem{Ness2021} G. Ness, M. R. Lam, W. Alt, Observing crossover between quantum speed limits, Sci. Adv. \textbf{7}, eabj9119 (2021).

    \bibitem{Murphy2010} M. Murphy, S. Montangero, V. Giovannetti, T. Calarco, Communication at the Quantum Speed Limit Along a Spin Chain, Phys. Rev. A \textbf{82}, 022318 (2010).

    \bibitem{Mandelstam1945} L. Mandelstam, The uncertainty relation between energy and time in nonrelativistic quantum mechanics, J. Phys. (USSR) \textbf{9}, 249 (1945).

    \bibitem{Fleming1973} G. N. Fleming, A unitarity bound on the evolution of nonstationary states, Nuovo Cimento A \textbf{16}, 232-240 (1973).

    \bibitem{Vaidman1992} L. Vaidman, Minimum time for the evolution to an orthogonal quantum state, Am. J. Phys. \textbf{60}, 182-183 (1992).

    \bibitem{Anandan1990} J. Anandan, Y. Aharonov, Geometry of quantum evolution, Phys. Rev. Lett. \textbf{65}, 1697 (1990).

    \bibitem{Margolus1998} N. Margolus, L. B. Levitin, The maximum speed of dynamical evolution, Physica D \textbf{120}, 188-195 (1998).

    \bibitem{Minganti2019} F. Minganti, A. Miranowicz, R. W. Chhajlany, F. Nori, Quantum exceptional points of non-Hermitian Hamiltonians and Liouvillians: The effects of quantum jumps, Phys. Rev. A \textbf{100}, 062131 (2019).

    \bibitem{Ashida2020} Y. Ashida, Z. Gong, M. Ueda, Non-hermitian physics, Adv. Phys. \textbf{69}, 249-435 (2020).

    \bibitem{Bender2007} C. M. Bender, Making sense of non-Hermitian Hamiltonians, Rep. Prog. Phys. \textbf{70}, 947 (2007).

    \bibitem{Roccati2022} F. Roccati, G. M. Palma, F. Ciccarello, F. Ciccarello, Non-Hermitian physics and master equations, Open Syst. Inf. Dyn. \textbf{29}, 2250004 (2022).

    \bibitem{Mostafazadeh2002b} A. Mostafazadeh, Pseudo-Hermiticity versus PT-symmetry III: Equivalence of pseudo-Hermiticity and the presence of antilinear symmetries, J. Math. Phys. \textbf{43}, 3944-3951 (2002).

    \bibitem{Bender2002} C. M. Bender, D. C. Brody, H. F. Jones, Complex Extension of Quantum Mechanics, Phys. Rev. Lett. \textbf{89}, 270401 (2002).

    \bibitem{Carmichael1993} H. J. Carmichael, Quantum trajectory theory for cascaded open systems, Phys. Rev. Lett. \textbf{70}, 2273 (1993).

    \bibitem{Deffner2013} S. Deffner, E. Lutz, Quantum speed limit for non-Markovian dynamics, Physical Review Letters \textbf{111}, 010402 (2013).

    \bibitem{Campaioli2018} F. Campaioli, F. A. Pollock, F. C. Binder, K. Modi, Tightening quantum speed limits for almost all states, Physical Review Letters \textbf{120}, 060409 (2018).

    \bibitem{Cao2023} K. Cao, S.P. Kou, Statistical mechanics for non-Hermitian quantum systems, Physical Review Research \textbf{5}, 033196 (2023).

    \bibitem{Sun2021b} S. Sun, Y. Peng, X. Hu, Y. Zheng, Quantum speed limit quantified by the changing rate of phase, Physical Review Letters \textbf{127}, 100404 (2021).

    \bibitem{Cui2012} X. D. Cui, Y. Zheng, Geometric phases in non-Hermitian quantum mechanics, Phys. Rev. A \textbf{86}, 064104 (2012).

    \bibitem{Srivastav2024} A. Srivastav, V. Pandey, B. Mohan, A. K. Pati, Family of exact and inexact quantum speed limits for completely positive and trace-preserving dynamics, arXiv preprint arXiv:2406.08584 (2024).

    \bibitem{Hasegawa2023} Y. Hasegawa, Unifying speed limit, thermodynamic uncertainty relation and Heisenberg principle via bulk-boundary correspondence, Nature Communications \textbf{14}, 2828 (2023).

    \bibitem{Yadin2024} B. Yadin, Quantum Speed Limit for States and Observables of Perturbed Open Systems, Phys. Rev. Lett. \textbf{132}, 230404 (2024).

    \bibitem{Zhang2015} Y. J. Zhang, W. Han, Y. J. Xia, J. P. Cao, H. Fan, Classical-driving-assisted quantum speed-up, Physical Review A \textbf{91}, 032112 (2015).

    \bibitem{Liu2016} H. B. Liu, W. L. Yang, J. H. An, Z. Y. Xu, Mechanism for Quantum Speedup in Open Quantum Systems, Phys. Rev. A \textbf{93}, 020105 (2016).

    \bibitem{Weidemann2021} S. Weidemann, M. Kremer, S. Longhi, A. Szameit, Coexistence of dynamical delocalization and spectral localization through stochastic dissipation, Nature Photonics \textbf{15}, 576-581 (2021).

    \bibitem{Impens2021} F. Impens, F. M. dAngelis, F. A. Pinheiro, D. Gu$\acute{\text{e}}$ry-Odelin, Time scaling and quantum speed limit in non-Hermitian Hamiltonians, Phys. Rev. A \textbf{104}, 052620 (2021).

    \bibitem{Campaioli2019} F. Campaioli, F. A. Pollock, K. Modi, Tight, robust, and feasible quantum speed limits for open dynamics, Quantum \textbf{3}, 168 (2019).

    \bibitem{Wong1967} J. Wong, Results on certain non-Hermitian Hamiltonians, Journal of Mathematical Physics \textbf{8}, 2039-2042 (1967).

    \bibitem{Jing2024} Y. Jing, J. J. Dong, Y. Y. Zhang, Z. X. Hu, Biorthogonal dynamical quantum phase transitions in non-Hermitian systems, Physical Review Letters \textbf{132}, 220402 (2024).

    \bibitem{Sun2022} G. Sun, J. C. Tang, S. P. Kou, Biorthogonal quantum criticality in non-Hermitian many-body systems, Frontiers of Physics \textbf{17}, 1-9 (2022).

    \bibitem{Mostafazadeh2002} A. Mostafazadeh, Pseudo-Hermiticity versus PT symmetry: The necessary condition for the reality of the spectrum of a non-Hermitian Hamiltonian, Journal of Mathematical Physics \textbf{43}, 205-214 (2002).

    \bibitem{Brody2013} D. C. Brody, Biorthogonal quantum mechanics, J. Phys. A: Math. Theor. \textbf{47}, 035305 (2013).


    \bibitem{Levitin2009} L. B. Levitin, T. Toffoli, Fundamental limit on the rate of quantum dynamics: the unified bound is tight, Phys. Rev. Lett. \textbf{103}, 160502 (2009).

    \bibitem{Wu2022} Y. Wu, L. Kang, D. H. Werner, Generalized PT symmetry in non-Hermitian wireless power transfer systems, Phys. Rev. Lett. \textbf{129}, 200201 (2022).

    \bibitem{Hao2023} X. Hao, K. Yin, J. Zou, R. Wang, Y. Huang, X. Ma, T. Dong, Frequency-Stable Robust Wireless Power Transfer Based on High-Order Pseudo-Hermitian Physics, Phys. Rev. Lett. \textbf{130}, 077202 (2023).

    \bibitem{Bender1998} C. M. Bender, S. Boettcher, Real spectra in non-Hermitian Hamiltonians having PT symmetry, Phys. Rev. Lett. \textbf{80}, 5243 (1998).

    \bibitem{Dorey2001} P. Dorey, C. Dunning, R. Tateo, Spectral equivalences, Bethe ansatz equations, and reality properties in PT-symmetric quantum mechanics, J. Phys. A: Math. Gen. \textbf{34}, 5679 (2001).

    \bibitem{Berry2004} M. V. Berry, Physics of nonhermitian degeneracies, Czechoslov. J. Phys. \textbf{54}, 1039-1047 (2004).

    \bibitem{Heiss2012} W. D. Heiss, The physics of exceptional points, J. Phys. A: Math. Theor. \textbf{45}, 444016 (2012).

    \bibitem{Bender2003} C. M. Bender, P. N. Meisinger, Q. Wang, Finite-dimensional $\mathcal{PT}$-symmetric Hamiltonians, J. Phys. A: Math. Gen. \textbf{36}, 6791 (2003).
\end{thebibliography}
\end{document}